\documentclass [12pt,a4,sfbold,openany]{book}
\usepackage{fancyhdr}
\usepackage{graphicx}
\usepackage{anysize}
\usepackage{subfigure}
\usepackage{setspace}
\usepackage{amssymb}
\usepackage{amsmath}
\usepackage{latexsym}
\usepackage[footnotesize]{caption2}
\usepackage{color}
 \fancypagestyle{plain}{%
\fancyhf{} 
\rhead{} \cfoot{} \rfoot{}

}%

\begin{document}
\newcommand{\YBCO}{ YBa$_{2}$Cu$_{3}$O$_{7}$ }
\newcommand{\YPBCO}{ Y$_{0.6}$Pr$_{0.4}$Ba$_{2}$Cu$_{3}$O$_{7}$ }
\newcommand {\LSMO} { La$_{2 / 3}$Sr$_{1 / 3}$MnO$_{3}$ }
\newcommand{\PBCO} { PrBa$_{2}$Cu$_{3}$O$_{7}$ }
\newcommand{\cuo}{CuO$_2$ }
\newcommand{\sto}{SrTiO$_3$ }
\newcommand{\twotheta}{2$\theta$ }
\newcommand{\onebar}{$(1\overline{1}0)$}

\pagenumbering{roman}%
\begin{center} \thispagestyle{empty}
{\huge {\bf Magnetic interactions and electron transport in hole-doped manganite-superconducting cuprate heterostructures.}}

\vskip 2.5cm

\emph{A Thesis Submitted}\\
in partial fulfilment of the requirements\\
for the degree of\\
Doctor of Philosophy

\vskip 2.5cm

by\\
\large{\bf{Soumen Mandal}}\\

\end{center}
\vskip 2.5cm

\begin{center}
to the\\
\large{\bf{DEPARTMENT OF PHYSICS}}\\
\large{\bf{INDIAN INSTITUTE OF TECHNOLOGY KANPUR}}\\
\large{\bf{SEPTEMBER, 2008}}\\
\end{center} \newpage \thispagestyle{empty} \mbox{} \newpage

\chapter*{\begin{center}Certificate\end{center}}
\addcontentsline{toc}{chapter}{Certificate}%
\begin{singlespace}It is certified that the work contained in this thesis entitled
{\bf ``Magnetic interactions and electron transport in hole-doped manganite-superconducting cuprate heterostructures''} by {\bf Soumen Mandal},
has been carried out under my supervision and that this work has not been submitted elsewhere for a degree.

\vskip 3cm

\hskip 10cm Dr. R. C. Budhani

\hskip 10cm Professor

\hskip 10cm Department of Physics

\hskip 10cm Indian Institute of Technology

\hskip 10cm Kanpur

September, 2008 \end{singlespace} \newpage \thispagestyle{empty} \mbox{} \newpage


\chapter*{\begin{center}SYNOPSIS\end{center}}
\addcontentsline{toc}{chapter}{Synopsis}%
\hrule \vskip 0.1cm \begin{center}
\begin{singlespace}

{{\bf Name of Student:} Soumen Mandal  \hskip 0.5cm {\bf Roll No.:} Y3809061}\\
\vskip 0.1cm
 {\bf Degree for which submitted:} Doctor of Philosophy \hskip 0.5cm {\bf
Department}: Physics \\ \vskip 0.1cm{{\bf \noindent Title:} Magnetic interactions and electron transport in hole-doped manganite-superconducting
cuprate heterostructures.} \\
\vskip 0.1cm {\bf \noindent Thesis supervisor:} Prof. R. C. Budhani \\ \vskip 0.1cm \noindent {\bf Month and Year of submission:} September,
2008
\end{singlespace} \vskip 0.5cm
\end{center}
\hrule

\vskip 0.5cm

The interplay between superconductivity (SC) and ferromagnetism (FM) is one of the fascinating fields of research in condensed matter physics. A
variety of prominent manifestations of the interplay result from the antagonism between these two macroscopic quantum states of matter. Over the
past decade, numerous reports have appeared on experimental studies of proximity effect in hybrid systems consisting of itinerant metallic
ferromagnets and $s$-wave superconductors based bilayers and multilayers. Such artificial structures have clear advantage over naturally grown
layered crystals due to the fact the thickness of magnetic and superconducting components can be tailored individually in hybrids, which is not
possible in naturally occurring compounds. As a result, it is possible to study the interesting predictions of the theory of FM-SC proximity
effect, like pair-breaking, $\pi$-phase superconductivity, inhomogeneous condensate etc. over a broader range of functionality of the
superconducting and magnetic orders.

The oxide based doped Mott insulators, like high T$_c$ superconductor YBa$_{2}$Cu$_{3}$O$_{7}$ (YBCO) and ferromagnetic manganite \LSMO (LSMO)
are of particular interest for making SC-FM hybrids. The high degree of spin polarization of the conduction bands in LSMO and the non-trivial
Fermi surface of YBCO bring about significant change in quasiparticle transport, magnetic coupling and proximity effect in this exotic FM-SC
hybrid as compared to that seen in systems involving band ferromagnets and $s$-wave superconductors. While earlier studies on this system have
revealed some interesting phenomena, most of these works have been done on films where the \cuo planes of YBCO are parallel to the plane of the
hybrid. Since superconductivity in YBCO lies in the \cuo planes (ab - plane), such structures do not allow direct injection of spin polarized
electrons from LSMO along the nodal or fully gapped directions of the Fermi surface of YBCO. In order to overcome this difficulty, it is
necessary to grow the YBCO film with crystallographic orientation such that the \cuo planes are normal to the plane of the hybrid.

In this thesis, we present a detailed study of the magnetic anisotropy in two polytypes [(001) and (110)] of LSMO thin films. We also present a
field (H) - temperature (T) phase diagram where we have shown the pinned and depinned states for both the polytypes. A stable recipe for growing
(110) FM-SC hybrids, where the \cuo planes of YBCO are in direct contact with the sandwiching FM layers, has been developed. Magnetization
measurements on these structures reveal that the coupling between the FM layers is stronger than that seen for (001) hybrids. Anisotropic
magnetoresistance (AMR) measurement on these hybrids have shown an unusually high value ($\sim72000\%$). In the last part of the thesis we
report the dependence of MR on the sample resistance. We find that the MR of the FM-SC hybrid is directly related to the ground state resistance
of the sample. The layout of the thesis is as follows.

\textbf{Chapter 1} is an introduction to the basic concepts of superconductivity, magnetism and their interplay in hybrid structures. The
structure and electronic phase diagram of the superconducting (YBa$_2$Cu$_3$O$_{7-\delta}$) and magnetic (La$_{1-x}$Sr$_x$MnO$_3$) component of
the hybrid structures have been reviewed. Here we also discuss the magnetic anisotropy in ferromagnets and its behavior in magnetic films. This
chapter also presents a contextual survey of the current experimental results and theoretical predictions concerning ferromagnet-superconductor
heterostructures and magnetic properties of simple and complex ferromagnets.

\textbf{Chapter 2} presents a detailed layout of the various experimental procedures used in this thesis. First, we have described the method of
preparation of thin films and trilayers using pulsed laser deposition technique. Then we discuss various techniques for characterization of
these thin films. Finally, the magnetization and transport measurement techniques are briefed.

\textbf{Chapter 3} presents our studies of electron transport and magnetism in two polytypes of LSMO films. We have focussed on isothermal
magnetoresistance [R($\theta$, H)] of (001) and (110) epitaxial films measured as a function of the angle ($\theta$) between current ($\vec{I}$)
and magnetic field ($\vec{H}$), both in the plane of the film, at several temperatures between 10 and 300 K. The magnetic easy axis of these
polytypes is intimately related to the orientation of Mn - O - Mn bonds with respect to the crystallographic axis on the plane of the substrate
and energy equivalence of some of these axes. A magnetization reorientation phase transition, which manifests itself as a discontinuity and
hysteresis in $R(\psi)$, where $\psi$ is the angle between $\vec{H}$ and the easy axis for the $\vec{H}$ below a critical value $\vec{H}^*$, has
been established. The boundary of the pinned and depinned phases on the H-T plane has been established. The highly robust pinning of
magnetization seen in (110) films is related to their uniquely defined easy axis. The isothermal resistances R$_\bot$ and R$_\|$ for $\vec{I}
\bot \vec{H}$ and $\vec{I} \| \vec{H}$ respectively for both polytypes follow the inequality R$_\bot >$ R$_\|$ for all ranges of fields ($0 \leq
H \leq 3500 Oe$) and temperatures (10 - 300 K). A full fledged analysis of the rotational magnetoresistance is carried out in the framework of
D\"oring theory for MR in single crystal samples. Strong deviations from the predicted angular dependence are seen in the irreversible regime of
magnetization.

\textbf{Chapter 4} presents our work on (110) LSMO-YBCO hybrids. We first discuss the growth of YBCO--LSMO heterostructures of (110) orientation
perpendicular to the plane of the film to allow direct injection of spin polarized holes from the LSMO into the \cuo superconducting planes. The
magnetic response of the structure at T $<$ T$_{sc}$ shows both diamagnetic and ferromagnetic moments with (001) direction as magnetic easy
axis. While the superconducting transition temperature (T$_{sc}$) of these structures is sharp ($\Delta$T$_{sc} \simeq$ 2.5 K), the critical
current density (J$_c$) follows a dependence of the type $J_c = J{_o}(1-\frac{T}{T_{sc}})^{\frac{3}{2}}$ with highly suppressed J$_o$ ($\simeq 2
\times 10^4$ A/cm$^2$) indicating strong pair breaking effects of the ferromagnetic boundaries. Further studies on the trilayer reveal a strong
coupling between the FM layers. The coupling is an order of magnitude higher than that seen in the case of (001) trilayers validating an earlier
prediction. We also report an unusually high ($\sim72000\%$) angular magnetoresistance in these trilayers.

\textbf{Chapter 5} presents some key results of our work on \LSMO-Y$_{1-x}$Pr$_x$Ba$_2$Cu$_3$O$_7$-\LSMO system with x = 0.4. Here the relevance
of pair-breaking by exchange and dipolar fields, and by injected spins in a low carrier density cuprate \YPBCO sandwiched between two
ferromagnetic LSMO layers is examined. At low external field ($H_{ext}$), the system shows a giant magnetoresistance(MR), which diverges deep in
the superconducting state. We establish a distinct dipolar contribution to MR near the switching field (H$_c$) of the magnetic layers. At
H$_{ext} \gg$ H$_c$, a large positive MR, resulting primarily from the motion of Josephson vortices and pair breaking by the in-plane field, is
seen.

\textbf{Chapter 6} presents a brief summary of important results of our experiments discussed in this thesis. Here we also identify some issues
which are potentially interesting for further studies.

\newpage \thispagestyle{empty} \mbox{} \newpage

\begin{spacing}{1.3}
\chapter*{\begin{center}Acknowledgements\end{center}}\addcontentsline{toc}{chapter}{Acknowledgement}
\hspace*{14pt} This thesis is the end of my long journey to obtain a PhD degree in Physics. The journey would not have succeeded without the
cooperation and encouragement I received all along.

The first person I would like to thank is my supervisor, Prof. R. C. Budhani. As MSc-PhD dual degree student of the institute I joined him in
2002 as an MSc project student. His constant guidance, enthusiasm and mission to achieve the best quality results have direct impact on the
final form and quality of this work. I owe him lots of gratitude for having me shown this way of research. I will remain indebted to him.

My sincere thanks are due to the members of my peer review committee, Dr. Amit Dutta and Dr. Zakir Hossain, whose constant advice and
encouragement have been beneficial for my thesis in some form or another.

I wish to express my sincere thanks to Dr Prahallad Padhan and Dr. Wilfrid Prellier of Laboratoire CRISMAT, France who assisted me with four
circle X-ray measurements. Their kind cooperation and assistance have been helpful for this thesis. I am grateful to Dr. Y. Zhu and Dr. Jiaqing
He of Brookhaven National Laboratory, USA for their assistance in high resolution transmission electron microscopy studies presented in this
thesis. My sincere thanks are due to Prof. G. U. Kulkarni of Jawaharlal Nehru Centre for Advanced Research, Bangalore, who assisted me in
Magnetic Force Microscopy studies.

I am grateful to my departmental staff for always being supportive both in research and official matters. Especially, I am thankful to the staff
of liquid nitrogen and helium plants for providing cryogenics indispensable for my research work. I acknowledge financial support from IIT
Kanpur.

I would like to thank my senior lab mates Dr. Kartik Senapati, Dr. Navneet Pandey and Dr. Rajib Rakshit. They spared their valuable time and
helped me in getting acquainted with various experiments in the lab. Special thanks is due to Dr. Kartik Senapati for his endless help and
encouragement extended to me over the years. It is through his patient training that I was able to pick up the nitty gritty of instrument
interfacing. My initial years as a novice in experimental condensed matter was spent with Dr. Rajib Rakshit. I thank him for the training he
imparted on pulsed laser deposition technique.

I would like to thank all my labmates; Saumyadip, Saurabh, Mudu, Gyanu, Prasanna, Ankur, Pooran and Akhilesh for their affection, friendly
company and cooperation. The distinct colors that they present make the lab a lively place to work. I will cherish the memories of all these
years spent together as labmates. Special thanks is due to Saumyadip for his constant supply of solutions to every problem.

I warmly thank Rajeev and Ashok for their help in the lab. Ashok had been very helpful in most workshop related jobs. Rajeev had been extremely
helpful in numerous way. All the SQUID measurements presented in this thesis had been carried out by him.

On a personal note, Rajeev is an extremely good friend whose friendship I will cherish forever. The numerous hours spent with him sitting on my
hostel roof will be some of the best times I spent during my last seven years of IIT Kanpur. Our regular trips on his bike (me driving off
course) to various dhabas had been a source of endless entertainment. His sincerity and devotion to his work has made a deep impression on me. I
wish, I meet more friends like him in this journey of my life.

This journey would not have come to an end without the constant encouragement and support I received from my friends. I would like to thank
Nitin, Partha, Ravi, Snigdha, Vishal and Dipanjan for their support and encouragement. In times when I was down and out, their belief that I can
succeed helped me more than they ever would know. I consider myself extremely lucky to have friends like them. In the past seven years I have
made friends without whom my stay at IIT Kanpur would have been stale. I acknowledge Sayak, Dhriti, Jishnu, Saikat, Devdeep, Subhadip, Subhayan,
Awnish, Anupam, Deepak, Anurag, Pankaj, Vijay, Deven, Sarvesh, Neeraj, Saptarshi, Surajit da, Santosh da, Andrew, Debashish for their support
and company. I acknowledge Andrew for careful proofreading of this thesis. There are also many friends and seniors, whom I have not named here,
whose support have enabled me to overcome many setbacks. I would like to thank them for their support.

My love for food has taken me to many places in this city of Kanpur in the past seven years. Subhayan and Subhadip had been a constant company
in all such endeavors. I thank them for their company. I would also take this opportunity to thank the staff of my hostel for making my stay at
IIT Kanpur comfortable. Special thanks are due to the mess staff for the numerous meals they served day in and day out.

In the end, I would say, whatever little I have achieved would not have been possible without the blessings and good wishes of my elders. I take
this opportunity to thank all my past teachers for imparting the knowledge that became the stepping stone for this PhD. I also thank Indranil
and Anuradha Mukherjee for their blessings, good wishes and constant encouragement. I will always remain grateful to Indranil Mukherjee for his
guidance in my initial years of science education. I thank Dr. Bimal Banerjee for his blessings and good wishes.

Above everything, this PhD would would not have been possible without the constant support, encouragement, blessings and good wishes of my
parents. I would never be able to repay their unconditional support and personal sacrifices they made, which made it possible for me to
accomplish this task.\vspace*{1in} \\ \hspace*{5in} Soumen Mandal \newpage \thispagestyle{empty} \mbox{} \newpage
\end{spacing}

\chapter*{}\thispagestyle{empty}
\vskip 2cm

\begin{center}
\begin{tabular}{ll}
&{\Huge {\it To my parents...}}\\
\end{tabular}
\end{center} \newpage \thispagestyle{empty}\mbox{}\newpage

\begin{singlespace}\chapter*{\begin{center}Publications\end{center}}
\addcontentsline{toc}{chapter}{Publications}%
\begin{enumerate}
\item Soumen Mandal, Saurabh K. Bose, Rajeev Sharma, R. C. Budhani,
Prahallad Padhan and Wilfrid Prellier, \textbf{Appl. Phys. Lett. 89, 182508},
November 2006, \textit{Growth of (110) La$_{2/3}$Sr$_{1/3}$MnO$_3$ -
YBa$_2$Cu$_3$O$_7$ heterostructures}

\item Pegcheng Li, Soumen Mandal, R. C. Budhani and R. L. Greene,
\textbf{Phys. Rev. B 75, 184509}, May 2007, \textit{Correlation between incoherent
phase fluctuations and disorder in Y$_{1-x}$Pr$_x$Ba$_2$Cu$_3$O$_{7-\delta}$
epitaxial thin films from Nernst effect measurements.}

\item Soumen Mandal and R. C. Budhani, \textbf{J. Magn. Magn. Mater.
\textit{(in Press)}}, \textit{Magnetization depinning transition, anisotropic
magnetoresistance and inplane anisotropy in two polytypes of La$_{2 / 3}$Sr$_{1
/3}$MnO$_{3}$ epitaxial films.}

\item Soumen Mandal, R. C. Budhani, Jiaqing He and Y. Zhu, \textbf{Phys. Rev. B 78, 094502}, September 2008,
\textit{Diverging Giant Magnetoresistance in Ferromagnet--\linebreak Superconductor--Ferromagnet trilayers}.

\end{enumerate}\end{singlespace} \newpage \thispagestyle{empty} \mbox{} \newpage

\begin{singlespace}
\tableofcontents%
\listoffigures \newpage \thispagestyle{empty} \mbox{} \newpage
\end{singlespace}%

\chapter[Introduction]{Introduction}
\pagenumbering{arabic}%
\setcounter{page}{1}%
\section{Introduction}
The antagonism between superconductivity (SC) and ferromagnetism (FM) has been a source of fascination to numerous researchers over the past
five decades since the Bardeen - Cooper - Schrieffer (BCS) theory of superconductivity came into being. It is true that both SC and FM emerge
from strong correlation between conduction electrons, but the microscopic origin of these correlations is very different. According to Bardeen,
Cooper and Schrieffer, superconductivity arises due to the condensation of electrons into a paired state below the single electron ground state
through phonon mediated attractive interaction between electrons of opposite spin and momentum \cite{Bardeen}. The paired electrons move in a
correlated fashion to compensate for the scattering loss of each other giving rise to lossless conductivity. In a magnetic field the ``zero
spin'' paired state tends to oppose the applied field upto a strength equal to the pairing energy. Ferromagnetism, on the other hand, is the
result of parallel alignment of electron spins by exchange interaction. The problem of coexistence between these two exotic states is
essentially the competition between the pair condensation energy (2$\Delta$) of the superconductor and the Zeeman energy ($\mu_B$E$_{ex}$) of
the ferromagnet \cite{Izyumov}. While the former tries to align the spins of two electrons antiparallel to each other, the later tries a
parallel configuration. This basic difference in the spin states of charge carriers leads to the antagonism between ferromagnetism and
superconductivity. Recent developments in thin film making processes have resulted in the fabrication of a variety of FM-SC heterostructures.
Numerous studies on such systems have revealed non-trivial dependence of superconducting transition temperature (T$_c$) and critical current
(I$_c$) on the thickness of the ferromagnetic layer. Apart from this, in multilayers of FM-SC structures, various types of magnetic order are
seen in the FM layer due to its indirect interaction with other FM layers through the SC layer. These are a few examples of the numerous
exciting problems that can be addressed in FM-SC hybrids.

There are a few naturally occurring compounds which exhibit both SC and FM orders, but the absence of tunability of these systems makes them
less attractive for the study of proximity effect and other related phenomena. One such system reported few years back in which
superconductivity and magnetic order coexist is RuSr$_2$GdCu$_2$O$_8$ (Sr-1212) \cite{Bernhard}. This compound orders magnetically at 132K and
has a superconducting transition temperature of 46K. The magnetic order seen in this system is canted antiferromagnetism. Crystal structure
examination reveals that the material is a superlattice of ferromagnetic Gd layers and superconducting Cu-O planes. The other compounds which
exhibit the coexistence of FM and SC orders are ReRh$_4$B$_4$ and ReMo$_6$S$_8$ where Re is a rare earth element. For example, ErRh$_4$B$_4$ and
HoMo$_6$S$_8$ are compounds belonging to these groups with T$_{c1}$ = 8.7 K and 1.8 K, respectively. T$_{c1}$ being defined as the temperature
at which superconductivity appears in the system. The magnetic ordering temperature for these compounds are 1.0 and 0.74 K respectively. On
decreasing the temperature further to 0.8 K and 0.7 K respectively, the superconductivity disappears and a ferromagnetic phase sets in. This
temperature at which superconductivity disappears is denoted as T$_{c2}$. Apart from these, rare earth boron-nickel carbides like HoNi$_2$B$_2$C
and TmNi$_2$B$_2$C also show coexisting phases \cite{Eisaki}. In these carbides ferromagnetism exists in the Ho-C or Tm-C layers and
superconductivity in the Ni$_2$B$_2$ layers. Because of the alternating nature of these layers, the carbides can be considered as naturally
growing analogs of FM-SC hybrids. But it is to be noted that artificial heterostructures have some distinct advantages over such naturally grown
heterostructures. For example, in natural crystals the distance between the FM and SC planes is predecided and the strength of these order
parameters are also fixed. These limitations can be overcome in artificial heterostructures where there is freedom on the choice of materials
and their individual layer thicknesses. With advances in thin film growth it has become possible to grow a variety of FM-SC heterostructures
which have resulted in a spate of activities in this area of research \cite{Izyumov, Goldman, Chien, Garifullin, Keizer, Pena, Senapati}.

While talking about FM-SC heterostructures, the main issue of interest is the extent of mutual influence that these states have on each other.
It is to be noted that in most of the cases the strength of magnetic ordering (E$_{ex}$) is an order of magnitude higher than the pairing energy
of the superconductor. This means that magnetism in a ferromagnetic layer is hardly affected by the proximity of a superconducting layer but the
reverse is not true. Earlier studies by Hauser et al. \cite{Hauser} carried out on the pair breaking effects of Fe, Ni and Gd layers on a
superconducting Pb layer revealed a considerable suppression of the superconducting order parameter in all cases as compared to
superconductor-normal metal systems. The suppression of T$_c$ was analyzed in the light of the de Genner-Werthamer \cite{Guyon, Werthamer}
theory of proximity effect incorporating the Abrikosov-Gorkov \cite{Abrikosov} theory of spin scattering at magnetic impurities. Further studies
have revealed similar suppression of T$_c$ in other type-I \cite{Wong, Koorevaar, Strunk, Westerholt} and type-II \cite{Prieto, Habermeier,
Sefrioui} superconductors.

Apart from pair breaking in the superconductor, the issue of the induced superconducting order parameter in ferromagnetic regions of FM-SC
hybrids has drawn considerable interest. This happens due to the diffusion of Cooper pairs into the ferromagnet through the FM-SC interface.
Though the diffusion process of the pairs through the interface with normal metals and ferromagnets is similar, barring Andreev reflection,
which happens at the clean FM-SC interface, the nature of the induced superconductivity is different. The Cooper pairs in a ferromagnetic region
acquire a net center of mass momentum which is zero in the case of normal metals. This results in oscillations of the order parameter in the
ferromagnetic layer, with alternate regions of positive and negative order parameter known as the ``0" and ``$\pi$" phases respectively.
Theoretical calculations by Radovi\'c et al. \cite{Ledvij} have shown that under suitable conditions the oscillatory order parameter may lead to
non-monotonic variation of the superconducting properties as a function of ferromagnetic layer thickness. Various researchers have seen such
non-monotonic variation in the superconducting properties in Fe-V multilayers \cite{Wong}, Nb-Gd multilayers \cite{Strunk, Jiang}, Fe-Nb-Fe
trilayer structures \cite{Westerholt} and \LSMO - \YBCO - \LSMO trilayers \cite{Senapram} as a function of ferromagnetic layer thickness.

Another important issue in this case is the dynamics of vortices in the presence of a ferromagnetic boundary. In the past people have tried to
control the vortex dynamics by increasing the pinning properties of superconductors. The most common method is to create defects in the
superconductor where the normal cores of the vortices can be accommodated. Methods like producing columnar defects by heavy ion irradiation in
high-T$_c$ superconductors \cite{Civale, Sahoo} or making samples with embedded second phase \cite{Meingast, Haugan} to enhance pinning are
quite common. The maximum pinning strength in these cases is limited to the loss of condensation energy in the normal core of the vortices given
by $(\Phi_0/8\pi\lambda_L)^2$, where $\Phi_0$ is the flux quantum [2.07 $\times$ 10$^{-7}$ G cm$^2$] and $\lambda_L$ is the London penetration
depth. Apart from these, one can try to pin the vortices by using localized magnetic moments. In fact this method is more effective
 \cite{Bulaevskii, Temst} than creating defects in a superconductor. This is because the magnetic defects pin the magnetic flux associated with
the vortex rather than the normal core. Since the volume of the flux associated with a vortex is much higher than the normal core, pinning
efficiency of a magnetic dot is higher.

Magnetic coupling in FM-SC hybrids is a relatively new area of research. This has been mainly due to the presence of oscillatory interlayer
exchange coupling (IEC) in multilayers of ferromagnets and normal metals \cite{Schreiber, Parkin}. In case of normal metal--ferromagnet
superlattices like Co-Cr, oscillations in saturation magnetoresistance and the magnetic exchange coupling as a function of spacer layer
thickness were clearly seen. One explanation of these oscillations is Ruderman-Kittel-Kasuya-Yosida (RKKY) coupling mediated by spin the
polarization of the Cr layers. This type of mechanism leads to the conclusion that as the spacer layer thickness is decreased, a ferromagnetic
state is achieved. But experiments on another NM-FM system, Co-Ru, show antiferromagnetic ordering when the Ru thickness is reduced to a few
layers \cite{Parkin}. In the light of the above facts, it was stated that some other mechanism may be responsible for the oscillations seen in
these multilayers though RKKY was used to explain the results in most cases. Now if the normal metal in these hybrids is replaced by a
superconductor, then the RKKY mechanism responsible for the IEC in the case of metallic spacers suffers a strong suppression. This is due to the
fact that the superconductor has an energy gap at the Fermi surface that opens up below T$_c$. The problem becomes even more interesting if the
superconducting spacer is a d-wave superconductor. de Melo \cite{deMelo, Melo}, in his paper, has calculated the exchange coupling through a
$d$-wave superconductor (\YBCO) along the $c$-axis in a trilayer geometry. There, he points out that the exchange coupling along the (110)
direction of the YBCO unit cell will be much higher than that along the (001) or (100) direction because of the presence of nodes in the
superconducting order parameter of \YBCO along the (110) direction. While studying these hybrids, it is also essential to understand the
individual properties of the elements involved.

In this introductory chapter, we outline briefly some highlights of the research in the field of ferromagnet-superconductor heterostructures.
The discussion starts with an introduction of the proximity effect. This will be followed by the origin of ferromagnetism and superconductivity
in materials which constitute the heterostructures studied in this thesis. We will also discuss magnetic anisotropy in ferromagnetic material
with a brief review of the work done so far. The chapter concludes with a statement of motivation for undertaking the present work and an
outline of the remaining chapters of this thesis.
\section{Proximity effect in Ferromagnet - Superconductor heterostructures}
\subsection{Proximity effect in superconductors}
The induced superconducting correlations in a normal metal placed closed to a superconductor is known as the proximity effect. This effect was
seen for the first time in 1960 by H. Meissner \cite{Meissner}. He measured the critical current between two crossed Sn wires separated by a
coating of noble metal like Au, Ag and Cu. He was able to observe critical currents through normal metal as thick as 10 microns in some cases.
These results were later confirmed by experiments by Smith et al. \cite{Smith}. In both the cases, the thickness of the separating normal metal
was too large to explain the results by direct tunneling phenomena. They interpreted it in terms of an induced superconducting correlation known
as proximity effect. A microscopic explanation of the phenomenon was proposed later by L. N. Cooper \cite{Cooper}. He suggested that in N-S
structures the electron-electron correlation should have a dependence on the absolute position of the electrons. As a result, the Cooper pair
sees an effective potential averaged over both sides of the N-S boundary, leading to a change in the coupling constant and the energy gap of the
overall structure. However, such an interpretation is valid only for thicknesses of normal layers less than the coherence length of the
superconductor. de Gennes et al. \cite{Guyon} further generalized the idea using Gorkov's Green's function approach where they introduced the
concept of a spatially varying energy gap in N-S structures. In the same year Werthamer \cite{Werthamer} provided a set of equations which
relate the transition temperature of the S-N-S structures (T$_c$), of the superconductor (T$_{cS}$) and of the normal metal (T$_{cN}$) with the
thicknesss d$_s$ and d$_n$ of superconducting and NM spacer layers respectively;
\begin{eqnarray}
\ln\left(T_{cS}/T_c\right) & = & \chi\left(\xi_s^2k_s^2\right) \\
\ln\left(T_{cN}/T_c\right) & = & \chi\left(-\xi_n^2k_n^2\right) \\
\mbox{and} \left[N\xi^2k\tan\left(kd\right)\right]_s & = & \left[N\xi^2k\tanh\left(kd\right)\right]_n \label{brac}
\end{eqnarray} where $\chi(z) = \Psi\left(\frac{1}{2} + \frac{z}{2}\right) -
\Psi\left(\frac{1}{2}\right)$ , $\Psi$ being the Digamma function, N the density of states at the Fermi level and $\xi$ is the effective
coherence length defined as $\left(\pi\hbar\rm{k_B}\sigma/6\rm{T_ce^2}\gamma\right)^{1/2}$. $\sigma$ and $\gamma$ are the low temperature
conductivity and the coefficient of normal electron specific heat, respectively. $\hbar$ and k$_B$ are Planck's and Boltzmann's constants
respectively. k$_s$ and k$_n$ are the wave numbers in the superconducting and normal metal layers. The subscripts s and n in eq. \ref{brac}
denotes superconductor and normal metal respectively.

The extent of this induced superconducting order parameter is dependent on the quality of the interface and the mean free path of electrons
($l_n$) on the normal side of the interface \cite{Deutscher_book}. In the clean limit i.e. $l_N > \xi$, the condensation amplitude decays
exponentially on the normal side of the interface over a distance of $\hbar v_f/2\pi k_BT$, where $v_f$ is the Fermi velocity on the normal
side. In this case the Cooper pairs survive over distance equal to at least the coherence length. In the other case, when $l_N < \xi_N$, the
process is dominated by diffusion. The coherence length $\xi_N$ in this case is given by $\left(\hbar v_fl_N/6\pi k_BT\right)^{1/2}$. A number
of experiments \cite{Jin} have been performed on superconductor--normal metal heterostructures since the first discovery of the proximity effect
\cite{Meissner}.
\subsubsection{The LOFF state}
The problem of induced superconductivity becomes more interesting if the normal side of the heterostructure is magnetically ordered. In such a
case one cannot ignore the effects of exchange field (E$_{ex}$) and spin polarization on the paired state. The exchange field tends to lift the
time reversal symmetry of the paired state and hence the Cooper pair entering the ferromagnetic region gains a finite momentum, $E_{ex}/\hbar
v_F$. This change in momentum also introduces a phase change ($\delta\phi$) which is given by $\delta\phi = E_{ex}x/\hbar v_f$, where $x$ is the
distance traversed in the ferromagnet. Under such conditions, the pair wave function oscillates as a function of distance from the FM-SC
interface. This was shown by Larkin and Ovchinnikov \cite{LO} and Fulde and Ferrell \cite{FF} independently. The modulation factor is given as
$\cos\left(E_{ex}x/V_f\right), x$ being the distance from the interface \cite{Demler}. This inhomogeneous superconducting phase is commonly
cited as the Larkin--Ovchinnikov--Fulde--Ferrell (LOFF) state. The LOFF state is generally observed in a narrow region on the parametric phase
space close to the normal state \cite{LO,FF} and is limited to the exchange energy below the Chandrasekhar-Clogston \cite{Chandrasekhar,
Clogston} limit ($\Delta/\sqrt{2}$) for the upper critical field. It is to be noted that while the LOFF state is very difficult to observe in
pure bulk superconductors, an FM-SC heterostructure is promising candidate for such exotic states since the proximity effect in these structures
allows a pair correlation even when the exchange energy is larger than the gap energy.
\subsubsection{The Andreev reflection picture}
An alternative explanation to the induced inhomogeneous superconducting phase near the normal metal-superconductor interface was given by A. F.
Andreev  \cite{Andreev} and is known as Andreev reflection. In this process a quasiparticle with energy $< \Delta$ incident on the NM-SC
boundary from the N side is retro reflected into the normal side as a quasi hole and a Cooper pair travels into the superconductor \cite{deJong}
since a particle with energy less than $\Delta$ cannot enter a superconductor. The hole thus formed travels into the normal metal increasing the
conductance of the NM-SC junction. This is in contrast with the case of specular reflection of the electron where the conductance vanishes below
the gap voltage \cite{Claughton, HuJong}. The reflected hole in turn carries the macroscopic phase information of the superconductor into the
normal metal inducing a pair-like correlation in the normal carriers \cite{HuJong}. Now let us change the situation slightly and replace the NM
with a ferromagnetic metal (FM). While for a nonmagnetic NM, it does not affect the interface since all energy levels in NM are doubly
degenerate with respect to the spin direction, however for FM this degeneracy is lifted due to the interaction of the spin with the spontaneous
moment of the ferromagnet. In the Andreev reflection process, the incident electron and the reflected hole have opposite spins. This change in
spin direction of the particle entering into the ferromagnet can lead to drastic changes in the properties of the FM-SC interface. Many
researchers have calculated the amplitudes of Andreev reflection probabilities at an FM-SC junction \cite{deJong, Zheng, Yamashita}. But what is
important here is to calculate the oscillating order parameter and see how it compares with the LOFF formalism.

Fominov et al. \cite{Fominov} have shown that this can be done by Feynman path integral formalism where the trajectories of quasiparticle pair
wave function in a ferromagnet in contact with a superconductor can be assumed to be consisting of both Andreev reflected and normal reflected
paths. They found an oscillatory behavior of the pair wave function, which can be written as
\begin{equation}F(x) \propto \cos\left(\frac{2E_{ex}d_f}{v_F}\right)\cos
\left(\frac{2E_{ex} (d_f + x)}{v_F}\right) \end{equation} which is similar to the expression obtained in the LOFF formalism \cite{Demler}.

The oscillating order parameter is responsible for stabilizing some regions of negative order parameter, known as the $\pi$--phase, inside the
ferromagnet, which leads to various exotic phenomena in FM-SC hybrids. For example Bulaevskii et al. \cite{Bulaevski1} have shown that the
Josephson current through an SC-FM-SC junction acquires a sign opposite to $\sin\phi$, where $\phi$ is the phase difference between the two
superconductors. The $\pi$--phase superconductivity has been experimentally verified by various experiments \cite{Ryazanov, Kontos, Shelukhin}.
\subsection{Transition temperature in FM-SC hybrid structures}
The suppression of the superconducting transition temperature in a FM-SC hybrid has been studied extensively \cite{Hauser, Jiang, Wong, Lazar,
Tagirov}. The earliest experiments on T$_c$ suppression in Pb-Cr system were explained by Hauser et al. \cite{Hauser} by the de
Gennes--Werthamer \cite{Guyon, Werthamer} theory of proximity effect along with the Abrikosov--Gorkov \cite{Abrikosov} model of Cooper pair
scattering at localized magnetic moments. They calculated the variation of T$_c$ by using a refined version of the original de Gennes
\cite{Guyon} equation, which did not describe a FM layer superimposed on a superconductor, and derived the following equation \cite{Hauser,
Hauser1};
\begin{equation} \ln\left(\frac{T_{cS}}{T_c}\right) =
\chi\left(\frac{\pi^2\xi_s^2}{4d_s^2}\right) \end{equation} Here $\xi_s$ and $d_s$ are the coherence length and thickness of the superconductor
respectively and $\chi(z) = \Psi\left(\frac{1}{2} + \frac{1}{2}z\right) - \Psi\left(\frac{1}{2}\right)$, $\Psi$ being the digamma function. It
is to be noted that this equation is independent of the properties of the magnetic layer. They noted that the relation remains unchanged when
the degree of dirtiness of the superconductor (scattering time $\tau_s$) is changed. Jiang et al. \cite{Jiang}, for Nb/Gd system, Wong et al.
\cite{Wong} for Fe/V system and Lazar et al. \cite{Lazar} for Pb/Fe system have reported an oscillating critical temperature as the FM layer
thickness is varied. Tagirov et al. \cite{Tagirov} reported a re-entrant behavior of the SC transition temperature in Fe--V--Fe trilayers at
V-thickness close to the critical thickness for superconductivity. In the Fe--V system studied by Wong et al. \cite{Wong}, the data were
analyzed by employing Ginzburg-Landau equations for the order parameter in a 2D film with the boundary condition $\psi = 0$ ($\psi$ is the order
parameter) at the interface and obtained the relation $(T_{c0} - T_c) \propto (1/d^2)$ was obtained, where d is the thickness of the
superconducting vanadium layer and T$_c$ and T$_{c0}$ are transition temperatures of the superlattice and a thick Vanadium film respectively. In
all the above experiments, the destructive effect of spin polarized conduction electrons of the ferromagnet was recognized.

The starting point of many theoretical approaches for the calculation of T$_c$ in hybrid films with dirty superconductors has been the Usadel
equations \cite{Usadel}, which are a simplification of the Eilenberger \cite{Eilenberger} equations in the dirty limit. Radovi\'c et al.
\cite{Ledvij} in their work solved the Usadel equations to calculate the transition temperature of FM-SC superlattices as a function of
thickness of the superconducting and magnetic layers. For the case of FM-SC-FM trilayers the transition temperature was given by;
\begin{equation} \ln\left(\frac{T_c}{T_{cS}} \right) = \rm{Re}\Psi\left(\frac{1}{2}
+ \frac{k_s^2\xi_s^2T_{cS}}{2T_c} \right) - \Psi\left(\frac{1}{2}\right)
\end{equation} where $\Psi$ is the digamma function, $\xi_s$ the coherence length
and $k_s$ the propagation momentum calculated using the transcendental equation;
\begin{equation} k_sd_s\tan(k_sd_s/2) = \frac{k_f\xi_s\sigma_f}{\sigma_s}\left(
\frac{d_s}{\xi_s}\right) \tanh(k_fd_f)
\end{equation} where $\sigma, k$ and $d$ are the normal state conductivity, wave number
and thickness of the ferromagnetic and superconducting components indexed as `f' and `s' respectively. The boundary conditions in these
calculation were assumed in the limit of perfect transparency of the FM-SC interface leading to conditions;
\begin{eqnarray}F_s & = & F_f  \\ \mbox{and   } \frac{dF_s}{dx} & = &
\eta\frac{dF_f}{dx}\end{eqnarray} where $F_f$ and $F_s$ are the quasiclassical Usadel Green's functions representing pair correlations on the
ferromagnetic and superconducting sides of an interface along the yz--plane. Subsequently many experimental results on the transition
temperature of hybrids were analyzed using this theory \cite{Koorevaar, Strunk, Jiang, Mercaldo}. It should be noted that the Radovi\'c
\cite{Ledvij} theory emerges from the competing `0' and `$\pi$' phase order parameters between two consecutive SC layers. Hence, bilayer and
trilayer FM-SC structures do not fall under the purview of this theory.

Subsequent works by Khusainov \cite{Khusainov}, Tagirov \cite{Tagirov1} and Fominov \cite{Fominov} used more realistic boundary conditions for
the quasiclassical Green's function \cite{Zaitsev, Kupriyanov, Lambert}. In these later studies, one sees a discontinuous jump of the pair wave
function at the FM-SC boundary in contrast with the continuous transition in the Radovi\'c theory. This discontinuity stems from the fact that
the diffusion constants in the FM and SC layers are different and are discontinuous at the interface. This formulation \cite{Fominov, Khusainov,
Tagirov1} is independent of any phase coupling between two superconducting layers as required by Radovi\'c theory \cite{Ledvij}.

For FM-SC-FM trilayers, the transition temperature is also dependent on the relative orientation of the magnetization vectors in the sandwiching
ferromagnetic layers. It has been observed that the transition temperature for antiferromagnetically oriented trilayers is higher than that of
ferromagnetically oriented films \cite{deGennes2, Buzdin2, Deutscher, Hauser2, You}. This effect was theoretically pointed out by de Gennes
\cite{deGennes2} and verified later experimentally on ferromagnet--insulator--superconductor--insulator--ferromagnet (F/I/S/I/F)
\cite{Deutscher}, ferromagnetic insulator--superconductor--ferromagnetic insulator \cite{Hauser2} and ferromagnet--superconductor--ferromagnet
\cite{You} hybrids. Here it is to be noted that the effect of magnetization orientation is considerable only when the thickness of the
superconducting spacer is of the order of the coherence length $\xi_s$.

While a considerable amount of theoretical work has been done to understand the properties of FM--SC--FM hybrid structures, in all these studies
the order parameter of the superconductor is assumed to be $s$-wave. For understanding proximity effects in high-T$_c$ cuprates it is essential
to work out these problems with an anisotropic order parameter ($d_{x^2 - y^2}$ wave or anisotropic $s$-wave). The other problem concerning
FM-SC-FM structures with high-T$_c$ cuprates is that the SC in this case is in the clean limit. So in this case Usadel's \cite{Usadel}
simplification of Eilenberger \cite{Eilenberger} equations may not apply. Nonetheless, these results can be used qualitatively to analyze data
involving FM-SC-FM structures of \YBCO since in this case $l_s/\xi_s \sim 1$.
\section{Magnetic coupling in FM-SC heterostructures}
The issue of exchange coupling between two FM layers separated by a normal metal spacer such as Cr has been studied extensively \cite{Schreiber,
Parkin}. For metallic spacers, the exchange coupling is driven by conduction electrons, hence it is dependent on the shape of the Fermi surface
along the direction of growth. The exchange coupling is an oscillatory function of the spacer layer thickness similar to the RKKY-interaction
between localized moments. A fundamental question to ask is ``What will happen to IEC given that it is a Fermi surface phenomenon, when the
spacer material undergoes superconducting transition and a gap opens up at the Fermi surface?''  This issue has been addressed theoretically by
\v{S}ipr and Gy\"{o}rffy \cite{Sipr}. For an $s$-wave superconductor, they found a strong suppression of exchange coupling at T $<$ T$_c$.
However, if the superconductor in the spacer has an anisotropic pair wave function like in the case of high-T$_c$ cuprates, then low and zero
energy quasiparticle excitations are possible. Based on these ideas, de Melo \cite{deMelo, Melo} proposed a theory of magnetic coupling across a
high-T$_c$ superconductor in FM-SC-FM trilayers. This theory \cite{Melo} reveals that even for high-temperature superconductor (HTSC) spacers,
an oscillatory magnetic correlation exists along the z-direction of the spacer as a function of the spacer thickness. The magnitude of the IEC
is found to be higher in the case of a $d$-wave superconductor as compared to the case of an $s$-wave superconductor. On the experimental side,
recent works by Senapati et al. \cite{Senapati, Senapram, Senapati1} have revealed some remarkable properties of \LSMO-\YBCO-\LSMO trilayers
with varying FM and SC thicknesses and temperatures. A more detailed description of the work on these trilayers by various groups will be
discussed in later section. Here it is to be noted that the discussion on IEC is not complete without a short description of the RKKY model or
the Quantum well model. In the following subsections, brief descriptions of these models are given.
\subsection[RKKY model]{Ruderman-Kittel-Kasuya-Yosida model}
The discovery of periodic oscillations in the coupling between two FM layers in Fe-Cr-Fe and Co-Ru-Co \cite{Parkin} as a function of the
nonmagnetic layer thickness initiated considerable interest in the problem of interlayer exchange coupling. The oscillatory behavior seen in
this case bears resemblance to the Ruderman-Kittel-Kasuya-Yosida (RKKY) interactions seen between two magnetic impurities. So, this model is an
excellent candidate for explaining the experimental data. But this theory when applied in its simplest form predicts a period of oscillation
approximately equal to one monolayer \cite{Yafet}, which is much shorter than the experimental observations. A general theory for IEC based on
the RKKY mechanism was given by Bruno et al. \cite{Brunorkky}.

Let us consider two FM layers (F1 \& F2) separated by a nonmagnetic spacer. Initially it is assumed that the FM layers are monolayers and only
the atoms near the interface interact with each other. The magnetic layers are assumed to consist of spins $\vec{S_i}$ located on atomic
positions $\vec{R_i}$ of the spacer near the interface, which essentially means that the spacer and the FM layers are coherent. The RKKY
interaction between two such spins can be written as \cite{Ruderman}
\begin{equation}H_{ij} = J(\vec{R_{ij}})\vec{S_i}.\vec{S_j} \end{equation} where $J(\vec{R_{ij}})$ is the exchange integral. The IEC is calculated
by summing $H_{ij}$ over all $i,j$. The coupling energy per unit area can be written as \begin{equation}E_{1,2} = I_{1,2}\cos\theta_{1,2}
\end{equation} where $\theta_{1,2}$ is the angle between the magnetization vectors of F1 and F2 and $I_{1,2}$ is the coupling constant given by
\begin{equation}I_{1,2} \sim \sum_{j\in F2} J(\vec{R_{0j}})\end{equation} where the label `0' defines a site in F1 which is taken as the origin.
Under the present sign convention, a positive value of $I_{1,2}$ corresponds to antiferromagnetic coupling. Under the free-electron
approximation of Yafet \cite{Yafet}, $I_{1,2}$ as a function of the distance $z$, assuming a continuous uniform distribution of spins in the
magnetic layer, is given by \begin{equation}I_{1,2}(z) \sim \frac{d^2}{z^2} \sin(2k_Fz), \rm{for\hspace{0.2cm}} z \rightarrow \infty
\end{equation} where $d$ is the spacing between the atomic planes of the spacer. This essentially means that the period of oscillation is $\lambda_F/2$
and the oscillation decays as $z^{-2}$. However, this free-electron approach does not yield the correct results for the oscillation period. A
more detailed calculation for IEC based on the RKKY mechanism which correctly predicts the period of oscillations can be found in
\cite{Brunorkky}.
\subsection{Quantum well model}
This model starts with a simple assumption that the interfaces of an FM-NM hybrid act as a potential capable of trapping electrons in the spacer
layer. The itinerant nature of the electrons in transition metals gives rise to spin-split band structure. So this will give rise to a spin
dependent reflectivity at the FM-NM interface. Let us consider a trilayer of FM-NM-FM with the magnetizations $\vec{M_1}$ and $\vec{M_2}$ of the
FM layers parallel to each other. In such a case electrons with parallel (antiparallel) spins will be weakly (strongly) reflected at the FM/NM
interface. This spin dependent reflection will give rise to quantum well states where one will see the emergence of standing waves for
particular spacer thicknesses. This also cause a spin-polarization in the spacer layer, since the spins parallel to the M vectors are free to
move in the entire stack but the antiparallel spins are confined to the spacer layer. The spin dependent interference effects seen here are used
to explain the oscillations in the exchange coupling.

The predominant contribution to the coupling comes from electrons with wave vectors $\vec{Q_i}$, defined as the vector perpendicular to the
interface connecting two sheets of the Fermi surface parallel to each other. The oscillation period in this case is given by $\Gamma = 2\pi/Q_i$
\cite{Stiles}. This essentially means that the coupling oscillations are determined by the electronic properties of the spacer layer. If we
define $J_1^i$ as the coupling constant due to electrons with wave vector $\vec{Q_i}$, then $J_1^i$ for spacer thickness $z$ is given by
\cite{Stiles} \begin{equation}J_1^i = J_0^i\sin\left(Q_iz + \phi_i\right)/D^2 \end{equation} Here $J_0^i$ includes the Fermi surface geometry
factor as well as the reflection probabilities. Experimentally, quantum well states have been observed by inverse photoemission on Cu/Co and
Ag/Fe hybrids \cite{Ortega}.
\section{Magnetic anisotropy in magnetic thin films}
The dependence of the internal energy of a magnetic material on the direction of its spontaneous magnetization results in magnetic anisotropy.
This phenomenon can be divided into various subcategories like uniaxial anisotropy, interface and volume anisotropy, single-ion anisotropy,
shape anisotropy, exchange anisotropy and magnetoelastic anisotropy. We will briefly discuss all these types of magnetic anisotropies in the
following subsections.
\subsection{Uniaxial Anisotropy}
The most common anisotropy effect is connected to the existence of only one easy direction of magnetization, and in the literature it is
referred to as uniaxial anisotropy. Let us consider a uniaxial ferromagnetic crystal in the absence of any external magnetic field. The
magnetocrystalline anisotropy energy, $E_k$, can be written in terms of $\sin\theta$ \cite{Asti}, $\theta$ being the angle between the
magnetization vector, $\vec{M_S}$ and the symmetry axis, $\vec{c}$. \begin{equation}E_k = K_1\sin^2\theta + K_2\sin^4\theta + K_3\sin^6\theta +
..... \label{uni}\end{equation} where $K_1, K_2, K_3, ....$ are the anisotropy constants having the dimensions of energy per unit volume
(J/m$^3$). The easy and hard direction are determined by extremizing eq. \ref{uni}. A minima in $E_k$ will mean easy axes while maxima will
define hard axes. \clearpage
\subsection{Interface and Volume Anisotropy}
Interface anisotropy is most common when we consider the case of thin film superlattices. In these films, the anisotropy comes from both the
bulk and the interface of the system. If we denote $K_s$ as the anisotropy originating from the interface per unit area and $K_v$ as the
contribution to the anisotropy per unit volume, the effective anisotropy can phenomenologically be described as \cite{Draaismaa, Engel}
\begin{equation}K_{eff} = K_v + \frac{2K_s}{t}\end{equation} Here $t$ denotes the ferromagnetic layer thickness and the
factor of 2 arises from the layer being bounded by two interfaces. The volume anisotropy $K_v$ consists of magnetostatic or demagnetization
energy, magnetocrystalline anisotropy and magnetoelastic energy. In bulk systems, the magnetocrystalline anisotropy of a system is dominated by
the volume term. In thin films and multilayers, however, the surface term can become more significant as $ t$ becomes small ($\sim
2K_{S}/K_{V}$). In most cases, the anisotropy in thin magnetic layers is dominated by the dipolar shape anisotropy, favoring in-plane moment
alignment.
\subsection{Single-ion Anisotropy}
Single-ion anisotropy (often referred to simply as ``magnetocrystalline anisotropy'') is caused by the spin-orbit interaction of the electrons.
The electron orbits are linked to the crystallographic structure and by their interaction with the spins they make the latter prefer to align
along well-defined crystallographic axes. This interaction is transferred to the spin moments via the spin-orbit coupling, giving a weaker
$d$-electron coupling of the spins to the crystal lattice. When an external field is applied the orbital moments may remain coupled to the
lattice whilst the spins are more free to turn. The magnetic energy depends upon the orientation of the magnetization relative to the crystal
axes. In a magnetic material, single-ion anisotropy is present in the entire volume contributing to $K_v$. The crystal orientation of the layer
decides whether this contribution will add up or subtract from $K_v$. Magnetocrystalline anisotropy is an important factor in deciding whether a
magnetic material can be made into a good hard magnet or a good soft magnet. While in transition metals this contribution is generally much
smaller than the shape anisotropy, in rare earth metals it can be comparable in magnitude.

For the calculation of the magnetocrystalline anisotropy constants, the starting point is the determination of the free energy density. For a
saturated crystal (i.e. a single domain), the free energy density ($F_d$) may be expressed as a series expansion in ascending powers of the
direction cosines $\alpha_i$ of the magnetization with respect to a set of rectangular Cartesian coordinate axes \cite{Darby};
\begin{equation} F_d = b_i\alpha_i + b_{ij}\alpha_i\alpha_j + b_{ijk}\alpha_i\alpha_j\alpha_k + ......... \end{equation} where
$b_i, b_{ij}, b_{ijk},...$ are the property tensors whose forms are dictated entirely by the requirements of the crystallographic and intrinsic
symmetry. It can be shown that, for cubic crystals, $F_d$ must have the form approximated to the fourth term as
\begin{equation}F_d = K_0 + K_1 \left(\alpha_1^2\alpha_2^2 + \alpha_2^2\alpha_3^2 +
\alpha_3^2\alpha_1^2 \right) + K_2 \left(\alpha_1^2\alpha_2^2\alpha_3^2 \right) + K_3 \left(\alpha_1^2\alpha_2^2 + \alpha_2^2\alpha_3^2 +
\alpha_3^2\alpha_1^2\right)^2 \end{equation} in which the $\alpha_i$ are referred to the cube edges and $K_1 , K_2$ and $K_3$ are called the
first, second, and third anisotropy constants, respectively. Putting the values of the $\alpha_i$s, one can find a relation between the free
energy and the angle between magnetization direction and the crystallographic axis. Expansions of the anisotropy energy appropriate to each of
the 32 crystallographic point groups have been derived and tabulated by D\"oring \cite{Doring1}.
\subsection{Shape Anisotropy}
\begin{figure}
\centerline{\includegraphics[height=4in]{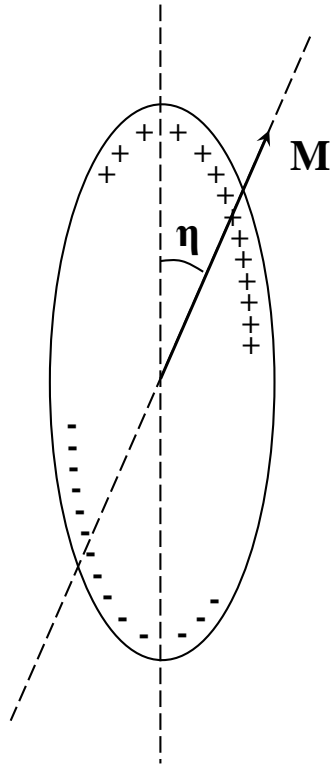}} \caption{Magnetized prolate spheroid.} \label{shapean} \end{figure}

The polarization in a long body is preferably aligned along the major axis of the object. This phenomenon is known as shape anisotropy. It is
mediated by dipolar interaction and is a long range effect. As a result it is predominantly dependent on the shape of the body. The simplest
example would be that of a magnetized prolate spheroid (Fig. \ref{shapean}).  Let the spheroid
be homogenously magnetized along a direction that makes an angle $\eta$ with the major axis as shown in fig. \ref{shapean}. The energy, $E$, of
the demagnetizing field for this case can be written as \cite{Zijlstra} \begin{equation}E = \frac{\mu_0}{2}\left\{ N_\|M^2 + \left( N_\bot -
N_\| \right)M^2\sin^2\eta \right\}V \label{spheroid} \end{equation} where $N_\|$ and $N_\bot$ are the demagnetization tensors parallel and
perpendicular to the major axis, respectively. The demagnetization factors perpendicular to the major axis will be identical. $M$ is the
magnetization and $V$, the volume of the spheroid. The second term in eq. \ref{spheroid} describes the shape anisotropy. For small values of
$\eta$ the effective anisotropy field is given by \begin{equation}H_A = \left(N_\bot - N_\|\right)M \end{equation} In the case of thin films,
all the tensor elements other than $N_\bot$ are zero. $N_\bot = 1$ for thin films. Hence the anisotropy energy contribution is equal to
\cite{Johnson} \begin{equation}E = \left\{ \frac{1}{2}\mu_0M_s^2\cos^2\eta \right\}V \end{equation} Here the magnetization is assumed to be
uniform with a magnitude equal to the saturation magnetization $M_s$, and subtends an angle $\eta$ with the film normal. V is the volume of the
film. According to this expression, the contribution favors an in-plane preferential orientation for the magnetization.
\subsection{Exchange Anisotropy and Exchange Interaction}
Exchange anisotropy arises from the exchange coupling between two magnetically ordered systems placed in proximity to each other. It was
discovered first by Meiklejohn and Bean in 1956 \cite{Meiklejohn, Meiklejohn1}. More specifically, it is seen in sandwiches where ferromagnetic
material and antiferromagnetic material are placed in close proximity to each other. The exchange interaction between the materials of different
magnetic phases gives rise to exchange anisotropy. When a hybrid FM-antiferromagnet is cooled in a static field from a temperature above the
N\'eel temperature (T$_N$) but below T$_c$ of FM (T$_N <$ T $<$ T$_c$), the hysteresis loop of the system below T$_N$ is shifted in the field
axis in the opposite direction with respect to the cooling field. The shift thus observed is known as the exchange bias. The coercive field of
the material also goes up after field cooling. These effects disappear once the temperature approaches the N\'eel temperature of the
antiferromagnet confirming that this effect is due to the presence of antiferromagnet material. An intuitive argument to explain this anisotropy
can be given as follows. Consider an FM-antiferromagnet couple at a temperature below T$_c$ but above T$_N$. The FM layer in the presence of a
field will be ferromagnetically aligned but the antiferromagnet layer will have a random orientation of its spins. Once the temperature is
cooled below T$_N$, the spins close to the interface will ferromagnetically align with respect to the FM spins and the other spin planes in the
antiferromagnet will follow the antiferromagnetic alignment with respect to the interface spins. Now, when the field is scanned at a particular
temperature, on reversing the field and for sufficiently high spin stiffness of antiferromagnet, the antiferromagnet spins remains unchanged but
the FM spins start experiencing a finite torque from the external applied field to align themselves in the field direction as well as a
microscopic torque from the interface spins of the antiferromagnet layer. As a result the spins in the FM layer align parallel to the applied
reverse field only when the torque applied by it is larger than that applied by the interface spins of antiferromagnet. By the same line of
argument, when the field is again reversed to its original direction, then the FM spins get aligned parallel to interface spins at a lower field
because of their interaction with interface spins. From these arguments it is clear that there is only one stable configuration for this couple
once the direction of AF alignment is decided. Hence this is also known as unidirectional anisotropy unlike the uniaxial anisotropy which has
energy minima at position 180$^o$ to each other. Experimentally this effect was observed for the first time by Meiklejohn and Bean
\cite{Meiklejohn, Meiklejohn1} in a nominal sample of Co nanoparticles. Later it was established that a fraction of nanoparticles were converted
to CoO which is antiferromagnetic, meaning that the particles had a ferromagnetic core of Co with an antiferromagnetic CoO shell.

The phenomenon responsible for aligning spins in a ferromagnetic material is known as exchange interaction. If we consider two spins $S_i$ and
$S_j$ on sites $i$ and $j$, the exchange energy will be proportional to $\vec{S_i} . \vec{S_j}$. The exchange energy in this case can be written
as \cite{Aharoni} \begin{equation}E = -{\sum_{i\neq j}}J_{ij}\vec{S_i}.\vec{S_j} \label{exchange}\end{equation} where $J_{ij}$ is the exchange
integral and it is positive for parallel alignment of $\vec{S_i}$ and $\vec{S_j}$. The exchange energy for neighboring spins is dependent only
upon the angle between them hence it does not give rise to anisotropy. But this is not the case when we consider magnetic multilayers separated
by non-magnetic spacers. In this case, the exchange energy term will have two parts: an isotropic exchange coupling and an anisotropic
Dzialoshinski-Moriya exchange coupling \cite{Xia}. The exchange energy in this case will be written as
\begin{equation}E_{ex} = -2J(z) \vec{M_1}.\vec{M_2}\end{equation} where $ J(z)$ is the
exchange coupling constant, $\vec{M_1}$ and $\vec{M_2}$ are the magnetization of the adjacent magnetic layers and $z$ is the non-magnetic layer
thickness. This term will have a maximum when the magnetization vectors are aligned parallel to each other. The anisotropic term of the exchange
energy will be \cite{Xia} \begin{equation} E_{aniso} = J_{DM}(z). \left(\vec{M_1} \times \vec{M_2}\right)
\end{equation} where $J_{DM}$ is the Dzialoshinski-Moriya
exchange constant. The cross product will result in a vector perpendicular to the layer magnetization. For positive $J_{DM}$, we will have
in-plane moment alignment while negative $J_{DM}$ will give rise to out-of-plane alignment.
\subsection{Magnetoelastic Anisotropy}
Magnetoelastic anisotropy arises from the change in lattice parameters due to strain in the lattice. We know that the spin moments are coupled
to the lattice via the orbital motion of electrons. Now if the distances between the magnetic atoms are altered due to strain, then the lattice
interaction energies are also changed and hence cause the anisotropy. In order to understand the dependence of magnetization on stress, let us
consider a stress of $\sigma Nm^{-2}$ acting on a ferromagnetic body. Let the direction cosines of this tension be
$(\gamma_1,\gamma_2,\gamma_3)$ and let $\theta$ be the angle between the magnetization and tensile stress direction. For the isotropic
magnetostriction we can write the magnetoelastic energy as \cite{Chikazumi} \begin{equation}E_\sigma = -\frac{3}{2}\lambda\sigma\cos^2\theta
\label{magela} \end{equation} where $\lambda$ is the magnetostriction constant. So when the magnetization rotates, the magnetoelastic energy
changes producing a kind of uniaxial anisotropy. In thin films, large mismatches between the lattice parameters of the substrate and the
material of the thin film can alter the easy axis of magnetization. For example, if $\lambda$ is positive, as in a metallic iron, the easy
magnetic direction or the direction of minimum energy will be along a direction of tensile stress.
\subsection{Anisotropic magnetoresistance}
Anisotropic magnetoresistance is defined as the difference in resistivity of a magnetic material depending on whether the sample magnetization
is parallel ($\rho_\|$) or perpendicular ($\rho_\bot$) to the current direction. In most cases $\rho_\| > \rho_\bot$ but exceptions are
plentiful \cite{Prinzjap, Prinzprb, Gorkom}. The resistivity varies continuously with the angle $\theta$ between the magnetization direction and
the current direction as \cite{Dienybook}
\begin{equation}\rho(\theta) = \frac{\rho_\| + \rho_\bot}{2} + \left[\cos^2\theta - \frac{1}{2} \right]\left(\rho_\| - \rho_\bot\right)
\label{rhotheta} \end{equation} Now the zero field resistance ($\rho_0$) for a truly randomly demagnetized sample is the statistical average of
the resistivity calculated over all directions of the magnetization with respect to the current direction and is given by \cite{Dienybook}
\begin{equation}\rho_0 = \rho_{av} = \frac{\int\limits_0^\pi {\rho(\theta)\sin\theta d\theta}}{\int\limits_0^\pi {\sin\theta d\theta}} =
\frac{\rho_\| + 2\rho_\bot}{3} \end{equation} So the anisotropic magnetoresistance (AMR) of a sample, defined as $\Delta\rho/\rho_{av}$, can be
written as \begin{equation} \frac{\Delta\rho}{\rho_{av}} = \frac{\rho_\| - \rho_\bot}{\frac{1}{3}\rho_\| + \frac{2}{3}\rho_\bot}
\end{equation} \clearpage \noindent But for simplicity in this definition the denominator is replaced by $\rho_\bot$ \cite{Velev}. On simplification, eq. \ref{rhotheta}
yields \begin{eqnarray}\rho(\theta) & = & \rho_\bot + \left(\rho_\| - \rho_\bot\right)\cos^2\theta \nonumber \\ \Rightarrow \rho(\theta) & = &
\rho_\bot + \Delta\rho\cos^2\theta  \end{eqnarray} This relation is the fundamental relation defining the magnetoresistance of the sample with
respect to the angle between the applied field and the current direction. Though in general one should see a $\cos^2\theta$ dependence in AMR,
it should be noted that deviations from this behavior are also observed \cite{Prinzjap, Prinzprb, Gorkom}.
\subsection{Experimental studies of magnetoresistance in simple ferromagnets}
Experimental studies in simple ferromagnets were carried out as early as 1857 by William Thomson when he discovered anisotropic
magnetoresistance in ferromagnetic metals \cite{Thomson}. But it was not until a century later that this effect was used in the recording
industry. Although there have been a variety of studies on simple ferromagnets, we will focus only on the anisotropy aspect of the problem. In
1951 Smit \cite{Smit} reported anisotropic magnetoresistance in iron. He could see a distinct difference in the resistance of the sample when
the applied field was rotated from a direction parallel to the current to one that was perpendicular. He attributed this anisotropy to spin
orbit coupling between the 4$s$ and 3$d$ electronic states. Van Elst \cite{Elst} measured the MR-anisotropy of a large number of nickel alloys
and found that in the alloys with V, Cr, Mo and W, the relative magnetoresistance anisotropy($\Delta\rho/\rho$) was smaller than in alloys with
Cu and Mn. The smaller $\Delta\rho/\rho$ has been attributed to the smearing out of the 3$d$ orbitals of the Ni atom due to the addition of non
magnetic components. Here $\Delta\rho/\rho$ is defined as; \begin{equation}\frac{\Delta\rho}{\rho} = 2 \left(\frac{\rho_{\|,0} - \rho_{\bot,0}}
{\rho_{\|,0} + \rho_{\bot,0}} \right)\end{equation} where $\rho_{\|,0}$ and $\rho_{\bot,0}$ are the values obtained by extrapolating to H=0 the
electrical resistances in parallel ($\rho_{\|}$) and transverse ($\rho_{\bot}$) strong magnetic fields.

These works of Smit \cite{Smit} and Van Elst \cite{Elst} marked the beginning of a period of many new experimental and theoretical studies
related to anisotropic magnetoresistance. Dahlberg et al. \cite{Prinzjap} showed how simple magnetotransport measurements can be used to
determine various anisotropy energies of epitaxial ferromagnetic films. They have studied the case of Fe film grown on (110) GaAs using
molecular beam epitaxy in great depth. In a recent work, van Gorkom et al. \cite{Gorkom} have measured the magnetization angle dependent
resistance of (110) Fe films grown on (11$\overline{2}$0) sapphire. They analyzed their data using D\"oring's \cite{Doring} formalism and found
that the dependence of MR on the magnetization direction is not just a simple $\cos^2\psi$ where $\psi$ is the angle between $\vec{M}$ and the
easy axis but rather higher order terms like $\cos^4\psi$ also contribute to the angular dependence. Some of the earliest works on anisotropy in
Ni thin films were those of Coren et al. \cite{Coren} and Marsocci \cite{Marsoccijap, Marsoccipr}. Coren et al. \cite{Coren} tried to reveal the
magnetic domain structure and anisotropy using magnetoresistance studies. They concluded that the anisotropy in their measurements was more or
less related to strain in their films. Marsocci \cite{Marsoccijap, Marsoccipr} measured the transverse magnetoresistance of (100) Ni deposited
on polished NaCl substrate. The data were analyzed in the light of Smit's \cite{Smit} explanation of anisotropy in magnetic materials. The
analysis presented in the work showed that spin-orbit interaction was the deciding factor in magnetoresistance measurements.
\subsection{Experimental studies on manganites}
Manganites are a class of compounds denoted by the general formula Re$_{1-x}$A$_x$MnO$_3$ where Re is a rare earth metal like La, Nd, Pr etc.
and A is an alkaline earth metal (Ba, Ca, Sr).  This class of doped perovskites has been extensively studied in the last two decades. Here, we
will briefly discuss some key results on the anisotropic MR of manganite thin films. As discussed in the previous section, extensive literature
exists on the phenomena of anisotropic magnetoresistance (AMR), rotational magnetoresistance (RMR) and orbital magnetoresistance (OMR) in
polycrystalline samples of ferromagnets like Fe, Ni and their alloys \cite{McGuire, Robert, Fert, Malozemoff}. However, extending the origin of
RMR in elemental ferromagnets to complex ferromagnetic oxides of poorly understood magnetocrystalline anisotropy, nature of the charge carriers,
their scattering mechanism and their coupling to lattice and spin degrees of freedom are not straightforward.

Most of the studies reported till date on anisotropy of magnetoresistance in manganites have been carried out on epitaxial films of the average
bandwidth compound La$_{1-x}$Ca$_x$MnO$_3$ (LCMO, x$\approx$0.3) deposited on (100) cut \sto. These ferromagnetic films have in-plane
magnetization with (001) as the magnetic easy axis and the magnetic ordering temperature is $\approx$270 K. For example, the measurements of
Eckstein et al \cite{Ecksteinapl1996} show a striking anisotropy and hysteresis in the low-field MR for $\vec{I}\|\vec{H}$ and
$\vec{I}\bot\vec{H}$ configurations, which they attribute to magnetocrystalline anisotropy and colossal magentoresistance. Ziese and Sena
\cite{Ziese}, and Ziese \cite{Ziesesing} have also measured the low-field resistance anisotropy for $\vec{I}\|\vec{H}$ and $\vec{I}\bot\vec{H}$
configurations in LCMO films at several temperatures. While the sign and magnitude of their AMR are consistent with the phenomenological $s-d$
scattering model of Malozemoff \cite{Malozemoff}, its temperature dependence needs interpretation. Infante et al \cite{Infante} have measured
the isothermal MR as a function of the angle $\theta$ between $\vec{I}$ and $\vec{H}$ at 180 K in (110) oriented LCMO epitaxial films. They
attribute the hysteretic angular dependence of MR seen at low field to the in-plane uniaxial anisotropy of these films. Hong et al. \cite{Hong}
have investigated the effect of injected charges, using a field effect geometry, on AMR of (100) oriented epitaxial La$_{0.7}$Sr$_{0.3}$MnO$_3$
films in order to separate the contributions of carrier concentration and disorder caused by chemical doping to AMR. It is to be noted that a
systematic study, especially of the low field regime, of AMR on different polytypes of \LSMO has still not appeared in the literature. While we
have talked about anisotropy studies in manganites it is important to discuss the temperature dependence of AMR. Ziese et al. \cite{Ziese}
studied the temperature dependence on epitaxial and polycrystalline thin films of LCMO. The AMR is negative throughout the range of measurement.
It was observed that there is a dip in the AMR near the Curie temperature of the sample for epitaxial films. The polycrystalline sample did not
show any well defined peak or dip. They argued that the presence of an AMR peak near the CMR peak essentially means that AMR is caused by same
scattering process as CMR. Later on, Amaral et al. \cite{Amaral} studied the temperature dependent AMR in (110) LCMO. They too found a dip in
the AMR near the T$_c$ of the LCMO film. While these temperature dependent AMR have been seen, a clear explanation to these effects is still
lacking.
\section{Heterostructures of high-T$_c$ cuprates and manganites}
To make heterostructures for the study of the proximity effect and magnetic anisotropy in thin films, it is essential that the following points
be taken into consideration;
\begin{enumerate}\item The pair breaking effects of the ferromagnet should not be
too large to suppress superconductivity completely. \item The transition temperature of the superconductor should be high. \item The lattice
parameters of both the compounds should be identical so that the films are not under stress. This would help avoid magnetoelastic anisotropy
effects. \item The interface between the FM and SC layers should be clean and free from any chemical mixing.
\end{enumerate} Some of the above requirements are satisfied by heterostructures of high-T$_c$ cuprates and manganites. The manganites are known
to have low exchange energy, for example, the exchange energy of LSMO is 2meV while for ferromagnets like Fe, Co and Ni it is 270, 360 and 160
meV respectively, which makes it possible for the superconductivity to survive in cuprate-manganite bilayers and trilayers. From the
experimental point of view, the high transition temperature of the superconductor makes it possible to perform experiments without requiring
very low temperature setups which are costly as well as technologically challenging to handle. From the fabrication point of view,
HTSC--manganites are most suited in the sense that their lattice parameters are close to each other with perovskite structures, the effect of
strain on the films are reduced to negligible levels. Due to these reasons, heterostructures of YBCO and LSMO (or LCMO) are extremely popular
with researchers studying proximity effects. In the following few sections and subsections I will discuss briefly some important works in this
area starting with a brief introduction to the materials of our choice, namely \YBCO and \LSMO.
\subsection{\YBCO : Structure and phase diagram}
\begin{figure} \centerline{\includegraphics[height=4in]{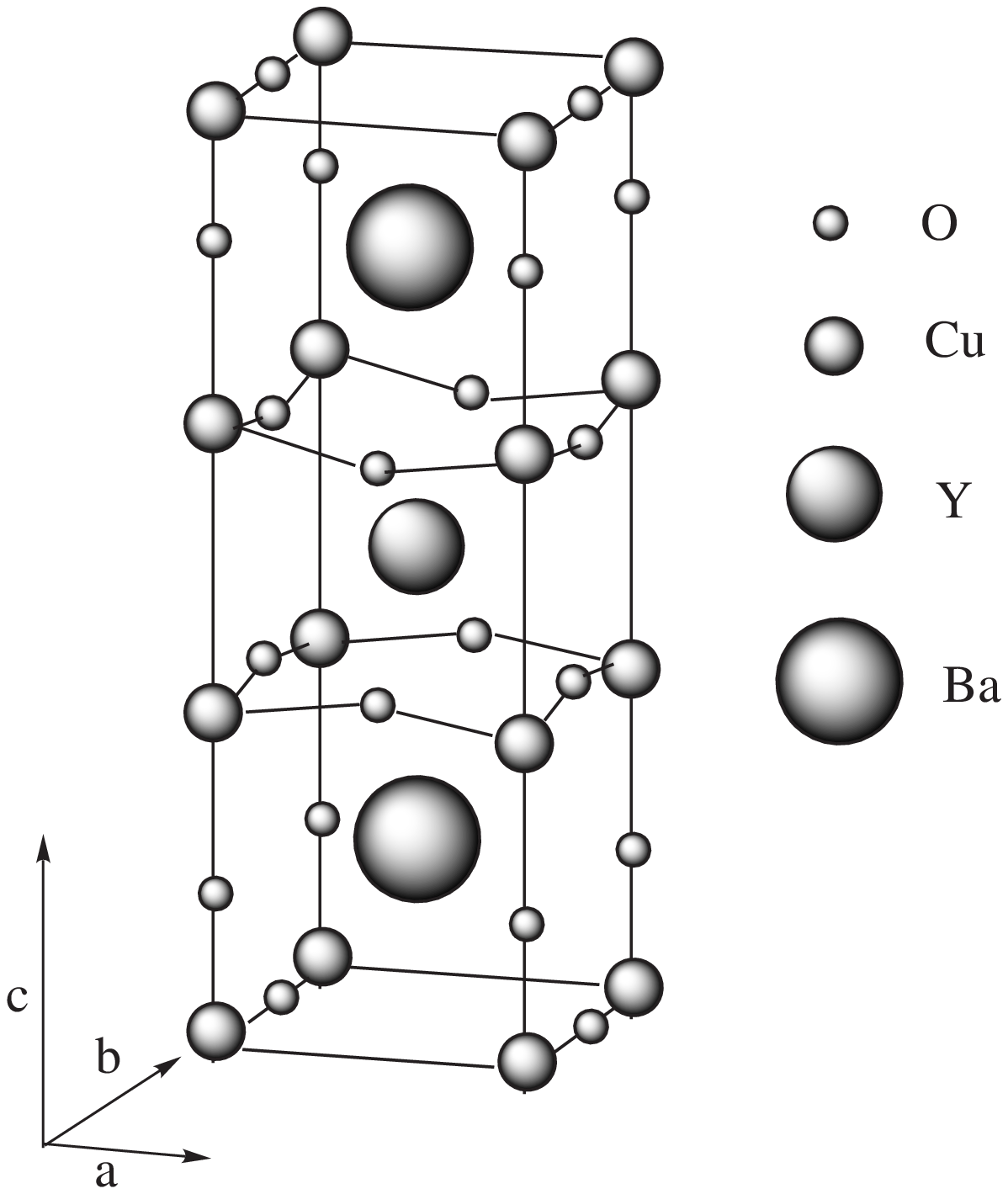}}
\caption{Crystal structure of \YBCO} \label{ybco} \end{figure}

Of all the cuprate superconductors, \YBCO is the most extensively researched compound. Discovered in 1987 by Wu et al. \cite{Wu}, \YBCO has a
T$_c$ of 90K. Although it has an orthorhombic structure, a transition into the tetragonal phase is observed above $\sim 600^oC$ \cite{Gurvitch}.
The lattice parameters $a, b$ and $c$ are 3.82 \AA, 3.88 {\AA} and 11.68 {\AA} respectively \cite{Goldman}. The structure has two Cu--O sheets
in the $ab$--plane and Cu--O chains along the $b$-axis (Fig. \ref{ybco}).  The structure of orthorhombic \YBCO derives from the stoichiometric perovskite by
elimination of rows of oxygen atoms parallel to the $a$-axis at the $z=0$ and $z=1/2$ levels. YBCO can be thought of as a stack of different
layers in the following order CuO--BaO--\cuo--Y--\cuo--BaO--CuO, as shown in Fig. \ref{ybco}. The separation between two consecutive sets of
\cuo planes is $\sim 8$ {\AA} along the $c$-axis. The Cu atoms in the \cuo plane are coordinated by five oxygen atoms forming a pyramid with the
apex pointing away from the Y atom. The apex oxygen in the pyramid is closer to the fourfold coordinated Cu in the CuO chain ($\sim 1.85$ \AA)
than the Cu of the \cuo plane ($\sim 2.3$ \AA) along the $c$-axis, showing the tendency of copper to adopt the square--planar coordination
within the layers. This results in the \cuo layers behaving as two dimensional planes perpendicular to the $c$-axis. This also explains the
anisotropy of YBCO between the $ab$-plane and the $c$-axis. The Cu atoms in the CuO layers are coordinated by four oxygen atoms in the square
planar geometry. The missing oxygen atom along the $a$-axis in the CuO layers accounts for the slight difference between the $a$ and $b$ lattice
parameters \cite{Sheahen, Santoro}.

The electronic phase diagram of YBa$_{2}$Cu$_{3}$O$_{7-\delta}$ is controlled by the oxygen occupancy in the CuO chains. The valency of the
copper atoms in a fully doped YBCO ($\delta \sim 0$) is $\sim 2.33$, due to a mixture of 2Cu$^{2+}$ and Cu$^{3+}$ ions \cite{Santoro}. On
increasing $\delta$, the oxygen aloms in the CuO chains are preferentially removed \cite{Beech} causing a hole depletion from the \cuo planes
until the compound YBa$_{2}$Cu$_{3}$O$_{6}$ is reached with no CuO chains at all. YBa$_{2}$Cu$_{3}$O$_6$ is an antiferromagnetic insulator with
$T_N \sim 500$ K \cite{Tvr}. Band structure calculations by Pickett \cite{Pickett} show that there is a considerable nesting of the Fermi
surface associated with the hybridization of the Cu$_{3d_{x^2-y^2}}$--O$_{2p}$ band in the \cuo plane. This makes the Fermi surface unstable and
an energy gap opens up making YBa$_{2}$Cu$_{3}$O$_6$ insulating. \begin{figure} \centerline{\includegraphics[height=3.5in]{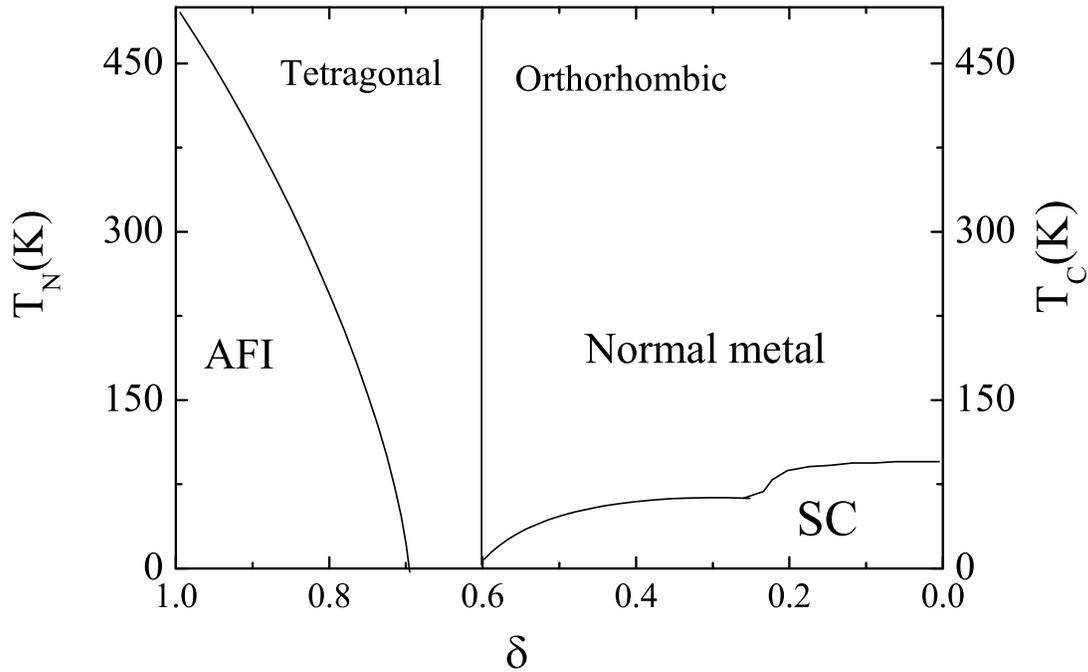}}
\caption[The electronic phase diagram of YBa$_{2}$Cu$_{3}$O$_{7-\delta}$]{The electronic phase diagram of YBa$_{2}$Cu$_{3}$O$_{7-\delta}$ as a
function of oxygen doping.} \label{ybcoo2} \end{figure} Fig. \ref{ybcoo2} shows the phase diagram of YBa$_{2}$Cu$_{3}$O$_{7-\delta}$ as a
function of oxygen doping. In the regime $0.7 \leq \delta \leq 1$, the compound  is insulating and antiferromagnetic with a steep decrease in
T$_N$ for lower $\delta$. At $\delta = 0.6$, the oxygen ordering in CuO chains gives rise to a transition from the tetragonal to the
orthorhombic phase \cite{Jorgensen}.
\subsubsection{Superconductivity in YBCO}
When we talk of superconductivity in YBCO, the first things that come to mind are the transition temperature and the order parameter of this
cuprate superconductor. The symmetry in this case is fundamentally different from the conventional BCS superconductor. It is now a well
established fact that superconductivity in these cuprates originates in the \cuo layers. Accordingly, the pairing symmetry also reflects some
underlying symmetry of the Cu--O plane \cite{Tsuei}. The pairing symmetry of high-T$_c$ superconductors is acknowledged to be highly anisotropic
with nodes along certain directions in the momentum space \cite{Scalapino, vanHarlingen}. \begin{figure}
\centerline{\includegraphics[height=3in]{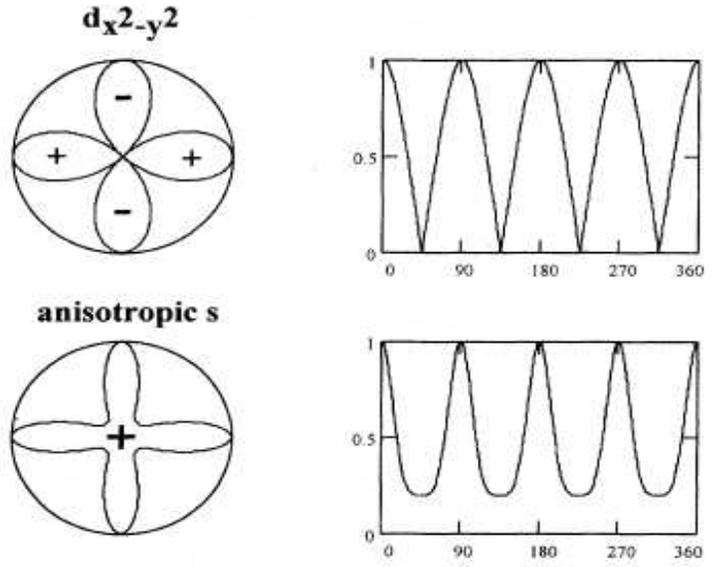}} \caption[$d_{x^2-y^2}$ and anisotropic $s$-wave symmetry of the pairing
state]{$d_{x^2-y^2}$ and anisotropic $s$-wave symmetry of the pairing state and the corresponding magnitude of the order parameter. (adapted
from ref. \cite{vanHarlingen})} \label{op} \end{figure} The shape of the pairing symmetry and the magnitude of the order parameter in the case
of $d_{x^2 - y^2}$ and anisotropic $s$ is shown in fig. \ref{op}. Phase sensitive tests employing SQUID interferometry \cite{Kirtley} and more
recently angle resolved electron tunneling \cite{Smilde} have established a preferential $d_{x^2-y^2}$ nature of pairing in YBCO.

A further scrutiny of the properties of high-T$_c$ cuprates reveals that unlike conventional  metallic superconductors in the normal state,
which can be described within the framework of Fermi-liquid theory, a different pairing mechanism is needed in this case. This is due to
compelling evidences for the non Fermi-liquid nature of the normal state in these cuprates. Secondly, in BCS superconductors, a hierarchy of
energy scales is maintained, i.e. $E_f > \hbar\omega_D > k_BT_c$, where $E_f$ and $\omega_D$ are the Fermi energy and the Debye frequency
respectively. But this is not the case in high-T$_c$ cuprates where the Debye temperature T$_D$ is comparable to T$_c$.

A closer scrutiny of the superconducting properties of high-T$_c$ demands a serious look into the pairing mechanism. One of the suggested
mechanism is through spin-charge separation. If holes are introduced into a 2D antiferromagnetic spin matrix, it causes a kinetic energy
frustration leading to the formation of quasi one dimensional ``hole-rich" and ``hole-poor" spin stripes \cite{Carlson}. In this case the spin
and charge excitations can be treated independently of each other as spin-solitons (spinons) and hole-solitons (holons) \cite{Maekawa}. A spin
pairing mechanism explained this way need not be affected by the strong charge repulsion (Coulomb repulsion).

Apart from that, phonon mediated pairing mechanisms in the presence of short antiferromagnetic correlations is also considered to be a potential
candidate in cuprates \cite{Tsuei}. Experimental evidences in support \cite{Hadjiev} and against \cite{Franck} the contribution of phonons in
the pairing mechanism of HTSC are present. As of now there is no consensus on the issue of pairing mechanism in cuprates.
\subsection{La$_{1-x}$Sr$_x$MnO$_3$ : Structure and phase diagram}
Electrical and magnetic properties of transition metal oxides are strong functions of the crystal field in which the ion resides. Transition
metal oxides showing fascinating physical properties, have in general perovskite type crystal structure with a general formula of ABO$_3$.
Perovskite manganites have a general formula of the type Re$_{1-x}$Ae$_x$MnO$_3$ where Re and Ae stand for trivalent rare earth and divalent
alkaline earth metals respectively. \begin{figure}[t] \centerline{\includegraphics[height=2.5in]{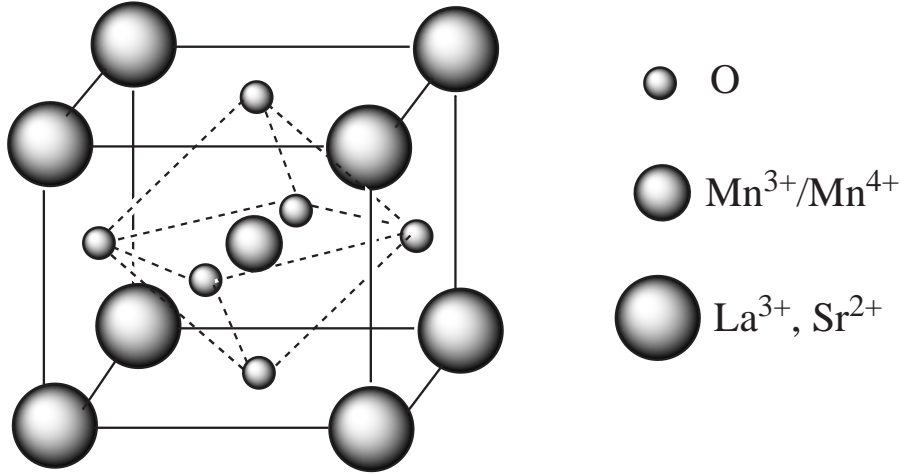}} \caption{Crystal structure of
La$_{1-x}$Sr$_x$MnO$_3$.} \label{lsmos} \end{figure} A schematic of a perovskite manganite unit cell is shown in Fig. \ref{lsmos}. The Mn atom
sitting at the center of the cube is coordinated by six oxygen atoms sitting at the face centers of the cube. The corners of the cube are
occupied by rare earth metal ions. The five fold degenerate $d$-orbital of the central Mn ion is split into $t_{2g}$ and $e_g$ orbitals under
the influence of the octahedral field of the negative oxygen ligands. The approximate separation between these levels is $\sim$ 1.5 eV.
\begin{figure} \centerline{\includegraphics[height=2in]{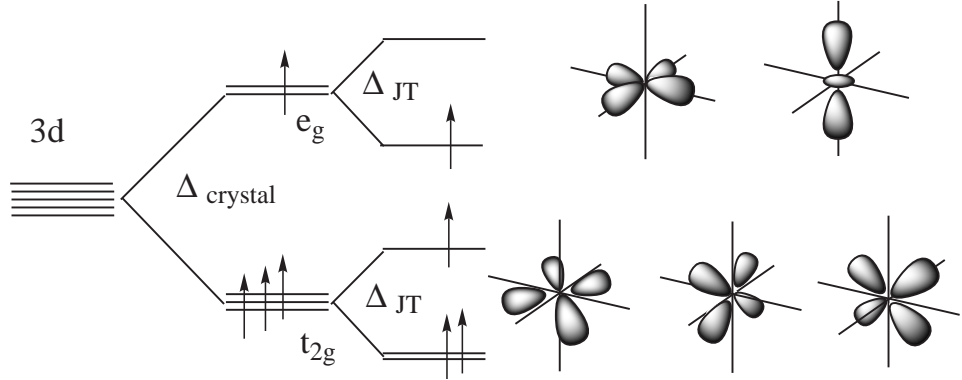}} \caption[Splitting of Mn 3$d$ orbital due to the octahedral crystal
field.]{Splitting of Mn 3$d$ orbital due to the octahedral crystal field of oxygen ions. $\Delta_{crystal}$ is the crystal field splitting and
$\Delta_{JT}$ is the splitting due to Jahn-Teller distortions.} \label{jt} \end{figure} Fig. \ref{jt} shows the relevant energy splitting
associated with the 3$d$ orbital of the Mn ion. The Mn$^{3+}$ ion has four 3$d$ electrons which are distributed as $t_{2g}^3e_g^1$, with only
one electron in the doubly degenerate $e_g$ orbital. As a result, the Jahn Teller distortion deforms the MnO$_6$ octahedron giving rise to a
slightly distorted cubic structure for the parent compound LaMnO$_3$. Another common source of lattice distortion in manganites is the relative
unit cell filling fraction of the Re and Mn sites quantified by the tolerance factor $f$, defined as $f = \left(r_{Mn} + r_0\right) /
\sqrt{2}\left(r_{Re} + r_0\right)$ where $r$ is the average ionic radius of the respective ions. A deviation of the tolerance factor f from the
ideal value of 1 results in a distortion of the Mn-O-Mn bond and deviation from ideal cubic structure. The crystal structure is very stable
against doping of the Re site. As a consequence, the complete phase space for carrier doping is accessible like in La$_{1-x}$Ca$_x$MnO$_3$. In
the case of La$_{1-x}$Sr$_x$MnO$_3$, stable structures are formed for $x < 0.7$.
\begin{figure} \centerline{\includegraphics[height=3.6in]{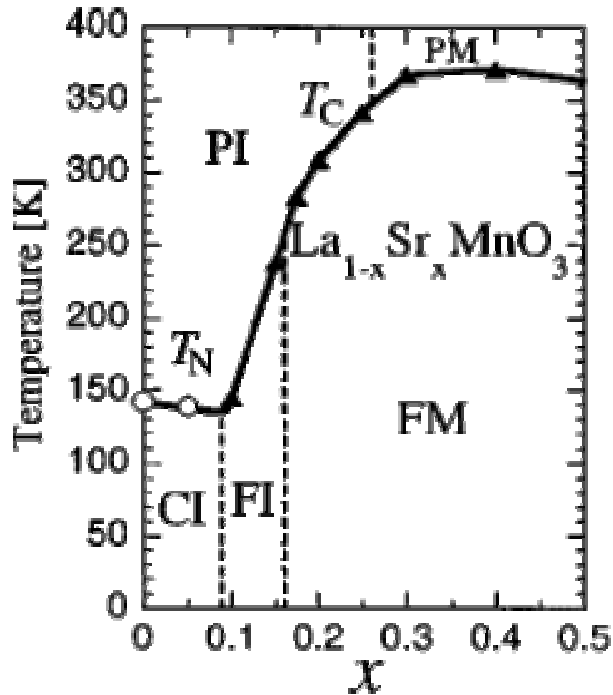}} \caption[Phase diagram of La$_{1-x}$Sr$_x$MnO$_3$]{Phase diagram of
La$_{1-x}$Sr$_x$MnO$_3$ with Sr$^{2+}$ concentration. States denoted by abbreviations: PI, paramagnetic insulating; PM, paramagnetic metallic;
CI, spin-canted insulating; FM, Ferromagnetic metal; FI, Ferromagnetic insulator.(Adapted from ref. \cite{Imada})} \label{lsmoph}\end{figure}

Fig. \ref{lsmoph} show the complete phase diagram of La$_{1-x}$Sr$_x$MnO$_3$ with respect to `x'. The different phases seen as we move from x=0
to x=0.5 result from the interplay between three primary energy scales; i) the one electron bandwidth W \footnote{W = 2z$t_{ij}$, where z is the
number of unpaired electrons on each site and $t_{ij}$ is the transfer probability between neighboring sites.} $\sim$ 0.2 eV, ii) the on-site
Coulomb repulsion U $\sim$ 5 eV, and iii) Hund's coupling energy ($J_H$) between $e_g$ and $t_{2g}$ electrons, which is $\sim$ 2 eV. Apart from
this, another energy in action is the crystal field splitting ($\sim$ 1.5 eV) of the Mn$^{3+}$ ion. The end member at zero doping (LaMnO$_3$) is
a Mott-Hubbard insulator which shows A-type antiferromagnetic ordering \cite{Imada}. For x $<$ 0.15, we have a spin-canted state which is
insulating \cite{Kawano}, although for x $>$ 0.10 the spin-ordered phase is almost ferromagnetic. The ferromagnetic insulating phase has a
narrow regime of existence between x = 0.1 and x = 0.17. In this region, the double exchange carrier is subject to (Anderson) localization but
can still mediate the ferromagnetic interaction between neighboring sites and give rise to the ferromagnetic state in a bond-percolation manner.
As x is further increased, a ferromagnetic phase appears below T$_c$ which steeply increases with x up to 0.3 and then saturates
\cite{Urushibara}. The effective transfer matrix $t_{ij}$ (and consequently the one electron band width W) depends on the relative orientation
of the relevant site. This means a ferromagnetic alignment will lead to considerable enhancement in the bandwidth and a metallic state can be
realized. \LSMO is therefore a ferromagnetic metal with a Curie temperature $\sim $360 K. Here we should also note that when we replace the rare
earth site with a divalent alkaline earth metal like Sr, we are forcing some of the Mn$^{3+}$ ions into Mn$^{4+}$ states with a orbital
structure $t_{2g}^3e_g^0$. Since the Hund's coupling energy J$_H$ is larger than the crystal field splitting, all spins in the Mn ion are
parallelly aligned. Hence increasing the hole filling by substitution of Sr reduces the average moment per Mn ion from 4$\mu_B$ to 3$\mu_B$.
\subsubsection{Magnetism in LSMO}
The correlation between ferromagnetism and metallicity in doped perovskite is explained by the Double exchange mechanism developed by Zener
\cite{Zener} which was further refined by Anderson and Hasegawa \cite{Anderson} and de Gennes \cite{Gennes}. This model couples localized spins
through indirect exchange mediated by conduction electrons. In LSMO, when the La site is doped with divalent Sr$^{2+}$, some of the Mn$^{3+}$ is
forced into the Mn$^{4+}$ state. The lone electron in the $e_g$ orbital of Mn$^{3+}$ now finds an empty $e_g$ orbital at a nearby Mn$^{4+}$
which is bridged through an oxygen ligand. The Mn$^{3+}$-O-Mn$^{4+}$ and Mn$^{4+}$-O-Mn$^{3+}$ states are degenerate and hence the transfer
probability of the electron becomes one. But it is to be noted that these two states are equivalent only in the framework of strong Hund's
coupling (J$_H$). So for metallic behavior it is essential that the neighboring spins are ferromagnetically aligned. Now we have seen earlier in
the phase diagram of LSMO that a spin canted state is also stable. But in this case below a certain critical angle the compound can be weakly
ferromagnetic but insulating. This is because the transfer probability is dependent on the relative angle between the neighboring spins, so
below a critical canting angle a finite magnetic moment can be seen but the effective transfer integral may be very small. This is the
explanation for the spin-canted insulating phase of manganites at certain doping concentrations.
\subsection{Experimental studies on cuprate - manganite heterostructures}
Advances in thin film synthesis and the structural similarity between high-T$_c$ cuprates and manganites combined with their relatively high
chemical inertness towards each other make it possible to realize FM-SC hybrids of these classes of materials with extremely well defined
interfaces. The earliest work on these hybrids was reported by Kasai et al. \cite{Kasai1}. They measured the I-V characteristics of
YBCO-LCMO-YBCO junctions and found measurable critical current through LCMO layers as thick as 500 nm. Later these results were interpreted as
evidence of the proximity effect by the same authors \cite{Kasai2}. They suggested that the coupling between YBCO is stronger than in
conventional Josephson junctions. Chahara et al. \cite{Chahara} studied the magnetoresistance effects of LCMO-YBCO-LCMO and LCMO films at 77K.
They found that the magnetoresistance of the trilayer was dependent on the thickness of the YBCO layer and that it was at least 1.5 times larger
than LCMO films when the YBCO thickness was below 2500 {\AA}. The authors have speculated that a long range coupling between the LCMO layers was
responsible for the enhanced MR. The logic behind this reasoning is that the enhanced MR is seen only when the YBCO layer is thinner than 2500
{\AA}. Jacob et al. \cite{Jacob} have studied the electrical properties and giant magnetoresistance (GMR) effect in c-axis oriented LBMO-YBCO
multilayers. They concluded from their magnetoresistivity study that GMR effect is compatible with the existence of high T$_c$ superconductivity
in the superlattices. Below T$_c$ both superconductivity in YBCO and magnetism in LBMO coexisted. Vas'ko et al. \cite{Vasko} reported strong
suppression of the critical current of a DyBCO layer in DyBCO-La$_2$CuO$_4$-LSMO structures due to a current flowing in the LSMO layer. They
attributed this to the pair-breaking effects associated with the injection of spin polarized carriers in the superconducting layer. Similar
experiments on supercurrent control through the injection of spin-polarized carriers were demonstrated on LSMO-YBCO structures by Stroud et al.
\cite{Stroud}, Yeh et al. \cite{Yeh}, Fabrega et al. \cite{Fabrega} and Chen et al. \cite{Chen} and in YBCO-STO-LSMO junctions by Plausinaitiene
et al. \cite{Plausinaitiene}. A review on the subject of spin-injection in cuprate manganite heterostructures has been published by Goldman et
al. \cite{Goldman}. Suppression of superconducting transition temperature and saturation magnetization were observed in LSMO-YBCO multilayers
\cite{Prieto, Habermeier} as well as bilayers \cite{Tan}. Sefrioui et al. \cite{Sefrioui, Sefrioui2} made a comparative study of T$_c$
suppression in LCMO-YBCO and PBCO-YBCO multilayers. Similarly Pang et al. \cite{Pang} reported a study on T$_c$ suppression in LSMO-YBCO
multilayers.

These studies discussed above clearly demonstrate two aspects of these hybrids:- i) Superconductivity in cuprates is suppressed by the proximity
of the manganite due to the pair-breaking effects of injected spin polarized carriers and ii) The cuprate-manganite heterostructures emerge as
potential candidates for spintronic applications due to the ability to control critical current through spin polarized carrier injection into
the cuprates.

The case of long range proximity effect in cuprate-manganite hybrids, which was first addressed by Kasai et al. \cite{Kasai1, Kasai2}, has
recently been revisited by Pe\~na et al. \cite{Pena, Pena2} using LCMO-YBCO superlattices. A GMR effect has also been reported in these
multilayers \cite{Pena2, Nemes} which is attributed to a spin dependent transport rather than vortex dissipation or AMR of the ferromagnetic
layer. Apart from these, an observation of the oscillatory transition temperature as a function of the FM layer's thickness has also been
reported in YBa$_2$Cu$_4$O$_8$--LCMO--YBa$_2$Cu$_4$O$_8$ \cite{Zhao}.

The discussion on experimental study is not complete without a short review of the work done on \YBCO-\LSMO hybrids in our group \cite{Senapati,
Senapram, Senapati1}. The first paper in the list \cite{Senapati} addresses the influence of the ferromagnetic boundary on the T$_c$ of
LSMO-YBCO-LSMO trilayers. The effect was attributed to the pair-breaking phenomenon near the F-S interface. The measured magnetization loops
showed clear signs of antiferromagnetic coupling between FM layers both below and above T$_c$. The exchange coupling between the FM layers
increased with a decrease in temperature but eventually saturated at low temperatures. The coupling decreased exponentially with increasing
spacer layer thickness. The second paper in the list \cite{Senapram} reported oscillations in critical current of LSMO-YBCO-LSMO trilayers as a
function of the LSMO thickness. The period of such oscillations was found to be $\sim$ 200 {\AA}. The oscillations were attributed to the
LOFF-like oscillatory superconducting order parameter near the FM-SC interface in the limit of weak exchange energy. The third paper
\cite{Senapati1} shows a comparative study on T$_c$ suppression in FM-SC-FM and NM-SC-NM hybrids. The NM in these studies in \PBCO. There it has
been said that all the effects of the FM boundaries on the mixed state are embodied in the critical thickness $\xi_c$, over which
superconductivity is quenched by pair breaking. 
\section{Motivation of the present work}
In this thesis we will mostly discuss electronic transport and magnetic properties of two polytypes of \LSMO [(001) and (110) oriented epitaxial
films], (110) oriented \LSMO-\YBCO-\LSMO trilayers and (001) oriented \linebreak \LSMO-\YPBCO-\LSMO trilayers. While a cursory thought at these
structures may suggest a disjointed problem, we will show that the magnetism, its anisotropy, magnetoresistance and superconductivity in these
structures are connected in fascinating ways. First of all our literature survey shows that a study on the hybrids of high-T$_c$ cuprates and
manganites is a very interesting and not so rigorously studied area. Since both manganites and cuprates are exotic, an understanding of the
phenomena arising from the proximity effects of these hybrids would facilitate the understanding of the origin of the ferromagnetism in former
and most importantly superconductivity in the latter. The research done thus on cuprate-manganite heterostructures is limited to systems where
the \cuo planes are parallel to the plane of the interface. Such structures do not allow injection of quasiparticles along the nodal or fully
gapped directions of the Fermi surface of the cuprate. In order to overcome this difficulty, it is necessary to grow the YBCO layer with
crystallographic orientation such that the \cuo planes are normal to the substrate. This can be achieved by growing either (100)/(010) or (110)
oriented YBCO films. While the (100)/(010) and (110) YBCO oriented films on lattice matched substrates have been deposited successfully
\cite{Marshall, Inam, Covington}, the growth of an FM - SC heterostructure or superlattice with the \cuo planes normal to the plane of the
substrate is quite non-trivial. The reason for non-triviality is as follows. The growth recipe for growing films with c-axis parallel to the
substrate plane involves a heterotemplate technique. In our case the template is PBCO. Same template is used for (100)/(010) YBCO growth. In
case we employ this growth recipe then the YBCO layer between FM layers has majority grains with c-axis perpendicular to the plane of the
substrate. This happens probably due to complete loss of information that allows the YBCO to grow with c-axis parallel to the surface in the
absence of intervening LSMO layer. This can be understood in this way that the lattice parameter of LSMO and YBCO (a and b axis) are close to
each other. So when there is a LSMO layer on which the YBCO has to grow then it takes the preferred growth direction which is the c-axis
perpendicular to plane of the substrate. For (110) growth this is not a problem since on (110) LSMO only three orientations for YBCO are
possible, (110), (103) and (013), which can be tuned by tuning the growth temperature of PBCO, LSMO and YBCO layers which is not possible for
(100)/(010) growth. One of the key results of this thesis is the successful synthesis of \LSMO-\YBCO hybrids such that the \cuo superconducting
planes are normal to the interface, thus allowing direct injection of spin polarized electrons into the superconducting planes. We compare the
properties of such structures with those where \cuo planes are parallel to the interface. One such example is of \LSMO-\YPBCO-La$_{2 / 3}$Sr$_{1
/ 3}$MnO$_{3}$. Here we address the fundamental problem of magnetoresistance in a spin-valve structure where the spacer material is a
superconductor. In order to take into account the anisotropy of the order parameter, we make structures which allow transport of spin-polarized
carriers from LSMO along different crystallographic directions of YBa$_{2}$Cu$_{3}$O$_{7}$. But, in order to study all these effects, it is
essential to understand the role of the anisotropies in LSMO. A systematic study of magnetic and magnetotransport anisotropies in \LSMO is
lacking. The major objectives of the thesis are listed below. \begin{description}
\item[First:] To develop a stable growth recipe for growing \LSMO-\YBCO hybrids where the \cuo planes are perpendicular to the interface.

\item[Second:] To study in detail the anisotropy of two polytypes of \LSMO namely (001) and (110), since their knowledge is important for our
understanding anisotropy in FM-SC hybrids.

\item[Third:] To investigate the superconducting and magnetic properties of $a,b$-axis \linebreak oriented \LSMO-\YBCO-\LSMO hybrids with an emphasis on the
GMR effects in these hybrids.

\item[Fourth:] To study the GMR effects in \LSMO-\YPBCO hybrids and compare the results with $a,b$ oriented hybrids.
\end{description}

The layout of the thesis is as follows:\\ \indent {\bf Chapter 1} is an introduction to the basic concepts of superconductivity, magnetism and
their interplay in hybrid structures. The structure and electronic phase diagram of the superconducting (YBa$_2$Cu$_3$O$_{7-\delta}$) and
magnetic (La$_{1-x}$Sr$_x$MnO$_3$) components of the hybrid structures have been reviewed. Here, we also discuss the magnetic anisotropy
associated with magnetic materials and their implications on this magnetic film. This chapter also presents a contextual survey of the current
experimental results and theoretical predictions concerning ferromagnet-superconductor heterostructures and the magnetic properties of simple
and complex ferromagnets.

{\bf Chapter 2} presents a detailed layout of the various experimental procedures used in this thesis. First, we have described the method of
preparation of thin films and trilayers using the pulsed laser deposition technique. Then we discuss various techniques for the characterization
of these thin films. Finally, the magnetization and transport measurement techniques are briefed.

{\bf Chapter 3} presents our studies of electron transport and magnetism in two polytypes of \LSMO films. We have focussed on the isothermal
magnetoresistance [R($\theta$, H)] of (001) and (110) epitaxial films measured as a function of the angle ($\theta$) between the current
($\vec{I}$) and the magnetic field ($\vec{H}$), both in the plane of the film, at several temperatures between 10 and 300K. The magnetic easy
axis of these polytypes is intimately related to the orientation of Mn - O - Mn bonds with respect to the crystallographic axis on the plane of
the substrate and the energy equivalence of some of these axes. The magnetization vector ($\vec{M}$) of the (001) and (110) type films is pinned
along the (110) and (001) directions respectively at low fields. A magnetization reorientation phase transition (MRPT), which manifests itself
as a discontinuity and hysteresis in $R(\psi)$ where $\psi$ is the angle between $\vec{H}$ and the easy axis for the $\vec{H}$ below a critical
value $\vec{H}^*$, has been established. The boundary of the pinned and depinned phases on the H-T plane has been established. The highly robust
pinning of magnetization seen in (110) films is related to their uniquely defined easy axis. The isothermal resistances R$_\bot$ and R$_\|$ for
$\vec{I} \bot \vec{H}$ and $\vec{I} \| \vec{H}$ respectively for both polytypes follows the inequality R$_\bot >$ R$_\|$ for all ranges of
fields ($0 \leq H \leq 3500$ Oe) and temperatures (10K - 300K). A full fledged analysis of the rotational magnetoresistance is carried out in
the framework of D\"oring theory for MR in single crystal samples. Strong deviations from the predicted angular dependence are seen in the
irreversible regime of magnetization.

{\bf Chapter 4} presents our work on (110) LSMO-YBCO hybrids. We first discuss the growth of YBCO--LSMO heterostructures of (110) orientation
perpendicular to the plane of the film to allow direct injection of spin polarized holes from the \LSMO into the \cuo superconducting planes.
The magnetic response of the structure at T $<$ T$_{sc}$ shows both diamagnetic and ferromagnetic moments with the (001) direction as magnetic
easy axis. While the superconducting transition temperature (T$_{sc}$) of these structures is sharp ($\Delta$T$_{sc} \simeq$ 2.5 K), the
critical current density (J$_c$) follows a dependence of the type $J_c = J{_o}(1-\frac{T}{T_{sc}})^{\frac{3}{2}}$ with a highly suppressed J$_o$
($\simeq 2 \times 10^4$ A/cm$^2$) indicating strong pair breaking effects of the ferromagnetic boundaries. Further studies on the trilayer
reveal a strong coupling between the FM layers. The coupling is an order of magnitude higher than that seen in the case of (001) trilayers
validating an earlier prediction. We also report an unusually high ($\sim72000\%$) angular magnetoresistance in these trilayers.

{\bf Chapter 5} presents some key results of our work on \LSMO-Y$_{1-x}$Pr$_x$Ba$_2$Cu$_3$O$_7$-\LSMO system with x = 0.4. Here the relevance of
pair-breaking by exchange and dipolar fields, and by injected spins in a low carrier density cuprate \YPBCO sandwiched between two ferromagnetic
\LSMO layers is examined. At low external field ($H_{ext}$), the system shows a giant magnetoresistance (MR), which diverges deep in the
superconducting state. We establish a distinct dipolar contribution to MR near the switching field (H$_c$) of the magnetic layers. At H$_{ext}
\gg$ H$_c$, a large positive MR, resulting primarily from the motion of Josephson vortices and pair breaking by the in-plane field, is seen.

{\bf Chapter 6} presents a brief summary of the important results of our experiments discussed in this thesis. Here we also identify some issues
which are potentially interesting for further studies.


\chapter{Detailed experimental layout}
\section{Introduction}
This chapter details the major experimental techniques used to carry out the present research. Epitaxial thin films and trilayer samples were
fabricated using a KrF pulsed excimer laser based deposition technique. A number of deposition runs were taken in order to optimize the
conditions for growth of good quality films. The magnetic properties of the samples have been probed using a DC-magnetization technique and for
electrical transport, we have used a four probe method. These measurements were carried out on three different sets of samples including single
layer and trilayer films. The four probe transport measurements were performed in a customized Advanced Research Systems Inc. closed-cycle
refrigerator (Model- DE-204S) based setup. This apparatus allows the study of transport properties between 4.2 and 100 K. An option for applying
a magnetic field of strength upto 4 kOe is also available. The angle of the applied field can be varied with respect to a fixed axis. DC
magnetization of the samples was measured as a function of the field and the temperature in a commercial Quantum Design Superconducting Quantum
Interference Device (SQUID) based magnetometer (MPMS-XL-5).
\section{Preparation of thin films and trilayers}
\subsection{Pulsed laser deposition (PLD)}
Magnetic and superconducting thin films and heterostructures were prepared using the pulsed laser deposition (PLD) technique which is a member
of a broad group of thin film deposition methods known as Physical Vapor Deposition (PVD). The various members of this group are Evaporative
Deposition, Electron Beam Physical Vapor Deposition, Sputter Deposition, Cathodic Arc Deposition, Pulsed Laser Deposition (PLD) etc. Out of all
these techniques, PLD is the most popular for the growth of multi-component oxide films. The main advantage of PLD over other members of the PVD
group lies in the mechanism of material removal from the source and its condensation on the substrate. PLD relies on the interaction of a high
energy density pulse of laser with a solid target of desired stoichiometry. The high energy density in conjunction with the short duration of
the pulse (typically a few nanoseconds) produces a local superheated spot on the surface of the target. The rate of heating overruns the
dissipation process and a shockwave explodes the superheated spot giving rise to a nascent plasmonic plume. Unlike thermal evaporation in which
the vapor composition depends on the vapor pressure of the individual components, the laser ablated plume maintains the original stoichiometry
of the target. For this reason PLD enables deposition of complex, multi-element compounds with ease. This is also the main reason behind the
popularity of this method.

Another advantage of PLD is the versatility in the deposition conditions offered by this technique. Thin films of pure metals, for example,
require a high vacuum during deposition. On the other hand, high T$_c$ superconducting oxides are best deposited in a high oxygen pressure
environment. In both cases, PLD has been successful in producing high quality films. The PLD technique is also cost effective in the sense that
one laser can serve a number of deposition vacuum chambers. However, along with these advantages, PLD also has some drawbacks. Notable among
them are: \begin{enumerate} \item Smaller area of uniform deposition due to the tight angular distribution of the ablated material. \item
Formation of defects in the film due to bombardment of high energy atoms and molecules. \item Evaporation of microscopic clusters from the
target known as ``splashing".\end{enumerate} In spite of these drawbacks, the robustness of the technique has considerably expanded its
applicability over the past decade. For example some of the limitations mentioned earlier can be minimized by a suitable tuning of the laser
energy density and the target-to-substrate distance. The problem of smaller area of uniformity, for example, has been dealt with techniques like
rastering the laser beam, moving the substrate with respect to the plume and inclined angle deposition. In fact, Hasegawa et al.\cite{Hasegawa}
have successfully prepared 10 meter long YBCO tapes using the inclined angle PLD. Similarly, there are some innovative attempts to mitigate the
effects of splashing. Most effective among these is the use of a `velocity filter' positioned in between the target and the substrate. A wide
variety of thin films, including pure metals, semiconductors and complex oxides and nitrides have been successfully synthesized using this
method.
\subsubsection{The excimer laser}
We have used commercial KrF excimer lasers (Lumonics PM - 800 \& Lambda Physik - Compex Pro) for PLD. The PM-800 is capable of producing pulses
at a maximum rate of 20 Hz and $\sim$ 650 mJ/pulse and Compex Pro has a maximum pulse rate of 10 Hz and $\sim$ 800 mJ/pulse. User parameters
like energy, repetition rate and the number of pulses are adjustable through a remote microprocessor control. Excimer lasers are gas lasers
emitting radiation typically at ultraviolet wavelengths. Lasing action in excimer lasers takes place by stimulated emission from the excited
dimers of the active gas components (Kr and F$_2$ in this case). The active elements are excited by an avalanche electric discharge inside the
laser cavity, in the presence of a background inert gas (Ne) pressure of $\sim$ 5 bar. At the molecular level, the ground state of KrF is
repulsive due to the inertness of the completely filled outer orbital of Kr. On the contrary, the excited state of the complex Kr$^+$+Fe$^-$ is
attractive. Hence, population inversion in excimer lasers is relatively more efficient.
\subsubsection{The deposition chamber}
\begin{figure}[!ht] \centerline{\includegraphics[width=4.5in]{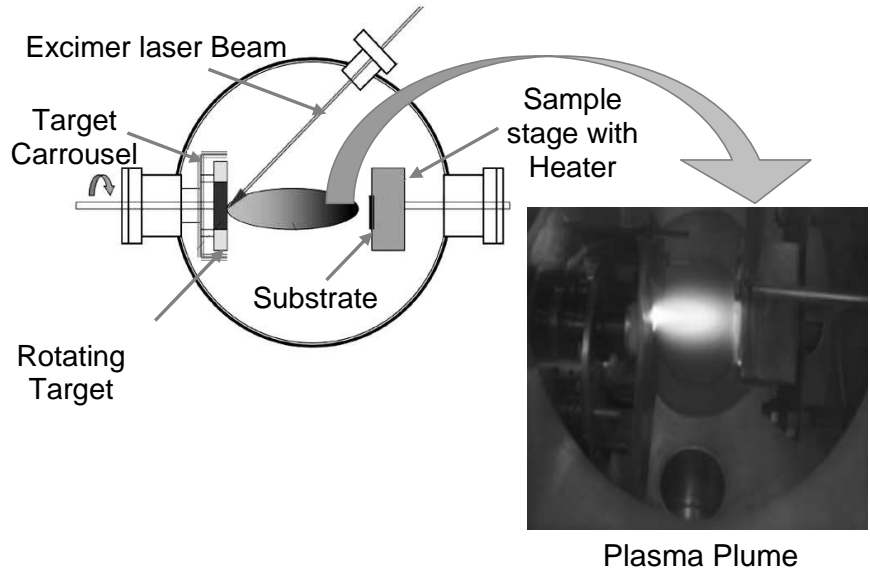}} \caption[{Schematic of PLD chamber.}]{Left hand panel shows
the schematic of the PLD chamber with its various components. The photograph of a plasma plume generated by the laser beam can be seen in the
bottom right corner.} \label{schematicpld}\end{figure} The main components of the deposition chamber are, a pumping system, a multitarget
carousal, and a substrate heater. The multitarget carousal supports upto three targets at a time. The position of the targets can be adjusted
remotely to facilitate the deposition of a desired material. This option enables the fabrication of heterostructures and multilayers. Fig.
\ref{schematicpld}(left) shows a schematic sketch of the deposition chamber. The right panel shows the photograph of a typical laser generated
plasma plume formed during deposition. The substrate heater is capable of maintaining a temperature of $\sim$ 800 $^0$C at an oxygen pressure of
0.4 mbar. The chamber attains a base pressure of $\sim 5 \times 10^{-3}$ mbar using a rotary pump. A liquid nitrogen trap has been installed
between the chamber and the pump in order to prevent the back streaming of oil vapors. The gas flow needed for the deposition of films was
maintained by an electronic mass flow controller.
\subsection{Preparation of targets for PLD}
Ablation targets of \YBCO(YBCO), \PBCO(PBCO), \LSMO(LSMO) and \YPBCO(YPBCO) were prepared through the standard solid state reaction route. High
purity powders (Aldrich) of Y$_2$O$_3$, Pr$_6$O$_{11}$, BaCO$_3$ and CuO were used as ingredients for the YBCO, YPBCO and PBCO targets. The LSMO
target was made from La$_2$O$_3$, SrCO$_3$ and Mn$_2$O$_3$ powders. The dried chemical precursors, taken in stoichiometric proportion, were
ground thoroughly to obtain a homogenous mixture. Before initiating the actual reaction process, the mixture was calcined at 800 $^0$C for 12
hours in the powder form. The calcination process removes the carbon content of the mixture by decomposing the carbonate group into CO$_2$ and
O. In addition, this step also helps removing any water content and volatile impurities. The grinding and calcination processes were repeated
three times to ensure carbon-free oxide mixture. The calcined powder was then pelletized under 50 kPa pressure using a hydraulic press. The
targets of ($\approx$ 22 mm diameter) YBCO, YPBCO and PBCO were sintered at 980 $^0$C, whereas the LSMO target was sintered at 1200 $^0$C.
\subsection{Substrate selection}
In order to attain high quality epitaxial growth of thin films and heterostructures, the selection and surface conditioning of the substrate are
of crucial importance. Particularly,  the dynamics of film growth and strain in the film are affected by the in-plane lattice parameter of the
substrate and the quality of its surface.

The initial stages of film growth are primarily governed by one of the two mechanisms, namely Volmer-Weber island growth and Frank-van der Merwe
layer-by-layer modes\cite{Horwitz}. The Volmer-Weber mode is favored when the adatom-adatom adhesion is stronger than the adatom-substrate
adhesion. It causes clustering of the atoms arriving at the substrate into small  islands. These islands slowly grow and merge with each other
at a latter stage, forming a continuous film. Clearly, this mechanism is unacceptable for the deposition of ultrathin films ($\leq$ 5 nm). On
the other hand, when the adatom-substrate bonding is stronger  than the adatom-adatom adhesion, the Frank-van der Merwe mode is the preferred
mechanism of growth. In this mode, the growth is essentially layer-by-layer leading to a very good epitaxy of the film. The lattice misfit
parameter, $\gamma$ (defined as $\frac{a-b}{b}$, where a and b are the lattice parameters of the film and the substrate respectively) plays an
important role in the selection of the growth mechanism. In general, a large lattice misfit parameter ($\gamma \geq 2\%$) may force Volmer-Weber
growth, bypassing the adatom-substrate adhesion energetics.

LaAlO$_3$ and \sto are the most popular candidates as substrates for the deposition of high-T$_c$ and Mn based CMR oxides. Both substrates are
cubic perovskites in structure. The lattice parameters of LaAlO$_3$ and \sto are 3.79 {\AA} and 3.91 {\AA} respectively. The average ab-plane
lattice parameter of YBCO is 3.85 {\AA} and that of LSMO is 3.89 {\AA}. In view of the matching of the lattice parameters, both the substrates
are suitable for layer-by-layer growth. However the LaAlO$_3$ crystals inherently contain a large density of twinning defects. These defects are
known to affect the superconducting properties of HTSC as well as magnetic anisotropy of CMR manganites. In view of these limitations of LAO, we
have selected (001) and (110) oriented \sto as the substrate for all samples used in this study.

\subsection{Deposition of thin films and heterostructures}
The high density ceramic targets, glued on aluminium holders, were positioned individually at the focal point of the laser beam inside the
chamber. A target-to-substrate distance of 5 cm is commonly used for the deposition of HTSC and manganite thin films. However, we have used a
distance of 6 cm to  avoid possible particulate deposition and to obtain a smoother surface. Prior to the deposition, the chamber was evacuated
to a base pressure of $\sim 5 \times 10^{-3}$mbar. The surface of the targets was cleaned by firing a few thousand shots of laser pulses before
initiating deposition on the substrate. Film deposition was carried out in an oxygen pressure of 0.4 mbar at 700, 750 and 800 $^0$C for PBCO,
LSMO and YBCO/YPBCO  films respectively. The typical rate of heating and cooling the substrates was 10 $^0$C/min. An energy density of $\sim$ 2
J/cm$^2$ was used to ablate the targets. With these parameters, a deposition rate of $\sim$ 1.6 {\AA}/sec for cuprates and $\sim$ 0.5 {\AA}/sec
for LSMO was realized. This rate was established using a stylus profiler (Tencore Alpha-Step 500), which has a topographic resolution of $\sim$
10 {\AA} on several test films. We have grown three sets of samples epitaxially on (001) and (110) \sto substrates.

\begin{tabular}{lp{5in}}
Series I:& (001) and (110) \LSMO films of thickness 600 {\AA}. \\
Series II: & Thin films of \YBCO sandwiched between 1000 {\AA} \LSMO on (110) \sto. The thickness of the YBCO layer was varied
from 100 {\AA} to 1000 {\AA}. \\
Series III: & Thin films of \YPBCO sandwiched between 300 {\AA} \LSMO on (001) \sto. The thickness of the YPBCO layer was varied from 100 {\AA}
to 1000 {\AA}.
\end{tabular}

\section{Measurement technique}
\subsection{Structural Characterization}
\subsubsection{X-ray Diffraction}
The determination of crystal structure is a very important part of any work involving epitaxial thin films. For that X-ray diffraction (XRD) is
the most powerful non destructive technique. XRD data can provide information like crystal structure, phase purity, lattice parameters, crystal
orientations, average grain size, crystallinity, strain and crystal defects to name a few.

$\theta - 2\theta$ measurements were done using a two-axis powder diffractometer (Model-Seifert, XRD-3000 P). For four circle measurements, the
sample were sent to our collaborators in France. Our two-axis diffractometer consists of a Cu--K$_\alpha$ source and a scintillating detector
mounted on the circumference of concentric vertical circles. A fixed sample stage is mounted at the center of the circle. A goniometer assembly
provides uniform rotation for both the source and the detector. The data collection in this machine is completely automated with the help of a
computer and company provided customizable software.

The four circle XRD measurements were done by our collaborators Dr. P. Padhan and Dr. W. Prellier in CNRS, France. Here we briefly describe the
salient features of a four circle geometry.
\begin{figure}\centerline{\includegraphics[width=4in]{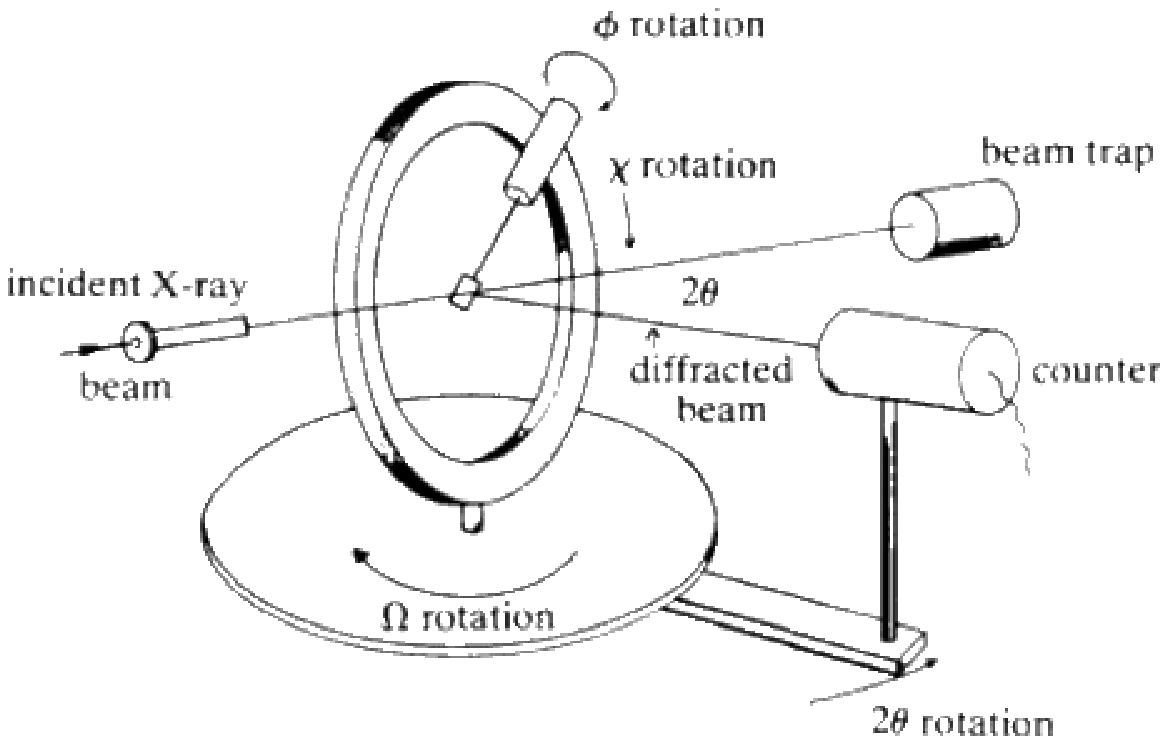}}
\caption[Schematic of four circle X-ray diffractometer.]{Schematic of four circle X-ray diffractometer.\\ \textit{Image courtesy:
http://serc.carleton.edu/research\_education/geochemsheets/techniques/SXD.html}} \label{fcxr} \end{figure} Figure \ref{fcxr} shows the schematic
of a four circle XRD setup.

\begin{description}
\item[The $\chi$ axis] The axis passing through the crystal and lying in the diffraction plane is the $\chi$ axis. The sample in a regular
diffractometer is mounted on the internal cylindrical surface of the ring associated with this axis. The sample is placed at the center of the
$\chi$ ring throughout the experiment.

\item[The $\phi$ axis] This is an axis about which the crystal along with the sample stage, to which the crystal is rigidly attached, can be
rotated. This axis is orthogonal to the $\chi$ axis. During the rotation about the $\phi$ axis the crystal remains at the center of the $\chi$
ring. The orientation of the axis of rotation about the $\phi$ axis is determined by the $\chi$ angle.

\item[The $\Omega$ axis] The $\Omega$ axis is basically the principal axis of the diffractometer. It passes through the plane of the $\chi$
ring and is perpendicular to the diffraction plane. There are in fact two circles in the $\Omega$ plane. As shown in the schematic (Fig.
\ref{fcxr}) the $\Omega$ motor rotates the $\chi$ and everything else mounted on it without moving the detector and the $2\theta$ motor rotates
the detector without changing the angle between the sample and the incident beam. In modern day x-ray diffractometers, the $\chi$ ring is kept
fixed in the $\Omega$ plane and the source is rotated by the $\Omega$ motor with respect to the $\chi$ ring.
\end{description}

\subsubsection{Scanning Electron Microscopy}
Scanning electron microscopy is an important tool for material research. It allows the study of surface morphology and the composition of
materials. In this process a beam of electrons is constricted by an assembly of magnetic lenses to give a thin probe beam. The probe scans over
the selected sample area. The interaction of probe electrons with the sample can give rise to secondary electrons, back-scattered electrons and
x-ray radiation. Based on the type of electrons detected, the technique is named as either Secondary Electron (SE) imaging or Back Scattered
Electron (BSE) imaging. The SE imaging technique works by detecting secondary electrons emitted by the sample because of the incident electron
beam. These are very low energy electrons and are useful in producing high resolution and high magnification images. The BSE imaging technique
uses electrons back-scattered by the sample. These have higher energy than the secondary electrons. Because of the higher energy of these
electrons, they can reveal not only the surface topography but the chemical profile of the sample to a certain depth can also be measured. The
X-ray radiation emitted by the sample is used for Energy Dispersive Spectrometry (EDS). The radiation emitted by the sample is characteristic of
the element with which the incident electrons are interacting. As a result, this technique is ideal for determining the elemental composition of
the sample. The SEM images in this thesis were taken by our collaborators Dr G. U. Kulkarni and his group at the Jawaharlal Nehru Centre for
Advanced Scientific Research, Bangalore using the BSE mode.
\subsubsection{Transmission Electron Microscopy}
As the name suggests, Transmission Electron Microscopy (TEM) is an imaging technique based on the detection of transmitted and refracted
electrons. The principle of operation is similar to that of an optical microscope with the only difference being that here the probe is a stream
of electrons rather than photons. The resolution achieved by this technique is of the order of 0.1nm. Hence, it is possible to resolve
individual atoms using this technique. An electron source in the microscope emits electrons which travel through the microscope column. These
electrons are focused into a thin beam with the help of magnetic lenses. The focused beam of electrons is allowed to pass through an
appropriately thinned sample. Depending on the structure and morphology of the sample, some electrons pass through the sample unhindered. The
rest are either reflected or refracted. These streams of electrons are projected on a fluorescent screen to form a shadow image depending on the
sample under investigation. Here we have shown high resolution TEM images of cross-sections of \LSMO-\YPBCO  and (110) \LSMO-\YBCO trilayer.
These images were taken by our collaborator Dr. Y. Zhu and his group at the Brookhaven National Laboratory, USA.
\subsubsection{Atomic Force Microscopy}
Atomic Force Microscopy (AFM) is a member of a large family of surface probes commonly known as Scanning Probe Microscopy. The family consists
of members, other than AFM, like Scanning Tunneling Microscopy (STM), Electric Force Microscopy (EFM), Lateral Force Microscopy (LFM) and
Magnetic Force Microscopy (MFM), to name a few. The AFM technique depends on the detection of extremely small forces of the order of 10$^{-12}$
- 10$^{-8}$ N. Invented by Binning et al. \cite{Binning}, it is one of the most powerful techniques to measure surface topography with very high
resolution. There are three basic modes of AFM which will be discussed in brief below:- \begin{description}
\item[Contact Mode:] As the name suggests, the contact mode AFM
operates by scanning the sample surface with the cantilever mounted AFM tip in close contact. The change in cantilever deflection is monitored
by a two-sided photodiode. As the tip scans the surface, the cantilever bends to adjust the change in the atomic force due to changes in the
surface morphology at the atomic level. But, in this mode there is a fair chance that the sample can get damaged due to excessive tracking force
exerted by the probe. The excessive tracking force is the result of air capillary forces due to the adsorbed gases on the sample surface which
is primarily made of 10-30 monolayers of water vapor and nitrogen.

\item[Non-contact Mode:] Sometimes, it is important that the sample is imaged without touching the surface. For that purpose, the non-contact mode AFM
was developed. In this mode the tip is made to hover over the sample at a distance of approximately 50-150nm. At this distance, the forces in
play are the van der Waals forces. The tip is scanned over the surface and the changes in the force strength are measured to detect the surface
topography. But in this case, the forces are considerably weaker than those in the contact mode AFM. Hence, for better detection, a small
oscillation is given to the tip so that AC detection techniques can be employed. By measuring the change in amplitude or frequency of
oscillations in response to the force gradients from the sample, one can measure the surface topography.

\item[Tapping Mode:] This technique was developed to achieve high resolution images without introducing damage to the sample due to frictional
forces associated with the AFM tip in the contact mode. In this technique, the tip is oscillated at or near its resonant frequency at an
amplitude greater than 20nm. The oscillating tip is then moved closer to the surface which than starts tapping the surface. The amplitude of
oscillation in this case is directly related to the surface topology. When the tip hits a hill, the amplitude is reduced and vice versa. To
overcome the tip-sample adhesion forces, it is necessary to keep the amplitude of oscillations above a certain critical level.
\end{description}
\subsubsection{Magnetic Force Microscopy}
Magnetic Force Microscopy (MFM) is done in three different modes viz. the non-contact mode, the tapping mode and the lift mode. For MFM, the tip
is coated with a ferromagnetic material. In the non-contact mode, the system operates by detecting the change in cantilever frequency due to the
interaction of the tip with magnetic domains. The image taken by a magnetic tip will contain both magnetic as well as topographic information.
The effect of magnetic forces persists at larger distances than the van der Waals force. Hence if the tip is too close to the surface, the image
will be essentially topographic. To separate out the topographic contributions and magnetic contributions, it is important to take the MFM data
for varying tip-to-sample distances. Similar consideration is required for the tapping mode as well. When the tip is tapping the surface it will
show the topography of the surface. Once the tip is lifted from the surface, the contributions from the topography and magnetic domains change.
At large distances, only information of the magnetic domains remain, giving rise to magnetic domain images. In general, the MFM tip is usually
composed of etched silicon sputter-coated with a ferromagnetic material.

The AFM and MFM data presented in this thesis were taken by our collaborators Dr G. U. Kulakarni and his group at JNCASR, Bangalore. The MFM
images were taken in the tapping mode with different lift heights to separate out the magnetic and topographical components. The MFM tip in this
case was CoCr-coated silicon.
\subsection{Measurement of anisotropic magnetoresistance and Resistivity}
\begin{figure}[!t]
\centerline{\includegraphics[width=4in]{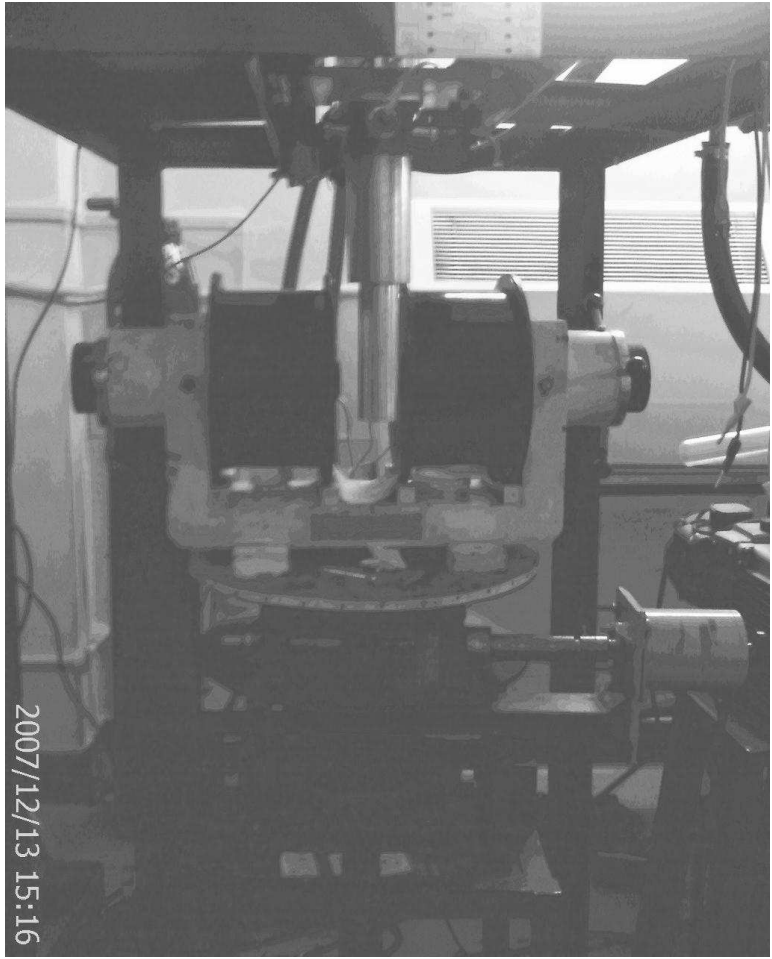}} \caption[{Photograph of close-cycle refrigerator.}]{Photograph of the close-cycle refrigerator
showing the cold head and the electromagnet mounted on a rotatable stage.} \label{closecycle} \end{figure} Galvanomagnetic properties of the
samples were measured using a close-cycle refrigerator based cryostat and an electromagnet which can generate a field of 4000 Oe. The sample was
mounted on a detachable sample platform, fabricated using oxygen free high conductivity (OFHC) copper, capable of orienting the sample in such a
way that the field remains in the plane of the film at all values of the angle between the applied field and the current direction. The stage
was equipped with two RuO$_2$ resistive sensors for accurate temperature measurement and control. The electro-magnet was set on a rotatable
platform capable of rotating the magnet with a resolution of 0.1$^o$. The rotation of the stage is controlled by a microprocessor based stepper
motor. The remnant field of the magnet is $\sim$ 100 Oe which can be completely eliminated by passing a reverse current through the magnet. A
photograph of the system is given in fig. \ref{closecycle}. The cold head can be moved in and out of the pole pieces as and when desired.

We have measured rotational magnetoresistance(RMR) on all three series of samples. We have also measured the isothermal magnetoresistance on all
the three sets of samples by scanning the field between -800 Oe to +800 Oe. Magnetoresistance have also been measured by varying the angle
between the field and current directions.

Samples are glued to the stage using Apiezone-N grease, which provides excellent thermal contact with the base. Measurements were performed
using the standard four terminal geometry. Electrical contacts on the samples were  made by soldering 25 micron gold wires with indium on
evaporated silver contact pads. \clearpage
\subsection{DC Magnetization}
DC-magnetization measurements are essential for the studies of equilibrium magnetic interactions. There are several techniques available for
such measurements, including torque magnetometry, magneto-optical Kerr effect (MOKE) and Superconducting Quantum Interference Device (SQUID)
based magnetometry. However, SQUID magnetometry is now the most widely used method, especially because of its extreme sensitivity and
simplicity. An absolute moment sensitivity of $\sim 10 ^{-8}$ emu is achievable by this technique.

At the heart of a SQUID magnetometer is a SQUID sensor consisting of a single Josephson junction (RF SQUID) or multiple Josephson junctions (DC
SQUID) mounted on a superconducting loop. The change in magnetic moment of the sample is fed inductively into the SQUID loop using a set of
superconducting pick-up coils. The SQUID loop, in principle, can detect a change in magnetic flux as low as 10$^{-7}$ Gauss/cm$^2$.

We have used a commercial Quantum Design SQUID magnetometer (MPMS-XL) for all measurements reported in this thesis. The operating range of
temperature and field in this system are 2 to 400 K and -5 to +5 T, respectively. A Reciprocating Sample transport Option (RSO) was used in
these measurements to achieve better sensitivity. All measurements were done in a parallel alignment of the film plane and magnetic field. This
geometry minimizes the demagnetizing effects of the sample, which otherwise may dominate the magnetization of thin samples. The samples were
rotated in the plane of the film and the MH loops recorded to determine the easy axis of the magnetic samples. Zero-field-cooled and
field-cooled magnetization was measured as a function of temperature for thin LSMO films. We measured the magnetization loops of the (001) and
(110) LSMO films at 10 and 200 K. We also measured magnetization loops of the LSMO-YBCO-LSMO and LSMO-YPBCO-LSMO trilayers at several
temperatures between 10 and 300 K, upto a maximum field of 1500 Oe.

\section{Summary}
We have made stoichiometric ceramic targets of \YBCO, \PBCO, \linebreak \LSMO and \YPBCO using the standard solid state reaction route. We have
successfully prepared (110) hybrid thin films of \LSMO-\YBCO-\LSMO on (110) \sto using the heterotemplate technique. Single layer (110) \LSMO
films were also grown on (110) \sto. (001) films of \LSMO and \LSMO-\YPBCO-\LSMO were prepared on (001) \sto. All these films were grown using
the pulsed laser deposition technique. The surface morphology and structure of the films were examined by scanning electron microscopy and X-ray
diffraction respectively. Magnetic domain structures were examined using atomic force microscopy. The transport and galvanomagnetic property
measurements on these samples were done using a home-made setup involving a close-cycle refrigerator. The temperature and field range of these
measurements were 4.2 - 300 K and 0 - $\pm$3 kOe respectively. For the measurement of magnetic properties, a commercial SQUID based magnetometer
was used.

\chapter[Anisotropic Magnetoresistance in LSMO]{Anisotropic Magnetoresistance
in \LSMO (001) and (110)}
\section{Introduction} A galvanomagnetic property of
fundamental interest in thin manganite films is their isothermal magnetoresistance (MR) measured as a function of the angle ($\theta$) between
the current $\vec{I}$ and applied magnetic field $\vec{H}$, both being in the plane of the film. This angle-dependent resistivity
($\rho(\theta)$) in polycrystalline films of metallic ferromagnets follows a dependence of the type \cite{Robert,McGuire,Fert,Malozemoff};
\begin{equation} \rho(\theta) = \rho_{\bot} + \left(\rho_{\|} - \rho_{\bot}\right)\cos^2\theta,\label{eqrho}\end{equation} where $\rho_\|$ and
$\rho_\bot$ are the resistivities for $\vec{I} \| \vec{H}$ and $\vec{I} \bot \vec{H}$ respectively. The resistivity $\rho(\theta)$, often called
the rotational magnetoresistance (RMR), derives contribution from two carrier scattering processes, one of which depends crucially on the spin -
orbit interaction; a magnetization($\vec{M}$) direction and strength dependent source of anisotropic scattering. While $\vec{M}$ may not be
necessarily collinear with $\vec{H}$ due to non-zero magnetocrystalline anisotropy, a grain averaging of $\rho$ in a polycrystalline film yields
eq. \ref{eqrho}. This contribution to RMR, which has been known as anisotropic magnetoresistance (AMR), saturates once the field intensity H
exceeds the saturation value H$_s$ beyond which the film becomes a single domain magnetic entity. Another contribution to RMR comes from the
trapping of mobile carriers in cyclotron orbits due to the Lorentz force. This localizing effect (a positive contribution to MR), which
increases with the carrier mean free path is called the orbital magnetoresistance (OMR). The magnitude of OMR varies as the square of the
magnetic induction $\vec{B}(=\vec{H} + 4\pi\vec{M})$, and shows a constant positive slope for $H > H_{s}$. While $\rho_\bot$ is always greater
than $\rho_\|$ for OMR, the relative magnitude $(\rho_\bot/\rho_\|)$ of these resistivities can be greater or less then unity for AMR. The sign
of the inequality between $\rho_\bot$ and $\rho_\|$ is intimately linked with the electronic band structure of the material under consideration.
In most of the $3d$ - transition metal alloys, $\rho_\| > \rho_\bot$ and the magnetoresistance $(\Delta\rho)/\rho) = (\rho_\| -
\rho_\bot)/\rho_{av}$ can be as large as $\simeq 30\%$ in some dilute Ni alloys \cite{Robert,McGuire,Fert,Malozemoff}. In low carrier density
ferromagnets such as the hole doped manganites \cite{Ecksteinapl1996,Ziese,Ziesesing,Infante,Hong}, GaMnAs \cite{Hamaya} and GdN \cite{Gns}, the
anisotropic magnetoresistance is small and negative.

The angular dependence of the resistivity as expressed by eq. \ref{eqrho}, does not hold in the case of single crystal samples with a non-zero
magnetocrystalline anisotropy. Here the magnetization is not necessarily collinear with $\vec{H}$ because its direction is decided by the
competition between the torques exerted on it by $\vec{H}$ and the crystalline  anisotropy field. The $\theta$ dependence of $\rho$ at low
fields, in particular when the torque on $\vec{M}$ is not strong enough to depin it from the easy axis, may deviate markedly from the behavior
predicted by eq. \ref{eqrho}. Often, the $\rho$ vs. $\theta$ curve has sharp jumps and hysteresis suggesting a first order transition. Such a
magnetization reorientation phase transition (MRPT) has been seen prominently in Fe film grown on GaAs \cite{Prinzjap,Prinzprb}.

It is evident that the measurements and analysis of the galvanomagnetic property RMR as a function  of field strength and temperature provide
rich insights into spin-orbit interactions, carrier mobilities, scattering length, the torques experienced by $\vec{M}$ from the applied field
$\vec{H}$ and various forms of magnetic anisotropies which pin $\vec{M}$ along a certain crystallographic direction.  While earlier studies
provide a wealth of information on resistance anisotropy in colossal magnetoresistance manganites, the issues which have remained unaddressed
are:-
\begin{enumerate}\item Although it is evident from the low field measurements of
O'Donnell et al. \cite{Ecksteinapl1996} on (100) LCMO and Infante et al. \cite{Infante} on (110) LSMO films that there is indeed a non-zero
in-plane anisotropy which pins the magnetization vector along the easy axis, the H-T phase space of the pinned phase is not established in these
studies. Moreover, the relative strength of the in-plane anisotropy, which is directed along different crystallographic axes for the (100) and
(110) films, is not known. There is also a need to understand the fundamental processes responsible for the anisotropy. \item At larger fields,
the magnetization rotates freely with the field. A systematic temperature dependence of $\rho(\theta)$ which would permit the calculation of
$\Delta\rho (=\rho_\bot - \rho_\|)$ is, however, lacking. A comparative study of $\Delta\rho$  and its temperature dependence in (001) and (110)
films would help in separating the band structure related contribution and the role of extrinsic effects such as disorder to anisotropic
magnetoresistance. \item While polycrystalline films of $3d$ ferromagnets show a simple $\cos^2\theta$ dependence of isothermal MR, in single
crystals this is generally not true. Here the orientation of both the current and the magnetization with respect to the crystal axis is
important. In the case of manganites, a full fledged analysis of $\Delta\rho(\theta)$ in terms of D\"oring's equations \cite{Doring} is lacking.
Such a study is desired to establish significant deviations from the $\cos^2\theta$ dependence of $\Delta\rho$ and the contributions of other
scattering processes to RMR.
\end{enumerate}

Here we present a rigorous study of RMR in two  variants of high quality epitaxial films of \LSMO; one with (001) and other with (110) axis
normal to the plane of the substrate. We have analysed the data in the light of D\"oring's equations \cite{Doring}. The field and temperature
dependence of the phenomenological coefficients between a selected range of temperature has also been reported for both types of film. We have
also drawn the H-T phase diagram for both films to show the pinned and depinned regions in the H-T phase space. The phase diagram clearly shows
that in the case of (110) film, the pinning of the magnetization vector is stronger than that of (001) film. A careful study of
magnetoresistance as a function of the angle between the applied field and the current direction in both (001) and (110) epitaxial films of
\LSMO is also presented. In both the cases, we have measured MR at 10 K and 300 K. While we see hysteresis in resistivity at 10 K in both cases,
but at 300 K only the (110) film shows hysteresis. This clearly demonstrates that the anisotropy energy of the (110) films are larger than that
of the (001) films. We base our arguments and explanation on the multidomain model given by O'Donnell et al. \cite{Ecksteinprb2}. \clearpage

\section{Results and Discussions}
Thin epitaxial films of \LSMO were deposited on (110) and (001) oriented \sto(STO) substrates using a multitarget pulsed excimer laser (KrF,
$\lambda$ = 248 nm) ablation  technique. The deposition temperature (T$_{d}$), oxygen partial pressure p$_{O_{2}}$, laser energy density
(E$_{d}$) and growth rate (G$_{r}$) used for the growth of 150 nm thick layers were 750$^{0}$ C, 0.4 mbar, $\sim$2J/cm$^2$ and 1.3{\AA}/sec
respectively. Further details of film deposition are given elsewhere \cite{Senapati}. The epitaxial growth in two sets of films with (110) and
(001) directions normal to the plane of the film was established with X-ray diffraction measurements performed in the $\Theta - 2\Theta$
geometry. For transport measurements, films were patterned in the form of a $1000 \times 100$ ${\mu}m^2$ bridge with photolithography and wet
etching such that the long axis of the bridge was parallel to the (001) and (100) direction for the (110) and (001) oriented films respectively.
The measurements of resistivity as a function of temperature, magnetic field strength and the angle$(\theta)$ between the field and current were
performed using a 4.2 K close cycle He - refrigerator with a fully automated home made setup for applying the field at varying angles between 0
and 2$\pi$ with respect to the direction of the current \cite{Patnaik}. The sample was mounted in a way to keep the field in the plane of the
sample for all values of the angle between $\vec{I}$ and $\vec{H}$. Isothermal magnetization loops (M-H) were measured for both the samples
using a commercial magnetometer (Quantum Design MPMS XL5 SQUID) by applying the field at various angles in the plane of the film.

\subsection{\LSMO thin films}
\begin{figure} \centerline{\includegraphics[height=4in, angle = -90]{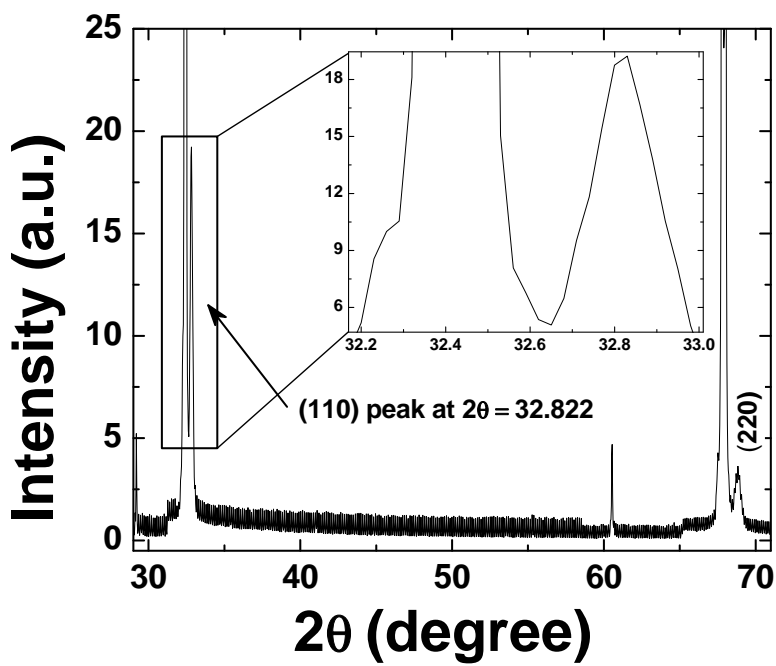}} \caption[{X-Ray data
of LSMO film grown on STO (110)}]{X-Ray data of LSMO film grown on STO (110). The lattice  parameter calculated from the (110) and (220) peaks
is 3.859{\AA}. The inset shows the zoomed in view of the (110) peak. The substrate peak can also be seen in the picture.}
\label{xraylsmo110}\end{figure}

\begin{figure} \centerline{\includegraphics[height=4in, angle=-90]{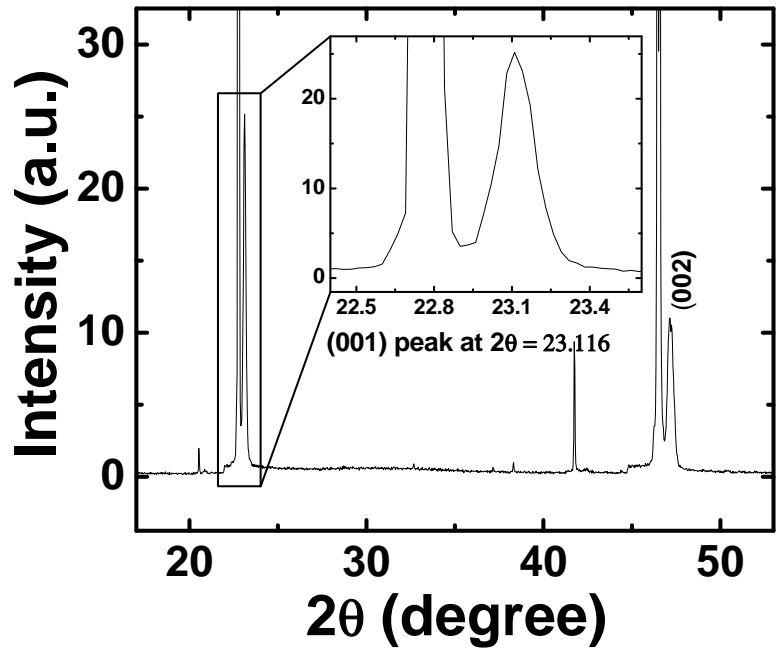}} \caption[{X-Ray data of LSMO film grown on STO
(001)}]{X-Ray data of LSMO film grown on STO (001). The lattice parameter calculated from the (001) and (002) peaks is 3.844{\AA}. The inset
shows the zoomed in view of the (001) peak. The substrate peak can also be seen in the picture.} \label{xraylsmo100}
\end{figure}

In fig. \ref{xraylsmo110}, we have  shown the $\theta - 2\theta$ X-ray data for a (110) sample grown on STO (110). The lattice parameter  calculated
from the position of the (110) and (220) peaks comes out to be $\approx 3.859 \pm 0.192${\AA}. In the inset we have shown a magnified view of
the (110) peak. The substrate peak can also be seen close to the LSMO (110) peak. Similarly fig. \ref{xraylsmo100} shows the X-ray
 data for a (001) LSMO sample grown on STO (001). The lattice parameter calculated from the (001)  and (002) peaks is $\approx 3.844
\pm 0.192${\AA}. The substrate peaks can also be seen near the LSMO peaks. These data clearly indicate that the films are epitaxial.

\begin{figure}[!h] \centerline{\includegraphics[width=2.8in]{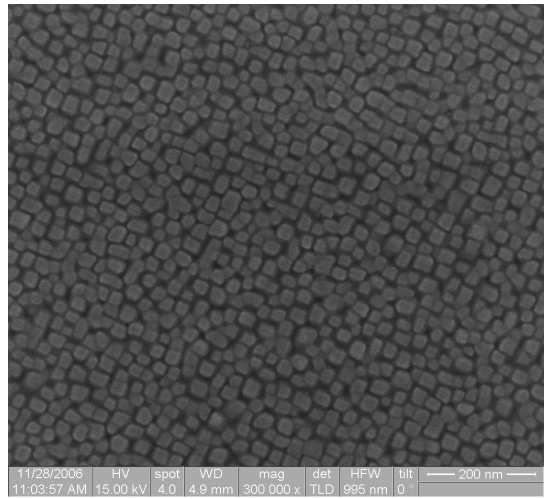}} \caption[{SEM image of the (001) \LSMO
film surface.}]{SEM image of the (001) \LSMO film surface. The surface texture is granular, with the grains being circular in nature. The
average grain size is $\sim$ 30nm.} \label{semlsmo100}\end{figure}
\begin{figure}[!h] \centerline{\includegraphics[width=2.8in]{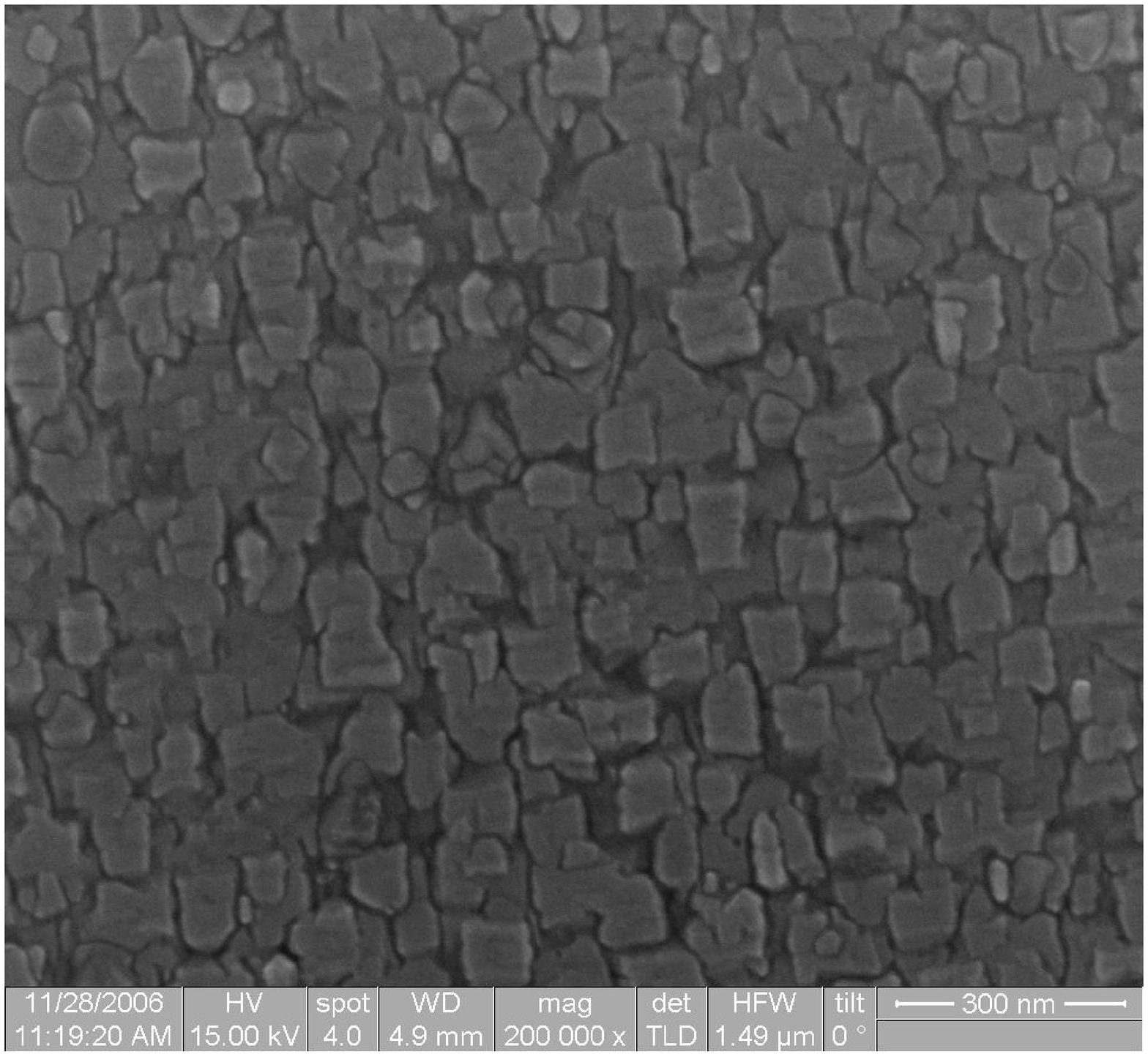}} \caption[{SEM image of the (110) \LSMO film surface.}]{SEM
image of the (110) \LSMO film surface. The surface texture is granular, with the grains being rectangular in nature. The typical grain size is
$\sim$70nm $\times$ 100nm.} \label{semlsmo110} \end{figure} Fig. \ref{semlsmo100} shows a high resolution  SEM image of a (001) LSMO surface.
The surface topography is granular in  nature with a typical grain diameter of $\sim$ 30nm. A similar micrograph for (110) LSMO is shown in fig.
\ref{semlsmo110}. In this case, the grains are rectangular with typical size being $\sim$ 70nm $\times$ 100nm. An earlier report on the surface
topography of (110) films shows a particular orientation of the longer side of the grains \cite{Suzuki}, our pictures do not show any such
orientation. Though the grains are oriented either along the \onebar or (001) direction, but there is no such preference as to which direction
will be along the length of the grains.

\begin{figure}[!h] \centerline{\includegraphics[height=6in, angle = -90]{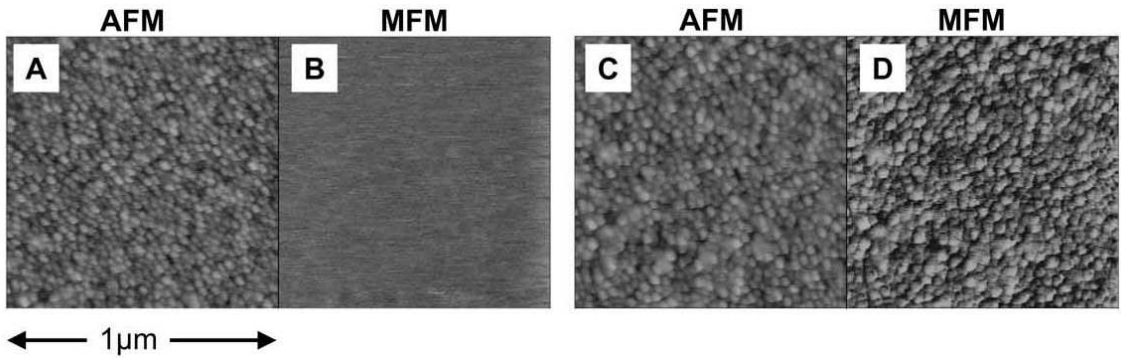}} \caption[{MFM image of (001) \LSMO
film.}]{AFM and MFM images of a (001) \LSMO film at two different lift heights. Panel B shows the MFM image with a lift height of 150nm while
panel D is the image with 20nm lift height. The left panel clearly shows the film being a single domain while in the right panel, the MFM image
is similar to the AFM image.} \label{mfm100}\end{figure} \begin{figure}[!h] \centerline{\includegraphics[height=6in, angle = -90]{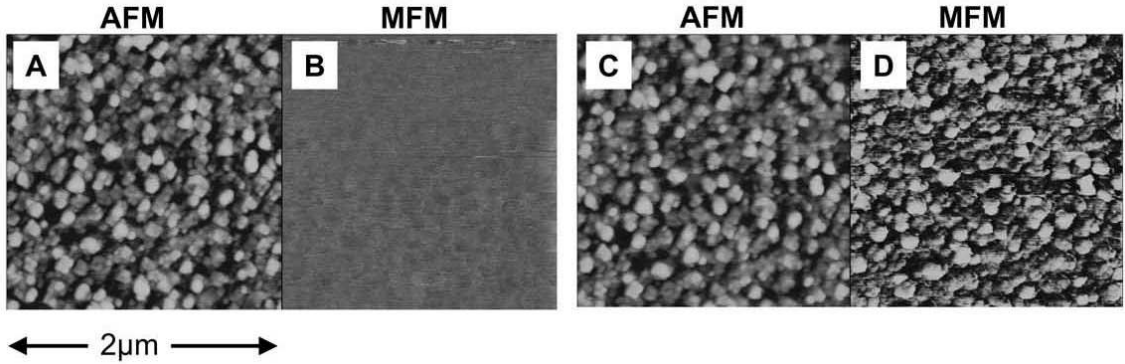}}
\caption[{MFM image of (110) \LSMO film.}] {AFM and MFM images of a (110) \LSMO film at two different lift heights. Panel B shows the image with
a lift height of 150nm while panel D is the image with 20nm lift height. The left panel clearly shows the film being a single domain while in
the bottom panel the MFM image is similar to the AFM image. In the AFM image, no clear directionality of the grains are visible \cite{Suzuki}.}
\label{mfm110} \end{figure}In fig. \ref{mfm100}(A, B, C and D) we have shown the AFM and MFM images of a (001) \LSMO film. Panels A and C are
AFM images and B and D are MFM images. Panels A and B are images for a lift height of 150 nm. From this picture it is clear that the film is a
single domain. Panels C and D show the image for a lift height of 20 nm. At this height, contributions from both the magnetic as well as the van
der Waals forces are present. Looking at the MFM image, we see a striking similarity between the AFM and MFM images. One may conclude that the
individual grains are acting as a magnetic domain. In that case, the magnetic domain structure should have been visible at large lift heights
which is not the case (panels A and B), so we can safely conclude that the samples are single domain samples. Similar micrographs for (110)
samples are shown in fig. \ref{mfm110} A, B, C and D. Panels A and B are images at a lift height of 150 nm and C and D are images at a lift
height of 20 nm. Giving similar arguments as we have given for fig. \ref{mfm100}, we find that the samples are single domain. Secondly the grain
size in the case of the (110) films is larger than that of the (001) films though no particular orientation of the grains is visible.

\subsection[Magnetization reorientation phase transition]{Magnetization reorientation
phase transition in (110) and (001) films} 
\begin{figure}[t] \centerline{\includegraphics[width=3in, angle =
0]{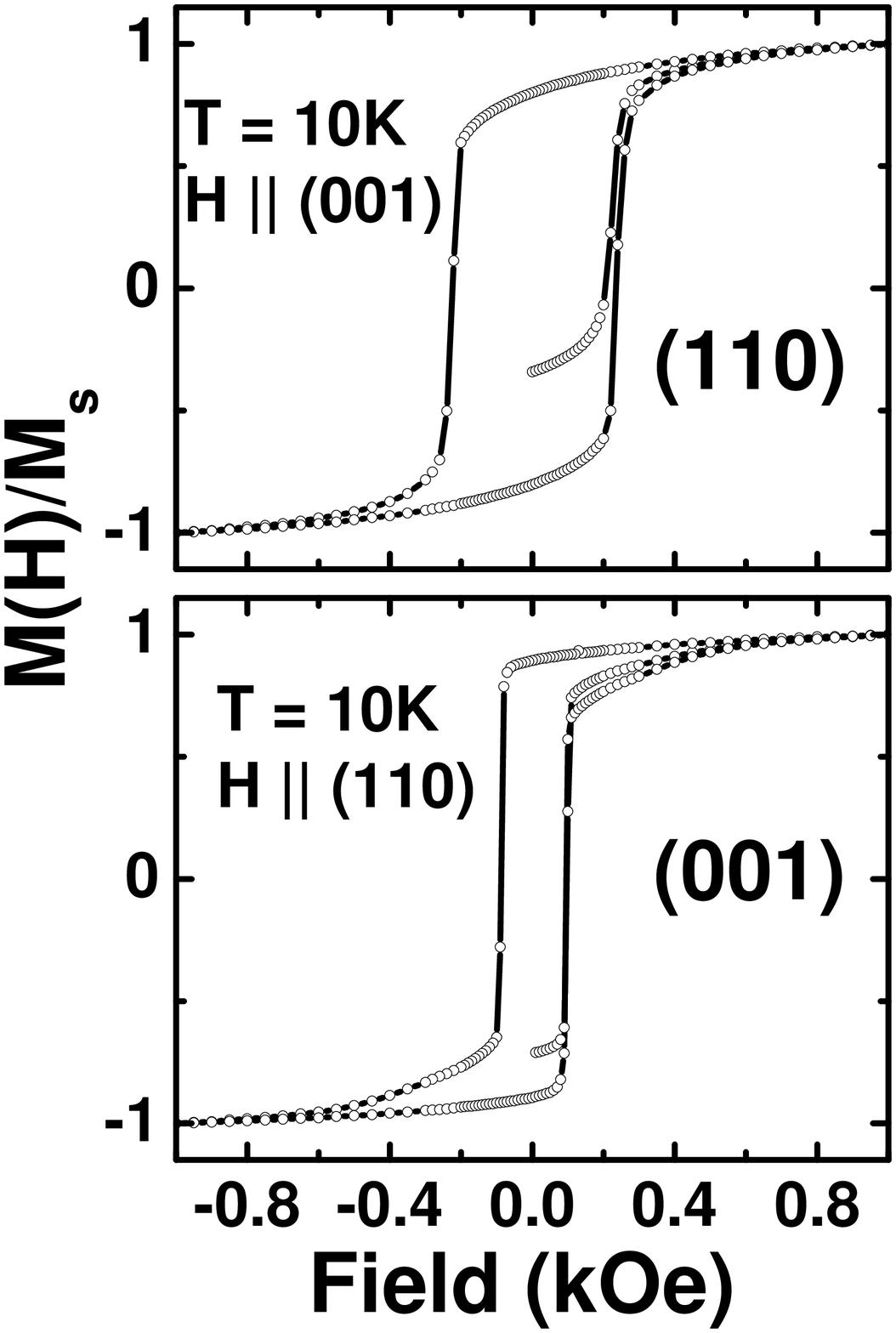}} \caption[Magnetic hysteresis loops of \LSMO films  at 10 K.]{Magnetic hysteresis loops of the (110) and (001) oriented
\LSMO films at 10 K. The measurements were done under the zero field cooled condition. The direction of the applied field in these measurements
was along the (001) and (110) crystallographic directions for the (110) and (001) oriented films respectively. The coercive field deduced from
these measurements is 230 Oe for the (110) and 90 Oe for the (001) samples respectively.} \label{mhloop}
\end{figure}

Fig. \ref{mhloop}  shows the magnetization vs. field (M-H) loops for the (110) and (001) epitaxial samples at 10 K in terms of the normalized magnetization
M(H)/M$_s$,  where M$_s$ is the saturation magnetization. The magnetic field in these measurements was applied along the easy axis which is
collinear with the (001) and (110) crystallographic axis for the (110) and (001) epitaxial films respectively \cite{Easy}. The coercive fields
for the (110) and (001) samples deduced from these measurements are 230 Oe and 90 Oe respectively. The marginally higher H$_c$ of the (110) film
seen here appears to be a common feature of such films \cite{Suzuki}.

\begin{figure} \centerline{\includegraphics[width=3in, angle = -90]{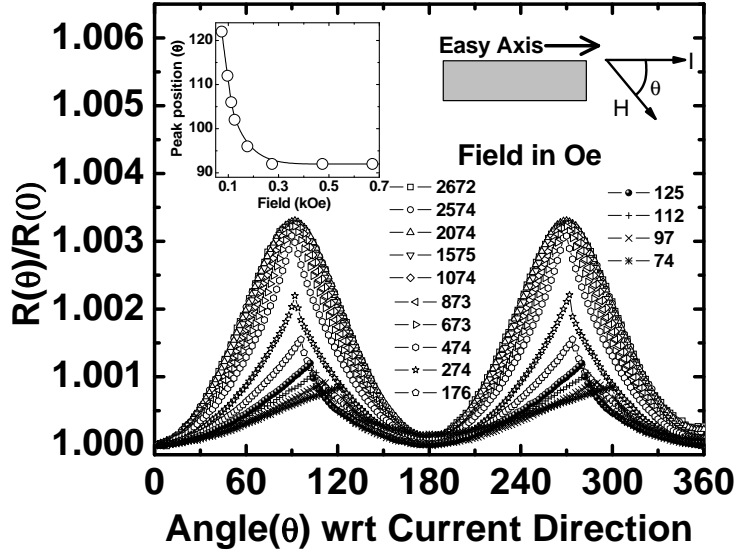}} \caption[Rotational magnetoresistance
R($\theta$) of the {(110)} film measured at 300 K]{Rotational magnetoresistance R($\theta$) of the (110) film measured at 300 K for different
values of the in-plane field. We observe a periodicity of $\pi$ when the external field H is $\geq$ 300 Oe. Below 300 Oe however, a distinct
deviation from this symmetry is seen, and the peak in resistance is now shifted to $\theta > 90^0$. One noticeable feature of the low field ($<$
300 Oe) data is a sudden drop in the resistance once the peak value is reached. The top left inset shows the variation of the position of the
first peak in the RMR data. A sketch of the sample geometry is shown in the top right hand corner of the figure.}\label{amr110300k}
\end{figure}

\begin{figure} \centerline{\includegraphics[width=3in, angle =
-90]{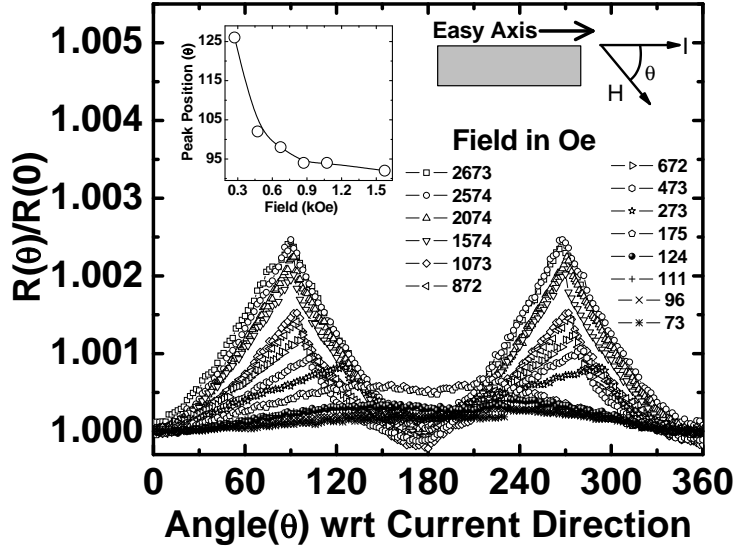}} \caption[The R$(\theta)$/R(0) vs. $\theta$ graphs of the {(110)} film at 10 K]{The R$(\theta)$/R(0) vs. $\theta$ graphs
of the (110) film at 10 K for several values of the in-plane field. Here we see deviations  from a symmetric dependence on $\theta$ at fields
lower than 1100 Oe. The inset at the top left hand corner shows the variation of the peak position with the field of the first peak in the RMR
data. The top right hand corner shows a sketch of the measurement geometry.} \label{amr11010k}
\end{figure}

The RMR of a (110) oriented LSMO film  at 300 and 10 K is shown in figs. \ref{amr110300k} and \ref{amr11010k} respectively, where we have
plotted the angle ($\theta$) between the directions of the current and the applied field along the x-axis and the resistance ratio
R($\theta$)/R(0) along the y-axis.

The relevant vectors in the plane of  the film are also shown in the right hand inset of the figures. For the measurements performed at 300 K
(fig. \ref{amr110300k}), we observe a symmetric R($\theta$)/R(0) curve about $\theta = \pi/2$ and $3\pi/2$ with a periodicity of $\pi$ when the
external field H is $\geq$ 300 Oe. Below 300 Oe however, there is a distinct deviation from this symmetry; the peak in resistance is now shifted to
$\theta > 90^0$. The variation of the peak position as a function of the field is plotted in the left inset of the figure.
  One noticeable feature of the low field ($<$ 300 Oe)
data is a sudden drop in the resistance once the peak value is reached. This suggests some kind of a depinning transition. For the
R($\theta$)/R(0) vs. $\theta$ curves at 10 K (fig. \ref{amr11010k}), this deviation from  the symmetric dependence persists upto $\simeq$1100
Oe. Here we also note that at fields below 200 Oe, the resistance has a negligible dependence on the angle between $\vec{I}$ and $\vec{H}$. Two
more noteworthy features of these data are the values of the resistance for the $\vec{I} \| \vec{H}$ and $\vec{I} \bot \vec{H}$ configurations.
Unlike the case of 3$d$ transition metal films, here $\rho_\bot > \rho_\|$. This is an interesting feature of the RMR in manganite thin films
\cite{Ecksteinapl1996,Ziese,Ziesesing,Infante,Hong}.

\begin{figure} \centerline{\includegraphics[width=3in, angle =
-90]{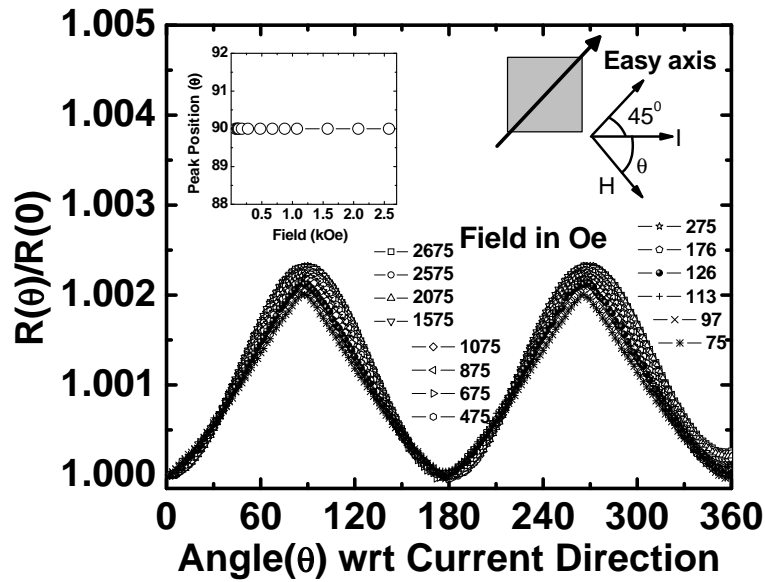}} \caption[Rotational magnetoresistance R($\theta$) of the {(001)} film measured at 300 K]{Rotational magnetoresistance
R($\theta$) of the (001) LSMO thin film measured at 300 K for different values of the in-plane field. We observe a periodicity of $\pi$ for all
external field H. The top left inset shows the variation of the position of first peak in the RMR data. A sketch of the sample geometry is shown
in the top right hand corner of the figure.} \label{amr100300K} \end{figure}

\begin{figure} \centerline{\includegraphics[width=3in, angle = -90]{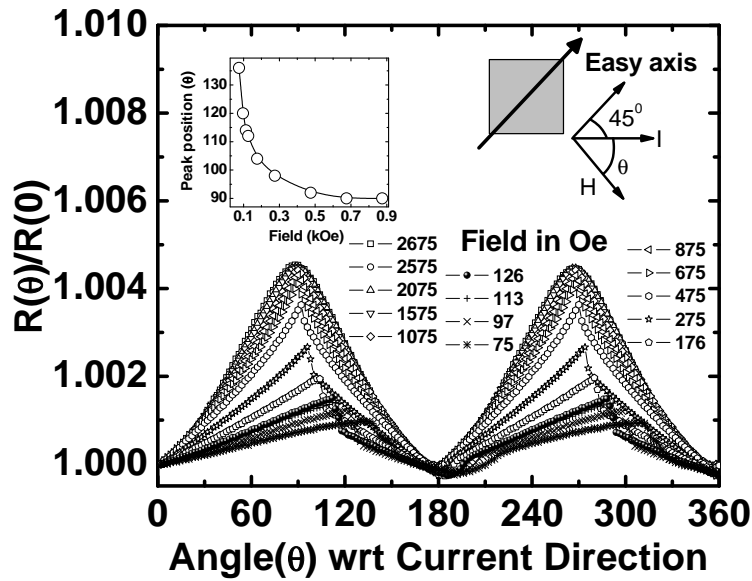}} \caption[The
R$(\theta)$/R(0) vs. $\theta$ graphs of the {(001)} film at 10 K]{The R$(\theta)$/R(0) vs. $\theta$ graphs of the (001) film at 10 K for several
values of the in-plane field. Here we see deviations from a symmetric dependence on $\theta$ at fields lower than 500 Oe. The inset at the top
left hand corner shows the variation of the first peak position with field in the RMR data. The top right hand corner shows a sketch of the
measurement geometry.} \label{amr10010K}
\end{figure}

\begin{figure}
\centerline{\includegraphics[width=3in, angle = -90]{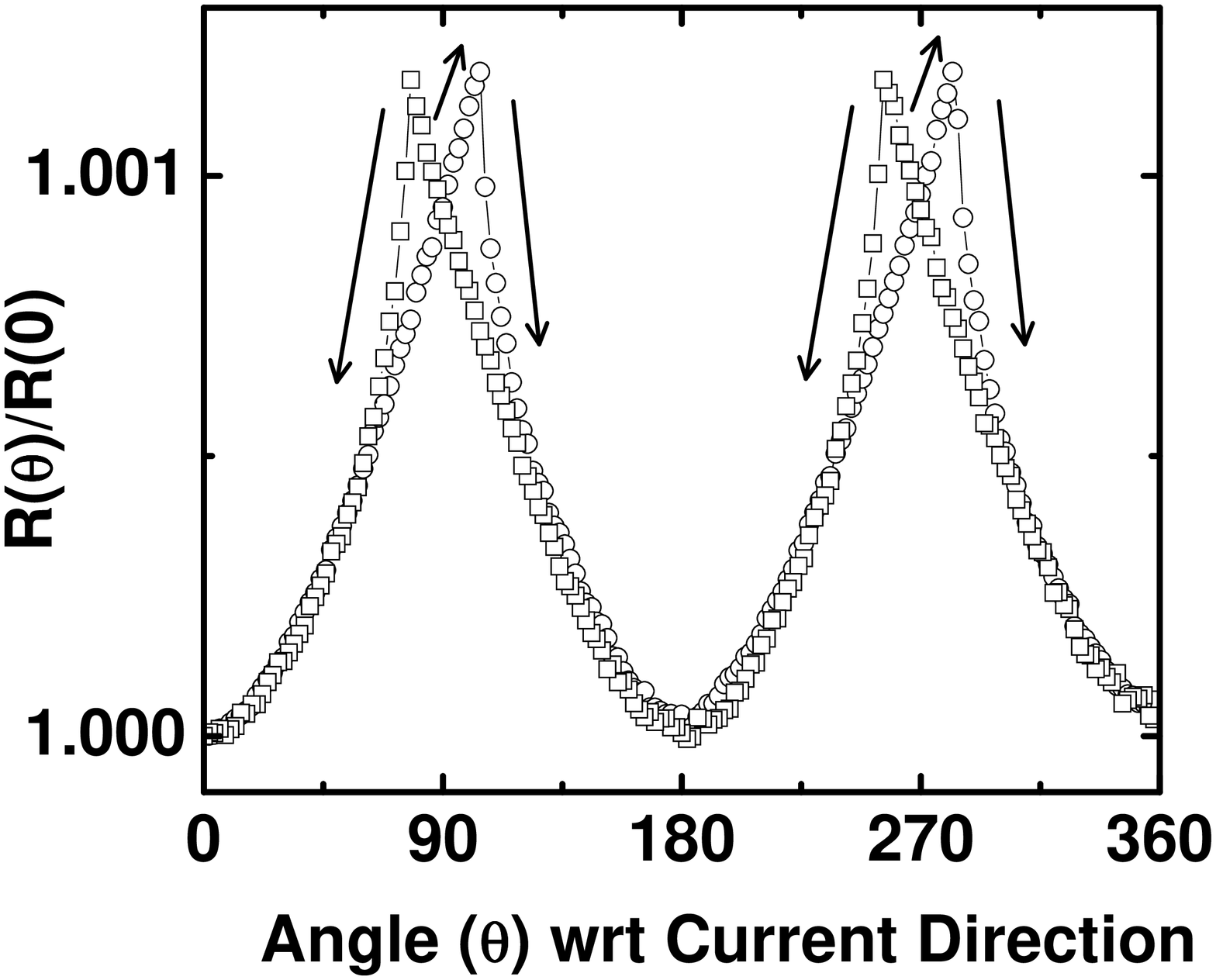}} \caption[Hysteresis in RMR for {(110)} LSMO]{Hysteresis in RMR for (110)
LSMO  film measured at 300 K and 120 Oe. The open circles and squares show the data when the field is varied in the forward (0-2$\pi$) and
reverse (2$\pi$-0) cycles respectively.} \label{amrhyst} \end{figure}

\begin{figure} \centerline{\includegraphics[width=3in, angle = -90]{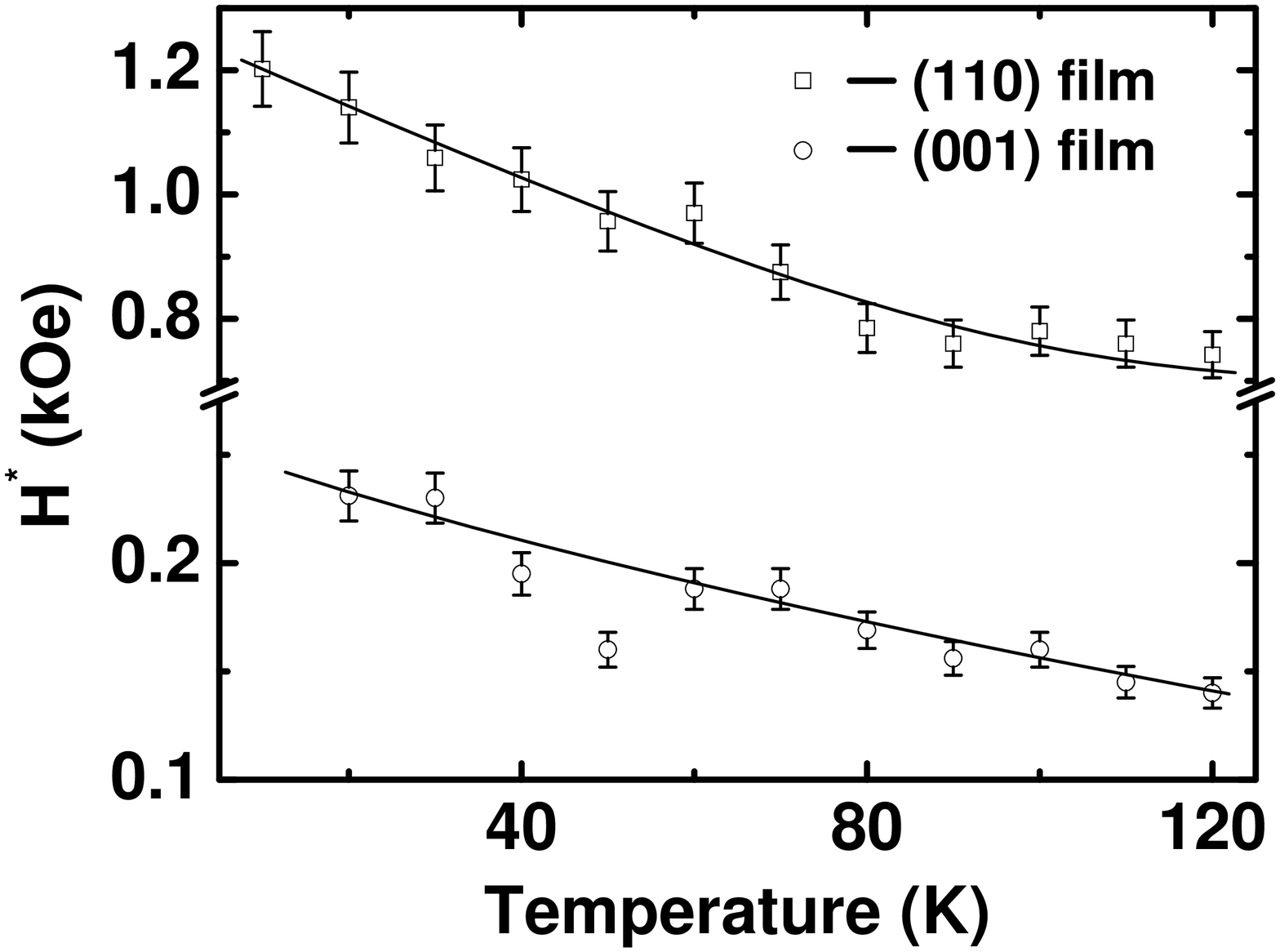}} \caption[H-T phase diagram for {(110)} and {(001)} LSMO
films.]{H-T phase diagram for (110)(open squares) and (001)(open circles) films. The solid lines are hand drawn to show the most probable
separation line between pinned and depinned states. These data clearly show that the pinning is much stronger in the case of the (110) films
than the (001) films. A change in y-scale emphasizes this point. H$^*$ is critical field for depinning demagnetization.} \label{phaseamr}
\end{figure}

In figs. \ref{amr100300K} and \ref{amr10010K} we have shown the RMR of the (001) oriented LSMO film  at 300 and 10 K respectively. For the
measurements at 300 K (fig. \ref{amr100300K}) we observe a symmetric angular dependence of the normalized resistance $R(\theta)/R(0)$ for all
values of the applied field, even at fields as low as 75 Oe.   At 10 K however (fig. \ref{amr10010K}), the $R(\theta)/R(0)$
values deviate from the symmetric behavior when the field is reduced below $\approx$ 500 Oe. A sharp drop in the resistance when the angle
$\theta$ is increased beyond $\theta_{peak}$ for these low field measurements is a remarkable feature of these data. This abrupt drop in
resistivity is accompanied by a hysteresis in the $R(\theta)/R(0)$ vs. $\theta$ plots when the angle is traced back from $2\pi$ to 0. A typical
hysteresis is shown in fig. \ref{amrhyst}.  The area under the hysteresis loop decreases with the field. In the left
inset of figs. \ref{amr110300k}, \ref{amr11010k}, \ref{amr100300K} and \ref{amr10010K} we have plotted $\theta_{peak}$ as a function of the
magnetic field strength. As noted  from these insets, the peak in RMR deviates rapidly from $\theta = \pi/2$ as the magnetic field is lowered
below a critical value H*. We have tracked the variation of H* with temperature between 10 and 120 K for the two types of films by measuring
$\rho(\theta)$ at several fields while the temperature is held constant. The result of such measurements is shown in fig. \ref{phaseamr}.

The discontinuous change in $\rho(\theta)$ below a characteristic field H* and the accompanying hysteresis indicate the existence of a
magnetization reorientation phase  transition (MRPT) \cite{Prinzjap, Prinzprb, Fisher} in the system driven by the torque of the $\vec{H}$ field
on $\vec{M}$. While a rigorous analysis of the MRPT carried out by minimizing the magnetization free energy functional allows the calculation of
the in-plane magnetocrystalline anisotropies, here we simply argue that the line in fig. \ref{phaseamr} separates the H-T phase space where the
magnetization is pinned along the easy axis and where it is free to rotate with the field. It is clear from fig. \ref{phaseamr} that the
magnetization in the (110) oriented film remains pinned along the easy axis over a much larger H-T phase space as compared to the (001) oriented
film.

\begin{figure}[t] \centerline{\includegraphics[width=3.5in, angle =
0]{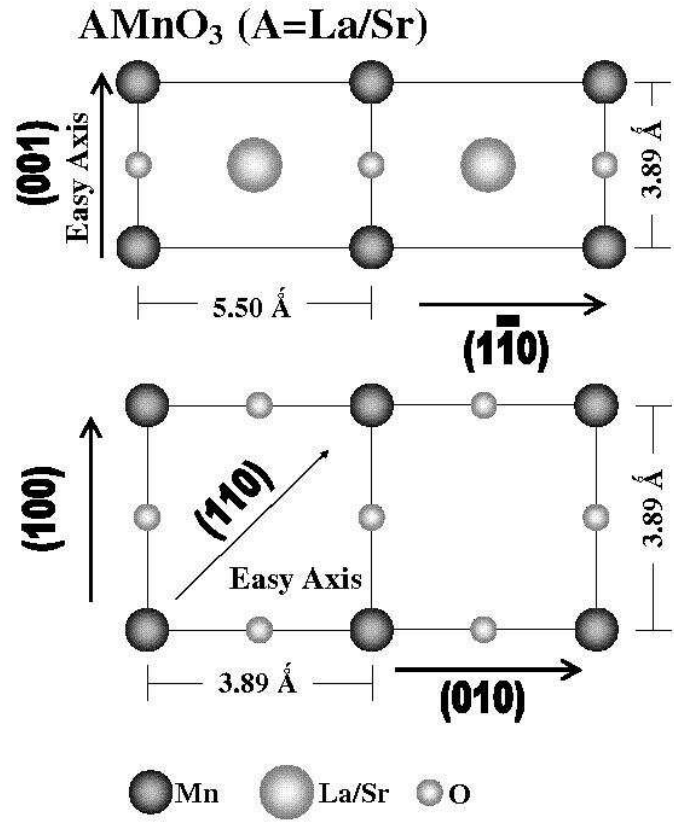}} \caption[Schematic of the ionic positions on the surface of {(110) and (001)}  films]{Schematic of the ionic positions on
the surface of (110) and (001) films are in the upper and lower panels respectively. For the (110) film, the two Mn ions along the (001)
direction are bridged by an oxygen ion. Hence the (001) direction acts as magnetic easy axis in these films. In case of (001) the (100) and
(010) directions are degenerate, hence the easy axis is along (110).} \label{schematic}
\end{figure}

The issue of why the magnetic easy axis in (001) and (110) films of manganites is different with a different degree of anisotropy energy as
suggested by the phase diagram of fig. \ref{phaseamr} has not been addressed in detail although several researchers have reported a difference
in the in-plane anisotropy axis of (001) and (110) LSMO films \cite{Suzuki,Tsui,Lecoeur}. It is generally agreed that while for the (001) films
the easy axis of magnetization is along (110) direction, the (110) films have uniaxial anisotropy with easy and hard directions along (001) and
\onebar respectively. This difference in (001) and (110) films can be understood if we consider the orientation of the Mn - O - Mn bonds on the
plane of such films. In fig. \ref{schematic},
 we sketch the atomic arrangement
on the top layer of the (001) and (110) cut \sto crystals and the way epitaxial registry is  maintained when the \LSMO film grows on the top. We
can see that in the case of the (001) oriented film, the Mn - O - Mn bonds are directed along the (100) and (010) directions making them
energetically degenerate. To avoid this degeneracy, the magnetization vectors prefer to lie along the (110) direction making it the easy axis.
The difference in the energy of the (110) and (100)/(010) states of magnetization is expected to be small. This is perhaps the reason why the
depinning field in this case is substantially lower. In the case of the (110) oriented films, the Mn - O - Mn bond, with a length of $\approx$
3.89\AA, is directed along the (001) direction, whereas along the \onebar direction the two Mn ions are separated by $\approx$ 5.5{\AA } without
any bridging oxygen atom. This makes the (001) direction the preferred direction for orientation of the magnetization vector. Furthermore, as
the (001) and {\onebar} directions are highly inequivalent, the pinning of $\vec{M}$ along (001) is expected to be robust, which is really the
case seen in the phase diagram of fig. \ref{phaseamr}. \clearpage

\subsection{Temperature dependence of RMR}
\begin{figure} \centerline{\includegraphics[width=3in, angle = -90]{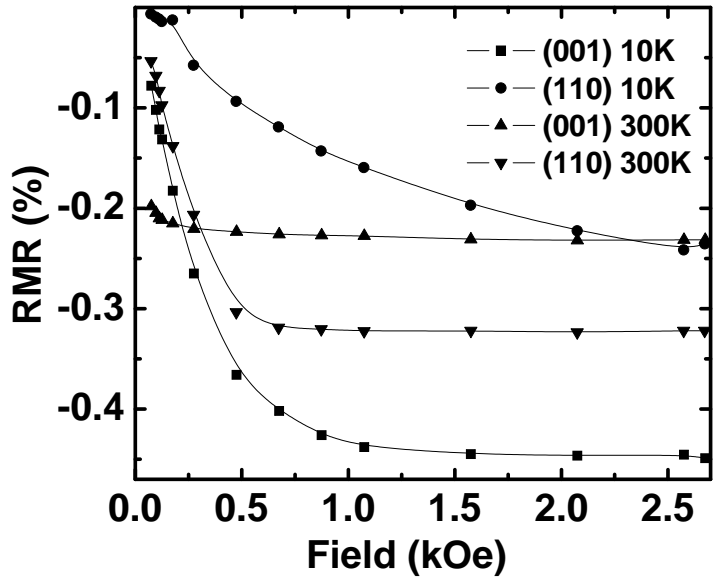}} \caption[Variation  of the RMR
percentage for {(110) and (001)} LSMO films]{Variation of the RMR percentage, defined as $\frac{R_\| - R_\bot}{R_\bot} \times 100$, with the
field for both (110) and (001) LSMO at 10 and 300 K. The RMR is negative for both films.} \label{amrper}
\end{figure}

In fig. \ref{amrper}  we plot the percentage RMR defined as $\left[100\left({\rho_\| - \rho_\bot}\right)/\left({\rho_\bot}\right)\right]$ at 10 and 300 K
for the two types of films as a  function of the field. The RMR is negative in both the cases. For the (001) film at 300 K, it is also nearly
constant at all fields. For the same film at 10 K, the magnitude of RMR first increases rapidly with the field and then acquires a saturation
value of $\approx -0.46\%$ at H $>$ 1 kOe. For the (110) film, the RMR at 300 K saturates to $\sim -0.32\%$ at H $\approx 0.5$ kOe. The same
film has the RMR of $\approx -0.2\%$ at 10 K. It is somewhat surprising to note that the RMR of the (110) film decreases while for (001) films
it increases as we go down in temperature.

\begin{figure} \centerline{\includegraphics[width=3in,angle=-90]{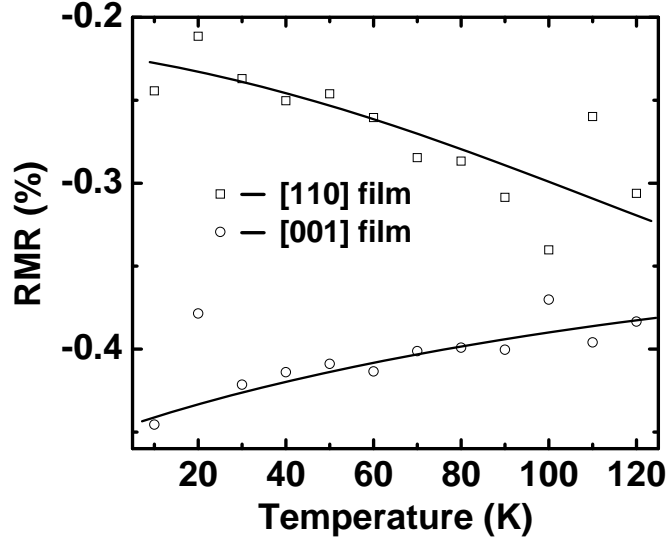}} \caption[{Temperature
dependence of RMR percentage between 10 and 120 K.}]{Variation of the RMR percentage with the temperature for (110)(open squares) and (001)(open
circles) LSMO films between 10 K and 120 K. The solid lines are hand drawn depicting the most probable trend in this temperature range. The
field applied in this case 2.5 kOe.} \label{rmrvartemp}
\end{figure}

In order to address this issue further, we have measured the 2500 Oe RMR of these films at several temperatures between 10 and 120 K. These data
are shown in fig. \ref{rmrvartemp}.   We note that the RMR of both the samples is negative and  its amplitude in the (001) case decreases monotonically as the temperature
is raised to 120 K. In fact the RMR deduced from $\rho_\bot(T)$ and $\rho_\|(T)$ measured between 120 and 240 K shows that this drop continues
till 240 K. For the (110) sample however, the magnitude of the RMR first increases with temperature till T $\simeq$ 200 K and then drops on
increasing the temperature further.

In the one band model of  Malozemoff \cite{Malozemoffprb} as applied to manganites by Ziese and Sena \cite{Ziese,Ziesesing}, the AMR is given
as; \begin{equation} \frac{\Delta\rho}{\rho_0} = -\frac{3}{4}\left[\frac{\lambda^2}{(H_{ex} - \Delta_{cf})^2} -
\frac{\lambda^2}{\Delta_{cf}^2}\right], \end{equation} where $\lambda$ is the spin - orbit coupling constant, $\Delta_{cf}$ the crystal field
splitting and $H_{ex}$ the exchange field. By putting in the values of $H_{ex}$, $\Delta_{cf}$ and $\lambda$ for a typical double exchange
manganite they find that $\Delta\rho/\rho_0 \approx -0.85\%$ in the limit of zero temperature. While a precise temperature dependence of the
parameters $H_{ex}, \Delta_{cf}$ and $\lambda$ is not known, in manganites of $T_c < 300$ K a significant enhancement in AMR near the Curie
temperature (T$_c$) has been observed. Herranz et al. \cite{Herranz,Bibes} have argued that as the Curie temperature is approached, the double
exchange mechanism is impeded by the enhanced Jahn-Teller distortion of the Mn-O octahedron with the concomitant unquenching of the orbital
angular momentum which enhances the spin-orbit interaction and hence the AMR. Our data, however, suggest that in these high quality films of
\LSMO where T$_c$ is $\simeq$ 360 K, the AMR decreases on warming to 300 K. While we have not been able to measure the AMR at T $>$ 300 K due to
experimental limitations, this is an important issue that needs to be addressed in the future.

A rigorous analysis of the RMR data of  our samples needs to be done using the approach of D\"oring \cite{Doring} which entails writing the
magnetoresistance of a cubic ferromagnet as a series in magnetization and current direction cosines $\alpha_i$ and $\beta_i$ respectively as
\cite{Doring},
\begin{eqnarray} {\frac{\Delta\rho}{\rho_0}} & = &
k_1\left(\alpha_1^2\beta_1^2 + \alpha_2^2\beta_2^2 + \alpha_3^2\beta_3^2 - \frac{1}{3}\right) \nonumber\\ && +
2k_2\left(\alpha_1\alpha_2\beta_1\beta_2 + \alpha_2\alpha_3\beta_2\beta_3 + \alpha_3\alpha_1\beta_3\beta_1\right) \nonumber\\ && + k_3(s - c) +
k_4\left(\alpha_1^4\beta_1^2 + \alpha_2^4\beta_2^2 + \alpha_3^4\beta_3^2 + \frac{2}{3}s - \frac{1}{3}\right) \nonumber\\ && +
2k_5\left(\alpha_1\alpha_2\alpha_3^2\beta_1\beta_2 + \alpha_2\alpha_3\alpha_1^2\beta_2\beta_3 +
\alpha_3\alpha_1\alpha_2^2\beta_3\beta_1\right)\label{doringeq}
\end{eqnarray} where $\rho_0$ is the resistivity at T = 0, the $k_i$'s
are phenomenological constants, $c$ is  a numerical constant depending on the easy axis direction, and $s = \alpha_1^2\alpha_2^2 +
\alpha_2^2\alpha_3^2 + \alpha_3^2\alpha_1^2$. For the situation when the magnetic domains are distributed equally among the easy axes in a zero
applied field, the constant c is 1/4 for the (110) easy axis and zero for the (100) easy axis \cite{Ziesesing}. \begin{figure}
\centerline{\includegraphics[width=4in,angle=0]{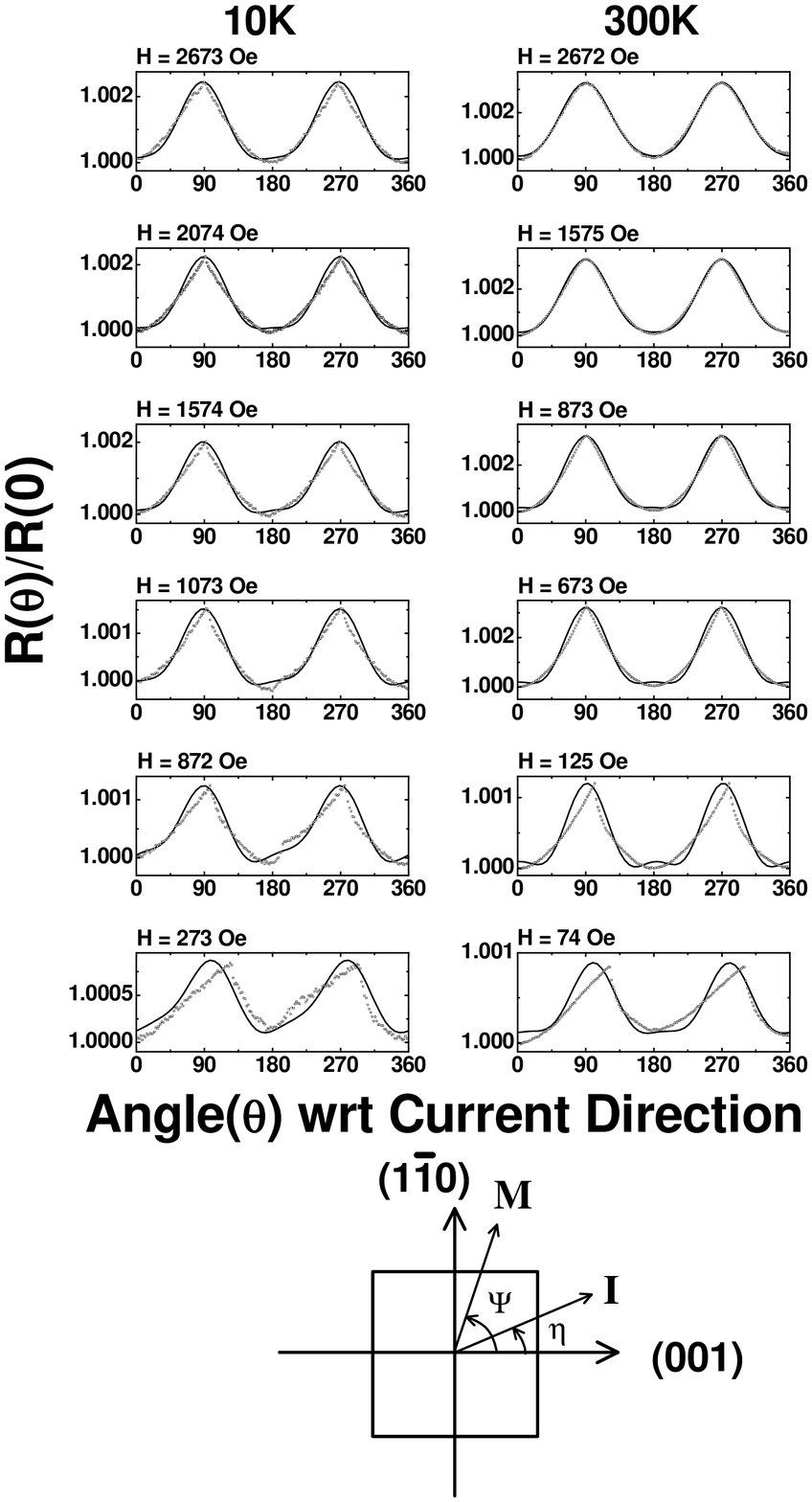}} \caption[{Fit of eq. \ref{eqnfin110} to the RMR data of the (110) film taken
at 10 and 300 K}]{Fit of eq. \ref{eqnfin110} to the RMR data of the (110) film taken at 10 and 300 K (left and right hand side respectively).
The dots are actual data and solid lines are the fitted curves. While at 300 K, a reasonably good fit to eq. \ref{eqnfin110} is seen down to
$\simeq$873 Oe, at still lower fields, the angular dependence is characterized by sharp drop of $R(\theta)/R(0)$ at $\theta > 90^0$. At these
fields the torque on $\vec{M}$ exerted by the external field is not strong enough for coherent rotation. For the 10 K data, the quality of fit
is poor even at the higher fields and worsens as it is reduced below 1574 Oe. A sketch of the three vectors $\vec{M}$, $\vec{I}$ and the unit
vector $\hat{n}$, and of the relevant angles is shown at the bottom of the figure. In our case $\vec{I} \| \hat{n}$ hence $\eta = 0$ and (001)
is the easy axis \cite{Easy}.} \label{fit110} \end{figure} For the purpose of our (110) films, the analysis is similar to that used by Gorkom
etal \cite{Gorkom} for (110) Fe with (001) easy axis. Following this work, we can write $\alpha_1 = -\alpha_2 =
\left(1/\sqrt{2}\right)\sin\psi$, $\alpha_3 = \cos\psi$, $\beta_1 = -\beta_2 = \left(1/\sqrt{2}\right)\sin\eta$ and $\beta_3 = \cos\eta$, where
$\psi$ is the angle between the magnetization and the (001) axis and $\eta$, the angle between the electrical current $\vec{I}$ and the (001)
direction. A sketch of the three vectors $\vec{M}$, $\vec{I}$ and the unit vector $\hat{n}$ pointing along the easy axis in a most generalized
situation is given at the bottom of fig. \ref{fit110}. On substituting these parameters in eq. \ref{doringeq} one gets,
\begin{eqnarray} \frac{\Delta\rho}{\rho_0} - \delta & = &
C_1\cos^2\psi + C_2\cos^4\psi + C_3\cos\psi\sin\psi + C_4\cos\psi\sin^3\psi \label{eqnfin110}
\;\;\;\;\;\;\;\;\;\;\;\;\;\;\;\;\;\;\;\;\;\;\;\;\;\;\;\;\;\;
\\ \mbox{where} && \nonumber\\ C_1 & = & k_1\left(\cos^2\eta -
\frac{1}{2}\sin^2\eta\right) - \frac{k_2}{2}\sin^2\eta + \frac{k_3}{2} + k_4\left(\frac{1}{3} -
\frac{1}{2}\sin^2\eta\right) \nonumber\\ &&+ \frac{k_5}{2}\sin^2\eta,\\
C_2 & = & - \frac{3k_3}{4} + k_4\left(\cos^2\eta + \frac{1}{4}\sin^2\eta - \frac{1}{2}\right) - \frac{k_5}{2}\sin^2\eta, \\ C_3 & = &
2k_2\cos\eta\sin\eta, \\ C_4 & = & k_5\cos\eta\sin\eta, \\ \rm{and} \;\; \delta & = & k_1\left(\frac{1}{2}\sin^2\eta - \frac{1}{3}\right) +
\frac{k_2}{2}\sin^2\eta + \frac{k_3}{4} + k_4\left(\frac{1}{4}\sin^2\eta - \frac{1}{6}\right)\end{eqnarray} In our case $\vec{I} \| \hat{n}$ and
hence $\eta = 0$. This result leads to $C_1 = k_1 + ({k_3}/{2}) + ({k_4}/{3}), C_2 = ({-3k_3}/{4}) + ({k_4}/{2}), C_3 = C_4 = 0 \mbox{  and }
\delta = ({-k_1}/{3}) + ({k_3}/{4}) - ({k_4}/{6})$. Since $\delta$ depends only on $k_1, k_2$ and $k_3$  which in turn are temperature dependent
coefficients, we can lump $\delta$ in $({\Delta\rho})/({\rho_0})$ for an isothermal measurement, and then, the right hand side of eq.
\ref{eqnfin110} can be written as $({R - R_0})/({R_0})$, where $R_0$ is the resistance at the peak position. \begin{figure}[t]
\centerline{\includegraphics[width=3in,angle=-90]{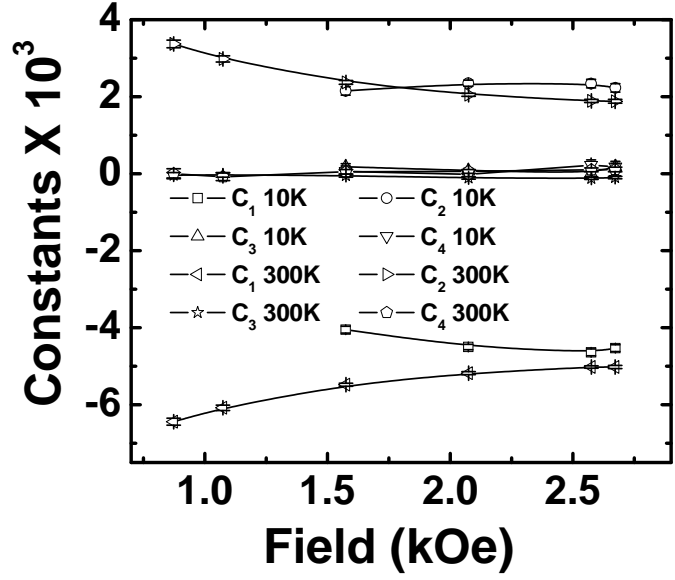}} \caption[{Field dependence of the coefficients  C$_1$, C$_2$, C$_3$ and
C$_4$}]{Field dependence of the coefficients  C$_1$, C$_2$, C$_3$ and C$_4$ obtained by fitting eq. \ref{eqnfin110} to the RMR data at 10 and
300 K. In both the, cases the parameter C$_3$ and C$_4$ stays close to zero. The result is remarkable as it validates the fitting, because we
already know that $C_3 = C_4 = 0$ due to $\vec{I} \| \hat{n}$ which makes $\eta = 0$. We also note that the ratio of the quadratic ($C_1$) and
biquadratic ($C_2$) term remains constant at 300 K in the field range of $\sim$800 Oe to $\sim$2700 Oe. At lower temperature, however, the term
appearing in fourth power of $\cos\psi$ is constant whereas the magnitude of the quadratic term increases with the field.} \label{const110}
\end{figure}

\begin{figure}
\centerline{\includegraphics[width=3in,angle=-90]{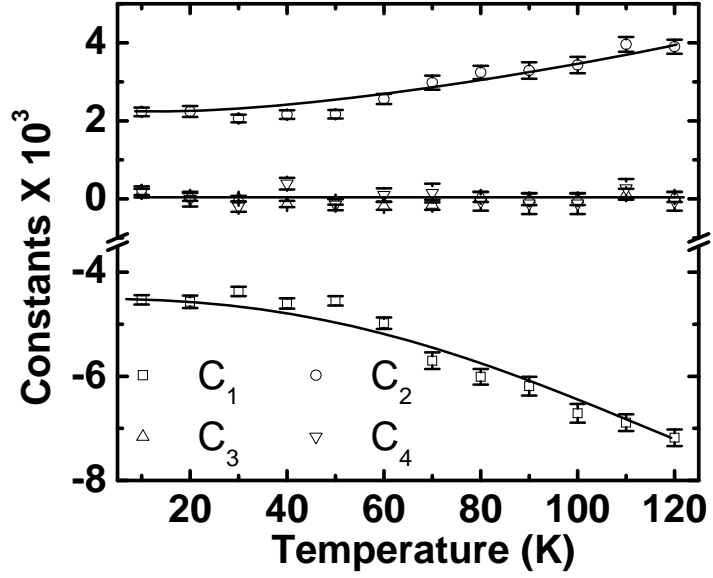}} \caption[{Temperature variation of the coefficients C$_1$, C$_2$, C$_3$ and
C$_4$.}]{Temperature variation of the coefficients  C$_1$, C$_2$, C$_3$ and C$_4$. The coefficients C$_3$ and C$_4$ remain close to zero at all
temperatures. The absolute value of C$_1$ and C$_2$ increase as we go up in temperature. The solid lines are hand drawn to indicate the most
probable trend.} \label{const110temp} \end{figure}

We have used eq. \ref{eqnfin110} to fit the RMR data for the (110) sample at 10 and 300 K, which were presented earlier in figs.
\ref{amr110300k} and \ref{amr11010k} respectively in a compact form. The quality of fit is shown for a representative set of data in fig.
\ref{fit110}. While we see a reasonably good fit to eq. \ref{eqnfin110}  down to $\simeq$ 873 Oe, the angular dependence at still lower fields
is characterized by a sharp drop in $\rho$ at $\theta > 90^0$ due to MRPT as discussed earlier. For the 10 K data, the quality of the fit is
poor even at high fields ($\approx 2000$ Oe) and it worsens when the field is reduced below 1574 Oe. Fig. \ref{const110}  shows the variation of
various fitting parameters with field at 10 and 300 K. In both the cases, the parameter C$_3$ and C$_4$ stays close to zero. This result is
remarkable as it validates the fitting, because we already know $C_3$ and $C_4$ are zero for our geometry ($\vec{I} \| \hat{n}$ \& $\eta = 0$).
We also note that the ratio of the quadratic ($C_1$) and biquadratic ($C_2$) coefficients remains the same at 300 K in the field range of $\sim$
800 to $\sim$ 2700 Oe. At lower temperatures, however, the term appearing in the fourth power of $\cos\psi$ remains constant where as the
magnitude of the quadratic term increases with the field. In fig. \ref{const110temp}, we have traced the variation of these coefficients with
temperature between 10 and 120 K at an applied field of 2.5 kOe. While C$_3$ and C$_4$ stay close to zero for all temperatures, the absolute
values of C$_1$ and C$_2$ increase with temperature. Before we discuss the significance of these coefficients (C$_i$'s), it is pertinent to
discuss the angle dependent data for the (001) epitaxial films. 

\begin{figure}
\centerline{\includegraphics[width=4in,angle=0]{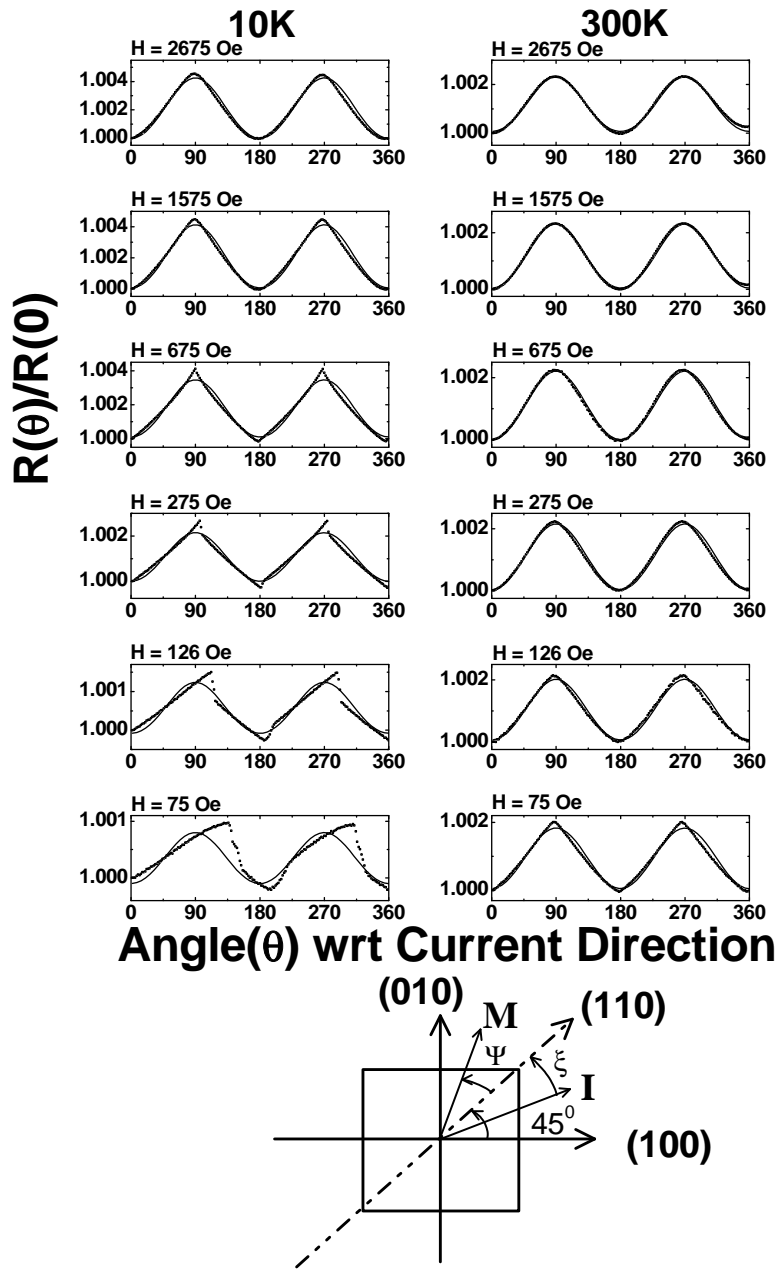}} \caption[{The fit of eq. \ref{eqnfin100} to the RMR data for the (001)
sample}]{The fit of eq. \ref{eqnfin100} to the RMR data for the (001) sample at 10 and 300 K.  It is evident that at 300 K, the model (eq.
\ref{eqnfin100}) correctly describes the behavior of the RMR down to fields as low as $\simeq$ 275 Oe. At still lower fields, although the
deviations become large, the peak and valleys of the data are correctly reproduced. The situation, however, is quite different at 10 K, here
even at the highest field deviation from the model are evident near the maxima. These deviations become prominent at lower fields. A sketch of
the three vectors $\vec{M}$, $\vec{I}$ and the unit vector $\hat{n}$ directed along the easy axis is shown at the bottom of the figure. In our
experiment, $\xi = \pi/4$ and $\psi = \theta - \pi/4$ where $\theta$ is the angle between the applied field and the current direction. The
current is flowing along the hard axis and the easy axis is (110) \cite{Easy}.} \label{fit100}
\end{figure}

For the (001) film, the easy axis is along (110) whereas the current is along the (100) direction. This makes the direction cosines of the
magnetization ($\alpha_i$'s) and the direction cosines of the current ($\beta_i$'s) with respect to the cubic axis as $\alpha_1 =
({1}/{\sqrt{2}})\left(\cos\psi - \sin\psi\right)$, $\alpha_2 = ({1}/{\sqrt{2}})\left(\cos\psi + \sin\psi\right)$, $\alpha_3 = 0$ and $\beta_1 =
({1}/{\sqrt{2}})\left(\cos\xi + \sin\xi\right)$, $\beta_2 = ({1}/{\sqrt{2}})\left(\cos\xi -\right.$ $\left. \sin\xi\right)$, $\beta_3 = 0$ with
$c = 1/4$. The final expression for the resistivity in this  case is'
\begin{eqnarray} \frac{\Delta\rho}{\rho_0} - \gamma & = &
A_1\cos^2\psi + A_2\cos^4\psi + A_3\sin\psi\cos\psi\label{eqnfin100}\\ \mbox{where}&& \;\;\;\;\;\;\nonumber
\\ A_1 & = & \frac{k_4}{3} -
k_2\left(1-2\cos^2\xi\right) -k_3,\\
A_2 & = & k_3-\frac{k_4}{3},\\
A_3 & = & -2\left(k_1+k_4\right)\cos\xi\sin\xi,\\ \rm{and} \;\;\gamma & = & \frac{k_1}{6}+k_2\left(\frac{1}{2} -
\cos^2\xi\right)+\frac{k_4}{12}\end{eqnarray} In this case $\xi = \pi/4$ and $\psi = \theta - \pi/4$ where $\theta$ is the angle between the
applied field and the current direction. A sketch of the three vectors $\vec{M}$, $\vec{I}$ and the unit vector $\hat{n}$ pointing along the
easy axis in a most generalized situation is given at the bottom of fig. \ref{fit100}. Since $\xi = \pi/4$, it
turns out that in this case $A_1 = -A_2$. Here we have assumed $\left[({\Delta\rho})/{\rho_0}\right] - \gamma \approx ({R - R_0})/{R_0}$, where
$R_0$ is the resistance when  the field is aligned along the (110), the easy axis. In fig. \ref{fit100}, \begin{figure}
\centerline{\includegraphics[height=3in,angle=-90]{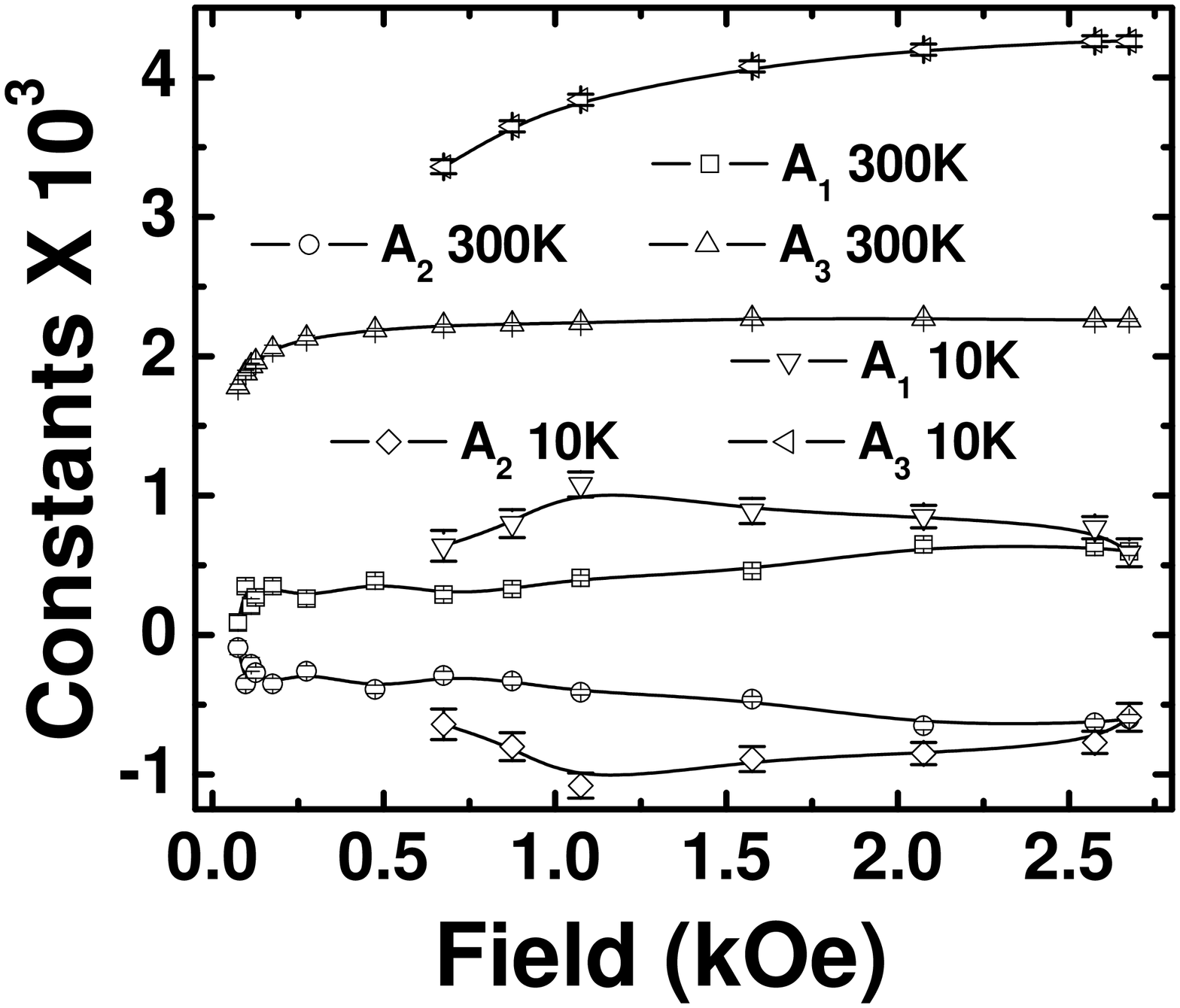}} \caption[{Field dependence of the phenomenological coefficients A$_1$,
A$_2$ and A$_3$}]{Field dependence of the phenomenological coefficients A$_1$, A$_2$ and A$_3$ obtained by fitting eq. \ref{eqnfin100} to the
RMR data. Here $A_1 = -A_2$ as expected from the model under the geometry of our measurements. The coefficient $A_3$ at both temperatures first
increases with the field and then becomes constant. It should be noted that $A_3$ in eq. \ref{eqnfin100} appears as a coefficient of
$\cos\psi\sin\psi$ which has extrema at $45^0, 135^0, 225^0$ and $315^0$.} \label{const100}
\end{figure} we have shown the fit
of eq. \ref{eqnfin100} to the RMR data for the (001) sample at 10 and 300 K. It is evident that at 300 K the model (eq. \ref{eqnfin100})
correctly describes  the behavior of the RMR down to fields as low as $\simeq$ 275 Oe. At still lower fields, although the deviations become
large, the peaks and valleys of the data are correctly reproduced. The situation, however, is quite different at 10 K; here even at high fields,
deviation from the model are evident near the maxima.

\begin{figure}
\centerline{\includegraphics[width=3in,angle=-90]{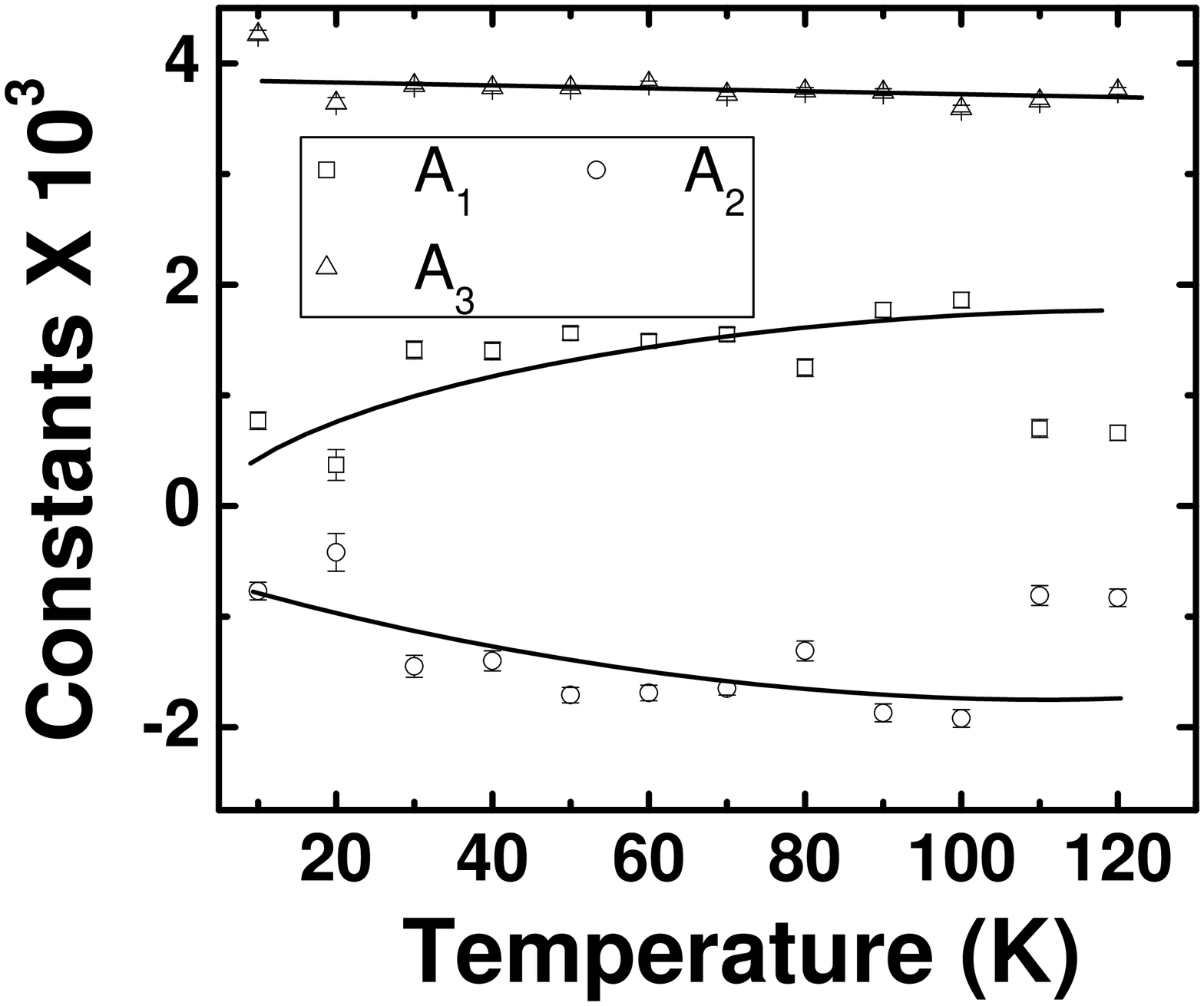}} \caption[{Temperature variation of phenomenological coefficients A$_1$,
A$_2$ and A$_3$.}]{Temperature variation of phenomenological coefficients A$_1$, A$_2$ and A$_3$. In this case A$_3$ remains almost constant
throughout the temperature range. The coefficient $A_1$ increases with temperature while $A_2$ decreases with temperature.}
\label{const100temp}\end{figure}

We now discuss the behavior of the coefficients $A_1, A_2$ and $A_3$ whose variation as a function of field is shown in fig. \ref{const100}.
First of all, we note that $A_1 = -A_2$ as expected from the model under the  geometry of these measurements. The coefficient $A_3$ at both
temperatures increases with the field and then becomes constant at a high field. Moreover, $A_3$ increases by a factor of 2 at 10 K. It should
be noted that $A_3$ in eq. \ref{eqnfin100} appears as a coefficient of $\cos\psi\sin\psi$ which has extrema at $45^0, 135^0, 225^0$ and $315^0$.
A higher weightage of $A_3$ will lead to large deviations from the $\cos^2\psi$ dependence. Fig. \ref{const100temp} shows the dependence of
coefficients $A_1, A_2$ and $A_3$ with temperature. We can clearly see that $A_3$ remains almost constant with temperature. The coefficient
$A_1$ increases with temperature while $A_2$ decreases with temperature. 

The data presented in figs. \ref{amr110300k}, \ref{amr11010k}, \ref{amr100300K} and \ref{amr10010K} clearly show that the RMR is both
temperature and film orientation dependent. From these data, we also conclude that the (110) films have higher anisotropy energy then the films
with (001) orientation. This becomes clear from the fact that at 300 K as well as at 10 K, the coherent rotation of magnetization with the
applied field which results in a $\cos^2\psi$ dependence of RMR appears at much higher fields for (110) films then for (001) films. Secondly, a
look at the AMR percentages calculated from these data shows that for the (110) films, the absolute value of the AMR increases with the
temperature while it is just the opposite for the (001) films. We believe that the magnetization vector of the (110) films at low temperature is
strongly pinned along the easy axis due to the large anisotropy energy. At higher temperatures, the thermal energy $k_BT$ helps in depinning and
the free rotation of $\vec{M}$ along with the external field $\vec{H}$. This leads to an enhanced RMR at higher temperatures. Of course, if the
field strength is increased further, free rotation would become possible at 10 K as well. The required fields, however, may be well beyond what
has been used in these experiments. A direct support to this argument comes from the non-saturating trend of the 10 K AMR of the (110) film as a
function of the field (fig. \ref{amrper}). A very interesting picture emerges from the value of the constants so calculated. While for the
(110), film the RMR is mostly dependent on even powers of $\cos\psi$, the dependence of the RMR for the (001) films is predominantly
$\cos\psi\sin\psi$ where $\psi$ is the angle between the applied field and the easy axis of film. A qualitative explanation for this observation
can be given if we refer to fig. \ref{schematic} where the direction of the Mn-O-Mn bond of the (001) STO surface is shown. As we have stated
earlier, the (110) direction is the easy axis because of the degeneracy of the (010) and (100) directions in a zero field. An external field
applied at 45$^0$ with respect to the (110) direction lifts this degeneracy, which perhaps is the reason why MR has a strong contribution from
the $\cos\psi\sin\psi$ term with its maximum at $\psi = 45^0$.

A discussion on the temperature and angular dependence of RMR would remain incomplete unless we address the role of electron localization in
cyclotron orbits. The orbital magnetoresistance resulting due to electron trapping is given as \cite{Gorkom},
\begin{equation}\left(\frac{\Delta\rho}{\rho}\right)_{OMR} = \left(\frac{eB_\bot\tau}{m^*}\right)^2 \label{omr} \end{equation} in the limit
$\left(eB_\bot\tau/{m^*}\right)^2 \ll 1$, where $\tau$ is the relaxation time, m$^*$ is the effective mass of the electron and B$_\bot$ the
component of the magnetic induction perpendicular to the current I. It is expected to be negligible when the carrier mean free path l is much
shorter than its cyclotron orbit (r$_c$). With the known hole density ($\approx 1.16\times10^{28}$/m$^3$), Fermi energy (1.8eV)
\cite{Stroudlsmo} of LSMO, 10 K resistivity and magnetic induction B (=H+4$\pi$M) of the (001) film, which are 0.11 m$\Omega$-cm and 7500 G for
(001) films respectively, we obtain $l \approx 2nm$ and $r_c \geq 6.8 {\mu}m$. Similarly for the (110) film, l and r$_c$ are $\approx 3nm$ and
$\geq 6.6 {\mu}m$ respectively. From these numbers, it can be concluded that the OMR will make negligible contribution to the RMR in these
films. Following Gorkom \cite{Gorkom}, the angular dependence of OMR can be written as;\begin{equation}
\left(\frac{\Delta\rho}{\rho}\right)_{OMR} = \kappa \left( \frac{\rho_{300}^2}{\rho^2} \right)\sin^2\theta, \label{rhoomr} \end{equation} where
$\kappa = \left(B/ne\rho_{300}\right)^2$ and $\rho_{300}$ is the resistivity at T = 300 K. Using the $\rho_{10 K}$ and $\rho_{300 K}$ data for
the (001) film and eq. \ref{rhoomr} we obtain $\kappa \approx 2.3\times10^{-10}$ and $\left(\Delta\rho/\rho\right)_{OMR} \approx
1.4\times10^{-5}\%$ which is much too small as compared to the measured RMR. From this analysis it can be concluded that the origin of the
angular magnetoresistance is these films is the spin-orbit coupling dependent AMR effect.

\subsection{Magnetoresistance in \LSMO films}
\begin{figure}[t] \centerline{\includegraphics[width=3in,angle=0]{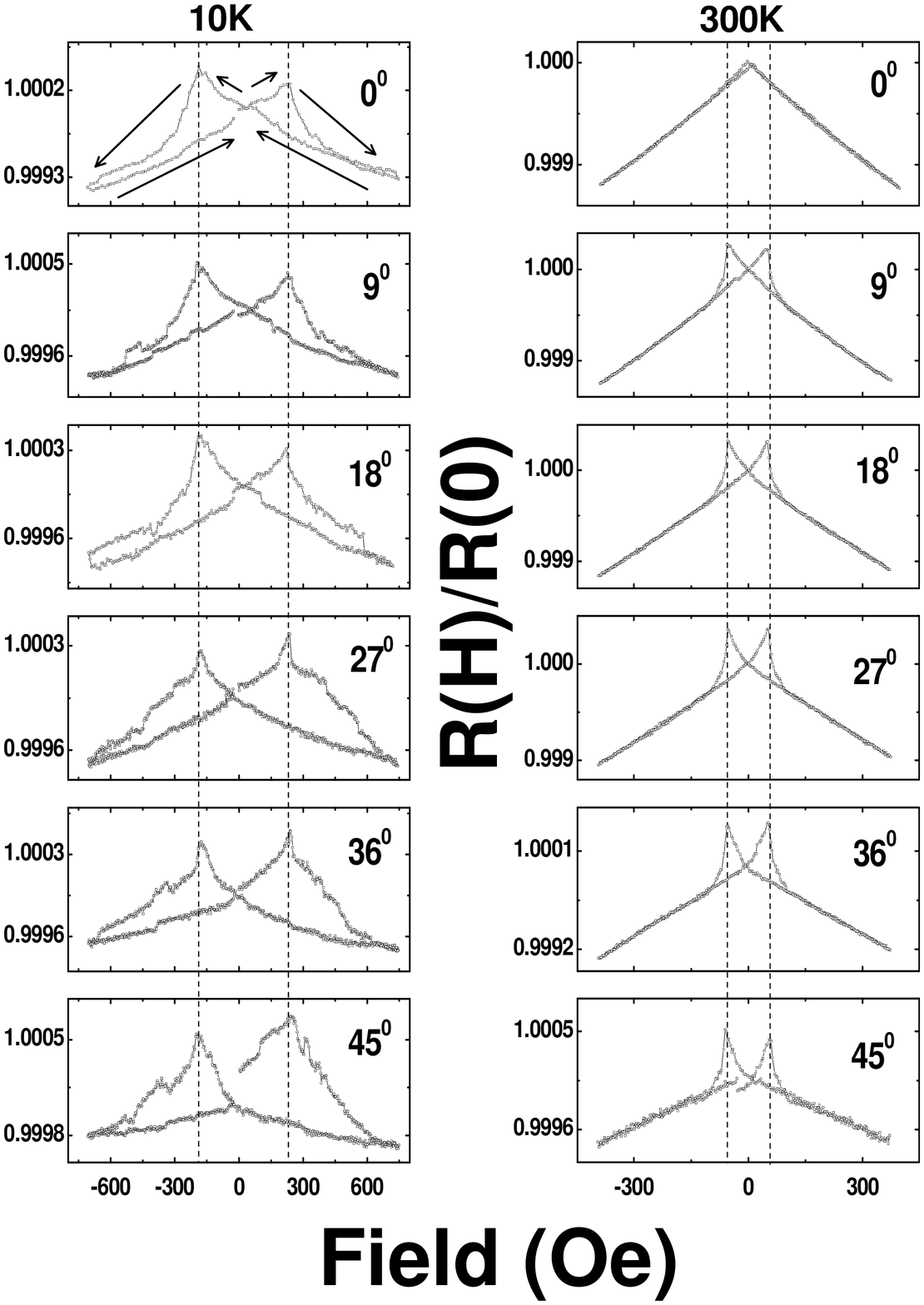}} \caption[{R(H)/R(0) vs. H data for the
(110) sample at 10 K and 300 K}]{R(H)/R(0) vs. H data for the (110)  sample at 10 K (left panel) and 300 K (right panel) taken at various angles
between the applied field and the current direction. The current in this case is flowing along the easy axis, and the hard axis is in-plane and
90$^0$ to the current direction. Arrows in the figure mark the trajectory followed by the resistance as the field is swept from positive or
negative extremities. At 10 K the resistance increases superlinearly as the field reduced from 700 Oe till it reaches a critical negative value
H$_c$. On increasing the field further in the reverse direction, the resistance drop, first rapidly and then gradually. The resistance profile
during -H$_{max}$ to +H$_{max}$ field sweep is a mirror image of the +H$_{max}$ to -H$_{max}$ sweep. A large hysteresis is evident in the figure
whose area increases with the angle $\theta$ between $\vec{H}$ and $\vec{I}$. However the critical field $\pm H_c$ remains the same for all
angles within the experimental error ($\pm$10 Oe).} \label{rh110}
\end{figure}

In fig. \ref{rh110}  we have shown the R(H)/R(0) vs. H data for the (110)
sample at 10 K (left panel) and 300 K right panel) taken at various angles between the applied field and the current direction. The current in
this case is flowing along the easy axis \cite{Easy} and the hard axis is in-plane and 90$^0$ to the  current direction. Arrows in the figure
mark the trajectory followed by the resistance as the field is swept from positive to negative extremities. At 10 K, the resistance increases
superlinearly as the field is reduced from 700 Oe till it reaches a critical negative value H$_c$. On increasing the field further in the
reverse direction, the resistance drops, first rapidly and then gradually. The resistance profile during --H$_{max}$ to +H$_{max}$ field sweep
is a mirror image of the +H$_{max}$ to --H$_{max}$ sweep. A large hysteresis is evident in the figure whose area increases with the angle
$\theta$ between $\vec{H}$ and $\vec{I}$. However the critical field $\pm H_c$ remains the same for all angles within the experimental error
($\pm$10 Oe), and also compares well with the coercive field deduced from the M-H loop(fig. \ref{mhloop}). For the measurement performed at 300
K, the R(H) curve is mostly reversible except for a narrow range of the field between $\pm H_c$ where twin peaks appear in the resistance for
non-zero values of $\theta$. The reversible part shows a $\rho \propto H^\tau$ dependence with $\tau$ nearly independent of the angle $\theta$.

\begin{figure}[t]
\centerline{\includegraphics[width=3in,angle=0]{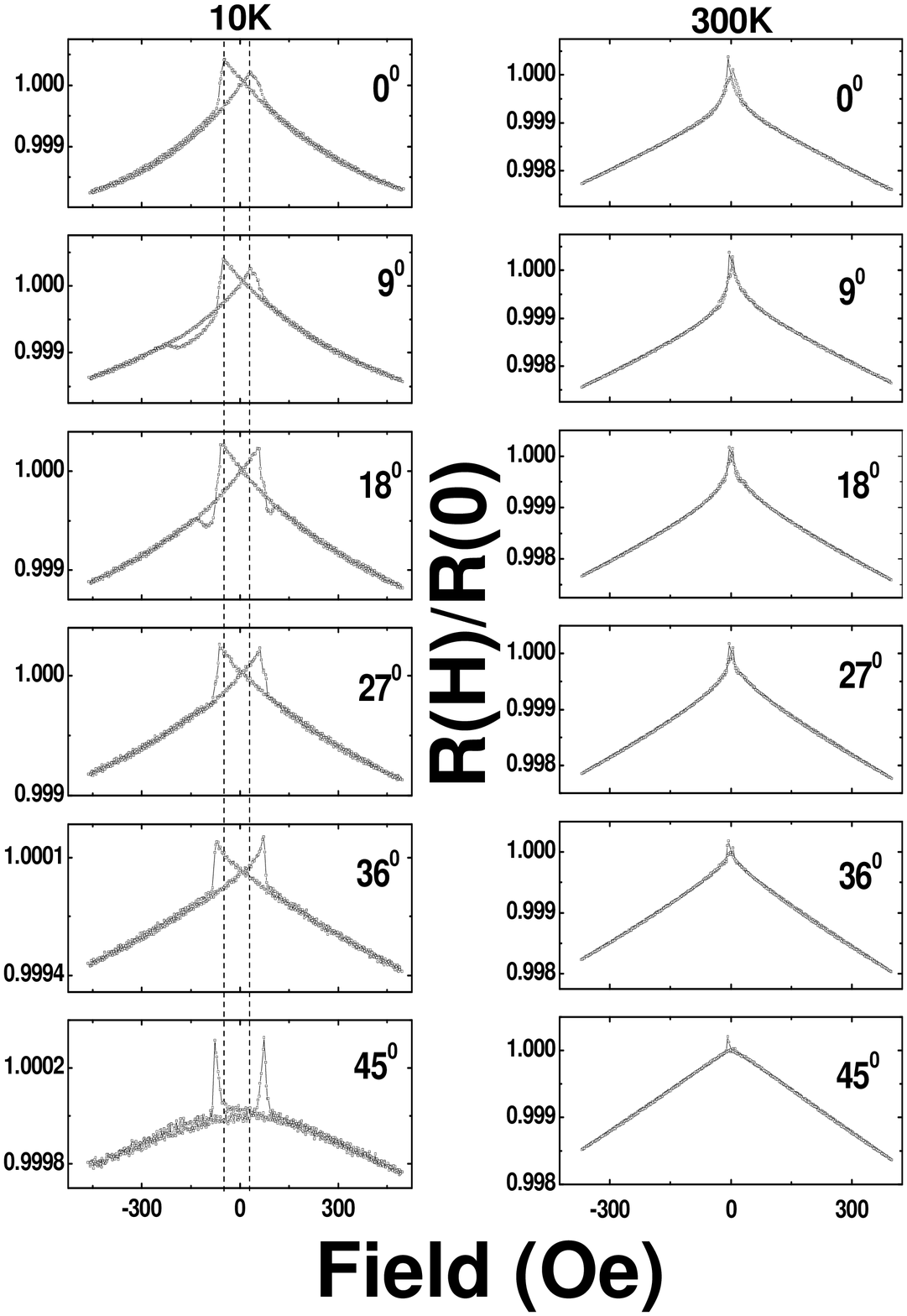}} \caption[{R(H)/R(0) vs. H data for the (001) sample at 10 K and 300 K}]{The
isothermal magnetoresistance at different angles between $\vec{I}$ and $\vec{H}$ for the films with (001) orientation for measurements performed
at 300 K (right panel) and 10 K (left panel). The current in this case is  flowing along the hard axis and the easy axis is 45$^0$ to the
current direction. At 10 K and $\theta = 0^0$, the R(H) curve is mostly reversible except for the twin peaks appearing at $\pm H_c$ which agrees
with the coercive field deduced from M-H measurements. On increasing the angle $\theta$ (moving away from the easy axis), two interesting
features emerge from the data. First, the critical field at which resistivity drops precipitously shifts to higher values, and second, the field
dependence on increasing field becomes superlinear to sublinear. At 300 K, $\rho(H)$ is devoid of detectable hysteresis. Also the field
dependence in this case remains linear at all angles.} \label{rh100} \end{figure}

The isothermal magnetoresistance at different angles between $\vec{I}$ and $\vec{H}$ for films with (001) orientation is shown in fig.
\ref{rh100} for measurement performed at 300 K (right panel) and 10 K (left panel). The  current in this case is flowing along the hard axis and
the easy axis \cite{Easy} is 45$^0$ to the current direction. At 10 K and $\theta = 0^0$ the R(H) curve is mostly reversible except for the twin
peaks appearing at $\pm H_c$ which agrees with the coercive field deduced from M-H measurements. On increasing the angle $\theta$ (moving away
from the easy axis), two interesting features emerge from the data. First, the critical field at which the resistivity drops precipitously
shifts to higher values and second, the field dependence on increasing field becomes superlinear to sublinear ($\rho \propto H^\tau, \tau = -8.8
\times 10^{-4} \mbox{ at } \theta = 0, \tau = -1.6 \times 10^{-4} \mbox{ at } \theta = 45^0$). At 300 K,  $\rho(H)$ is devoid of detectable
hysteresis. Also the field dependence in this case remains linear at all angles.

The primary factors that contribute to the low field MR in
these systems are the colossal magnetoresistance effect and the tunneling magnetoresistance if the system has a non zero granularity. The
explanation in the case of a non-granular film lies in the multidomain configuration model proposed by O'Donnell et al. \cite{Ecksteinprb2}. At
high fields, the magnetization is aligned in the direction of the applied field. But as the field is  lowered and then applied in the opposite
direction, the magnetization has to reverse at some point. Considering that the reversal will be rapid, we would expect a change in resistivity
$\Delta\rho$ due to colossal magnetoresistance (CMR) given as \cite{Ecksteinprb2}\begin{equation} \Delta\rho \approx \rho\left(M_0 + \chi
H_{sw}\right) - \rho\left(M_0 - \chi H_{sw}\right)\label{cmrm}\end{equation} where $H_{sw}$ is the switching field (the field at which the
magnetization reversal occurs), $M_0$ is the spontaneous magnetization and $\chi$ is the susceptibility. This change will be seen for a
transition in magnetization from anti-parallel $\left(M \approx M_0 - \chi H_{sw}\right)$ to parallel $\left(M \approx M_0 + \chi H_{sw}\right)$
to the applied field. As pointed out by O'Donnell et al. \cite{Ecksteinprb2}, this simple model cannot explain some features in these data.
Considering the case when we treat the film as a single domain, the magnetization either flips directly from antiparallel to parallel alignment
or it comes to the parallel alignment following a two step process, passing from antiparallel, to transverse, to parallel. When $\vec{M}$ is
parallel or antiparallel to the applied field we have,\begin{equation}\left|\vec{M}\right| \approx M_0 \pm \chi
H\label{singledomainm}\end{equation} and for $\vec{M}$ perpendicular to the applied field, $\left|\vec{M}\right| \approx M_0$ which is
independent of $H$. Below T$_c$ and at very low fields, we can assume that $\chi H \ll M_0$ for a single domain sample. So in this case the
linear approximation of eq. \ref{singledomainm} is correct. So the CMR below T$_c$ is a first order expansion in the small parameter $\chi H$
about $\rho\left(M_0\right)$. From these arguments and taking into account that the CMR is linear, the change in resistivity upon flipping of
the magnetization from antiparallel to parallel expressed by eq. \ref{cmrm} can be approximated to \begin{equation}\Delta\rho = 2\chi H_{sw}
\left.\frac{d\rho}{dM}\right|_{M_0}\end{equation} The single-domain model also fails to explain the deviation from linearity. To overcome the
shortcomings of the single domain model, O'Donnell et al. \cite{Ecksteinprb2} proposed a multidomain-model. In this model, the magnetization
reversal proceeds via motion of the domain walls. \clearpage \noindent The resistivity for a sample with the applied field along the easy axis
can be written as \cite{Ecksteinprb2}
\begin{equation}\rho \approx x\rho_{par} + y\rho_{antipar} + z\rho_{transverse}\label{multidomain}\end{equation} where $\rho_{par},
\rho_{antipar} \hbox{ and } \rho_{transverse}$ are the resistivities of the domains parallel,  antiparallel and transverse to the applied field.

Now we try to analyze our MR data in the light of the multi-domain model and take into account that our (110) LSMO films show uniaxial and (001)
films show biaxial anisotropy. Looking at the data for (110) samples at 300 K (right hand panel of fig. \ref{rh110}), we do not see any
discontinuous change in resistivity when $H \| J$. So it looks like that as soon as the field is reversed, the magnetization also reverses. For
other angles, we see a clear discontinuous change. This behavior can be understood by arguing that as the field is slowly increased from zero,
the moments align along the hard axis that is perpendicular to the current and as a result, the resistivity increases. Once the field crosses a
threshold limit, the magnetization flips abruptly towards the applied field direction. Since T=300 K is very close to $T_c$ for LSMO, it is very
weakly ferromagnetic at that temperature. Hence it is possible that for fields aligned along the easy axis we may not see any discontinuous
jump. The data at 10 K (left hand panel of fig. \ref{rh110}) show large hysteretic loop in MR. Secondly, the transition is not very sharp. This
can be attributed to the deviation from the square hysteresis loop that this sample shows at 10 K (top panel of fig. \ref{mhloop}). In this
case, a transition is also seen for $H\| J$ since at 10 K, LSMO is strongly ferromagnetic.

The R(H)/R(0) curves for the (001) sample reveal some interesting facts as well. If we look at the data at T = 300 K (right hand panel of fig.
\ref{rh100}) we see linear R-H behavior when $H$ is parallel to the easy axis (i.e. $45^0$ to the current direction) \cite{Easy}.  Some
non-linearity sets in once the applied field direction moves away from the easy axis but the non-linearity is not hysteretic. The observed
non-linear dependence presumably arises from the slow rotation of the magnetization vector with the increasing external field.

The data at 10 K (left hand panel of fig. \ref{rh100}) show some features of first order transition but the explanation is a little different in
this case since (001) samples show biaxial anisotropy. In this case, the muti-domain model is a proper explanation of the effect. So, when the
applied field is along the current direction at low fields, domains aligned in the direction of the easy axis start appearing. Since these
domains are aligned away from the current direction the resistance increases. At a particular field, the moments switch to the direction of the
applied field and we see the step in the resistivity. But in this case, the step is also present for $H \|$ to the easy axis that is 45$^0$ to
the current direction. This can be attributed to the fact that for the field to flip completely, it has to cross two hard directions. So a
minimum magnetic energy is required to completely flip the moments.

\section{Conclusion}
We have carried out a comparative study of the isothermal magnetoresistance of (001) and (110) epitaxial \LSMO films as a function of the angle
between the current  and the coplanar magnetic field at several temperatures between 10 and 300 K. The magnetic easy axis of the (001) and (110)
films is along the (110) and (001) directions respectively. In view of the similar texture of these two types of films, which can otherwise
contribute to shape anisotropy, we conclude that the easy axis is fundamentally related to the orientation of the Mn-O-Mn bonds on the plane of
the substrate. The isothermal resistance $\rho_\bot$ and $\rho_\|$ for $\vec{I} \bot \vec{H}$ and $\vec{I} \| \vec{H}$ configurations
respectively of these two type of films obey the inequality $\rho_\bot > \rho_\|$ for all fields and temperatures. However, $\rho(\theta)$ shows
deviation from the simple $\cos^2\theta$ dependence at low fields due to pinning of the magnetization vector $\vec{M}$ along the easy axis. This
effect manifests itself as a discontinuity in $\rho(\theta)$ at $\theta > \pi/2$ and a concomitant hysteresis on reversing the angular scan. We
establish a magnetization reorientation phase transition in this system and extract the H-T phase space where $\vec{M}$ remains pinned. A robust
pinning of magnetization seen in (110) films suggests strong in-plane anisotropy as compared to the (001) films. We have carried out a full
fledged analysis of the rotational magnetoresistance of the two types of epitaxial LSMO films in the framework of the D\"oring theory
\cite{Doring} of anisotropic magnetoresistance in metallic ferromagnet single crystals. We note that strong deviation from the predicted angular
dependence exists in the irreversible regime of magnetization. A simple estimation of orbital MR in these films suggest that the RMR is
dominated by spin-orbit interaction dependent anisotropic magnetoresistance.

\chapter[LSMO-YBCO-LSMO heterostructures]{\LSMO-\YBCO-\LSMO heterostructures}
\section{Introduction}
The transport of quasiparticles and paired electrons across superconductor-ferromagnet (SC-FM) proximity effect junctions provides valuable
information on the degree of spin polarization in the ferromagnet, exchange-field-induced inhomogeneous superconductivity at the SC-FM
interface, symmetry of the SC order parameter and a plethora of other effects arising from the antagonism between superconductivity and
ferromagnetism \cite{Izyumov, Goldman, Chien, Garifullin, Keizer, Demler, Tagirov, Buzdin, Pokrovsky, Bergeret}, some of which are
technologically important \cite{You, Eom}. Investigations in heterostructures of ferromagnetic manganites and hole doped cuprates present a rich
field of research due to the $d$-wave symmetry of the SC  order parameter and a high degree of spin polarization in manganites. One of the
systems of interest for such studies is the \LSMO-\YBCO-\LSMO [LSMO - YBCO - LSMO] trilayer. While magnetotransport and magnetic ordering in
such manganite - cuprate heterostructures and superlattices have been studied in detail \cite{Goldman, Pena, Senapati, Dong}, the YBCO in all
such studies has its c-axis perpendicular to the plane of the substrate. Since superconductivity in YBCO lies in the \cuo planes (ab - plane),
the c-axis oriented structure does not allow injection of quasiparticles along the nodal or fully gapped directions of the Fermi surface.

In order to overcome this difficulty, it is necessary to grow the YBCO layer with crystallographic orientation such that the \cuo planes are
normal to the substrate. This can be achieved by growing either (100)/(010) or (110) oriented YBCO films. While the (100)/(010) and (110) YBCO
oriented films on lattice matched substrates have been deposited and studied successfully \cite{Marshall, Inam, Covington, Trajanovic}, the
growth of an FM - SC heterostructure or superlattice with the \cuo planes normal to the plane of the substrate is quite non-trivial.
\begin{figure} \centerline{\includegraphics[width=4.5in,angle=0]{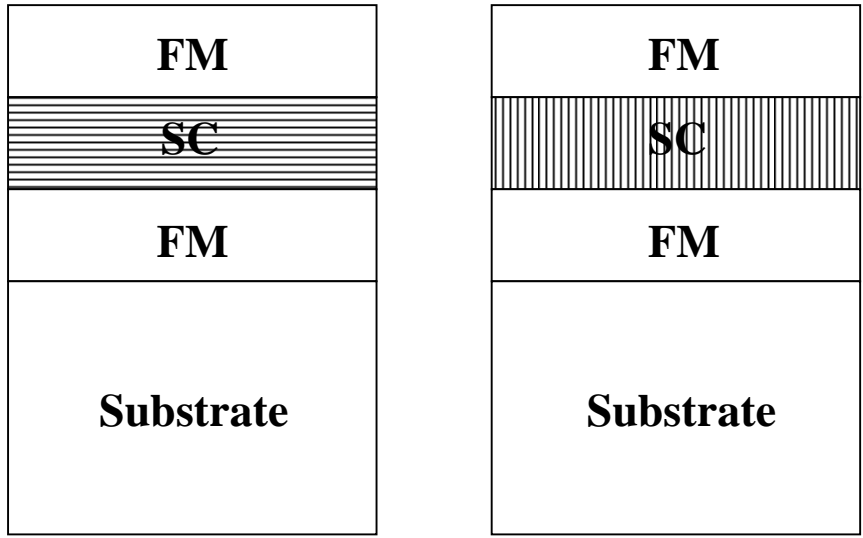}} \caption[Orientation of \cuo planes in (001) and (110)
hybrids]{Orientation of \cuo planes in (001) heterostructures (left panel) and (110) heterostructures (right panel). We can see that in the case
of (110) hybrids, the two FM layers are directly connected to each other through the superconducting \cuo layers.} \label{mot110}
\end{figure} Fig. \ref{mot110} shows the orientation of \cuo planes when the film
is (001) oriented (left panel) and (110) oriented (right panel)

With this background in mind, in this chapter we will first discuss the growth recipe for trilayers where the \cuo planes are normal to the
plane of the heterostructure such as (110) oriented growth, and then present some fascinating results on such trilayers.

\section{Results and Discussion}
\subsection{Growth and characterization of (110) trilayers}
The growth of a (110) trilayer requires a heterotemplate technique where the template layer is \PBCO. It has been found earlier that 100\% (110)
oriented YBCO grows at $\sim$600 $^o$C \cite{Poelders} on bare (110) STO. But the T$_c$ of the films thus grown is much lower than the reported
T$_c$ of 90 K for YBCO. As a result, a self template method was employed in which a layer of YBCO, $\approx$50 nm thick was deposited at
$\sim$600 $^o$C and then another layer of the desired thickness was deposited at 800 $^o$C. These films showed a remarkable improvement in T$_c$
but the 100\% (110) oriented film had a T$_c$ close to 70 K. In the next step, the YBCO template was replaced by PBCO and in this case 100\%
(110) showed a T$_c$ of 89 K even when the template was deposited at 700 $^o$C. Based on the above facts, we chose the heterotemplate technique
for our (110) thin films. Thin films of PBCO - LSMO - YBCO were deposited using the pulsed laser deposition (PLD) technique \cite{Senapati}. The
optimized growth temperature (T$_{d}$), oxygen partial pressure pO$_{2}$, laser energy density (E$_{d}$) and growth rate (G$_{r}$) used for the
deposition of the 500 {\AA} thick PBCO template were 700 $^{0}$C, 0.4 mbar, $\sim$2 J/cm$^2$ and 1.6 {\AA}/sec respectively. After the
deposition of the PBCO layer, the substrate temperature was raised to 750 $^{0}$C keeping pO$_2$ constant. The growth of 300 {\AA} thick LSMO
was carried out at T$_d$ = 750 $^{0}$C, pO$_2$ = 0.4 mbar, E$_{d}\sim$2 J/cm$^2$ and G$_{r} \simeq$ 0.5 {\AA}/sec. Once the growth of the LSMO
layer was complete, a 1000 {\AA} YBCO film was deposited on top of the PBCO - LSMO bilayer at T$_d$ = 800 $^0$C, pO$_2$ = 0.4 mbar, E$_{d}\sim$2
J/cm$^2$ and G$_{r} \simeq$ 1.6 {\AA}/sec. After completion of this layer, the deposition chamber was filled with O$_2$ to atmospheric pressure
and then the sample was cooled to room temperature with a 30 minutes holdup at 500 $^0$C to realize full oxygenation of the structure. These
deposition parameters were established after taking a series of calibration runs where the crystal orientation, high T$_{sc}$ of the YBCO and
low coercivity (H$_c$) of the LSMO layer were important factors in deciding the best condition.

\begin{figure}[t]
\centerline{\includegraphics[height=5in,angle=-90]{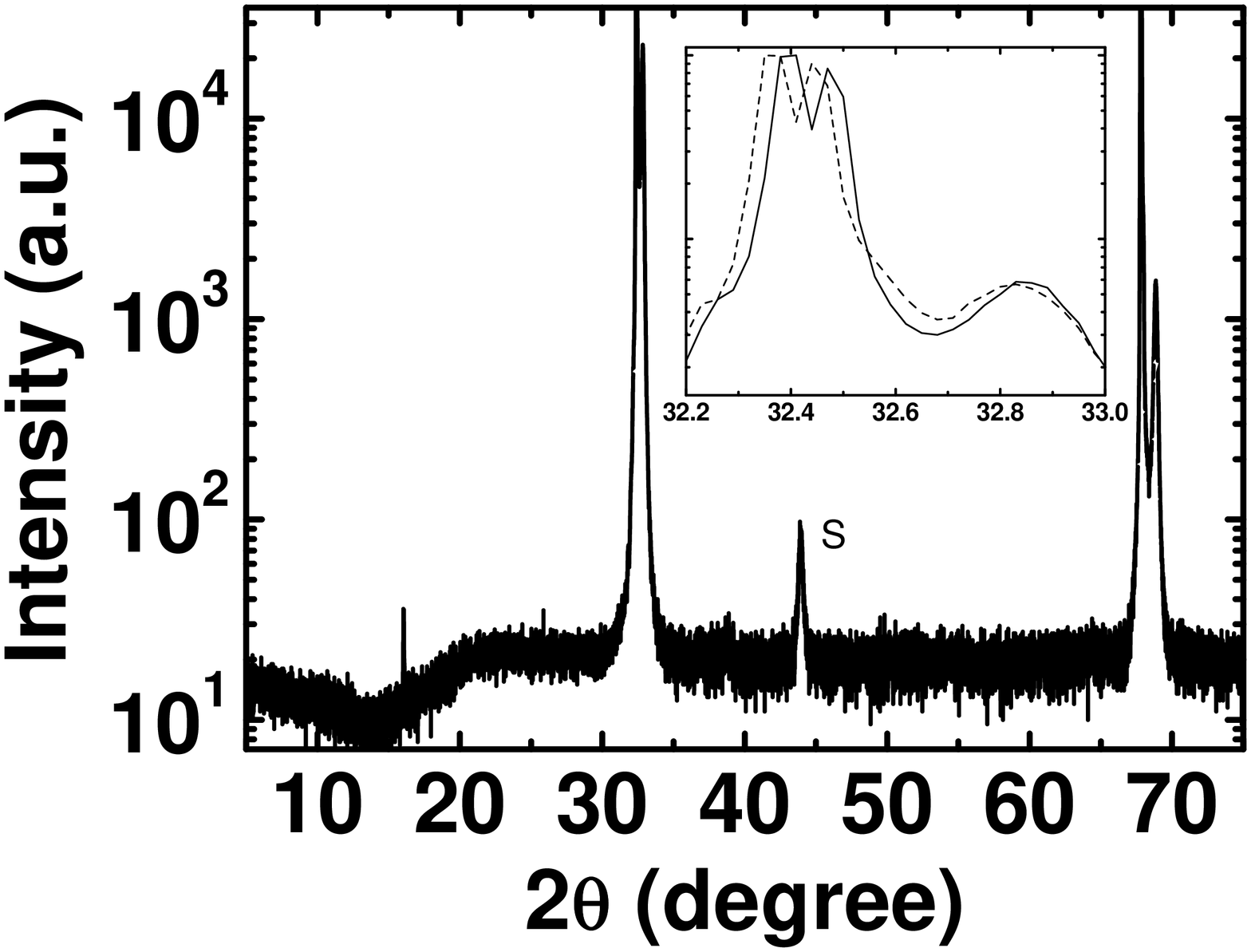}} \caption[$\theta - 2\theta$ X - ray diffraction pattern of PBCO - LSMO - YBCO
trilayer.]{$\theta - 2\theta$ X - ray diffraction of PBCO - LSMO - YBCO heterostructure grown on (110) STO. The inset presents a magnified view
of the curve between $2\theta$ = 32.1$^o$ and 33.0$^o$ for PBCO - LSMO - YBCO (solid line) and for PBCO - YBCO (broken line) heterostructures.
The splitting of the substrate peak in the inset is due to the k$_{\alpha1}$/k$_{\alpha2}$ component of the incident X-rays.}
\label{xrdt110}\end{figure} Epitaxial growth of the heterostructures was established with X-ray diffraction [XRD] used in $\theta - 2\theta$ and
$\phi$ scan modes. In Fig. \ref{xrdt110} we have shown $\theta - 2\theta$ scans of a \PBCO - \LSMO - \YBCO heterostructure. Two intense doublets
located at $2\theta \simeq$ 32.5$^o$ and 69$^o$ are seen of which the lower angle component is due to the (110) and (220) reflections of the
substrate. The manganite has cubic symmetry with lattice parameter a = 3.85 {\AA}. Fully oxygenated YBCO and PBCO are orthorhombic with a = 3.83
{\AA} and 3.82 {\AA}, b = 3.89 {\AA} and 3.88 {\AA}, and c = 11.66 {\AA} and $\sim$ 11.79 {\AA} respectively. The weaker component of the
doublets which appear at the higher \twotheta can be identified with the scattering vector of the heterostructure normal to the plane of the
substrate. The observation of a single peak instead of three separate peaks corresponding to each component of the heterostructure suggests
coherent epitaxy. This is also evident from the blownup view of the first doublet shown in the inset along with the diffraction pattern for a
PBCO - YBCO bilayer on (110) STO. However, the observations of these peaks in $\theta$-\twotheta scattering geometry does not confirm the (110)
phase purity as the reflections from the (103) and (013) oriented phases also fall at the same \twotheta value.

\begin{figure}[t]
\includegraphics[width=4in]{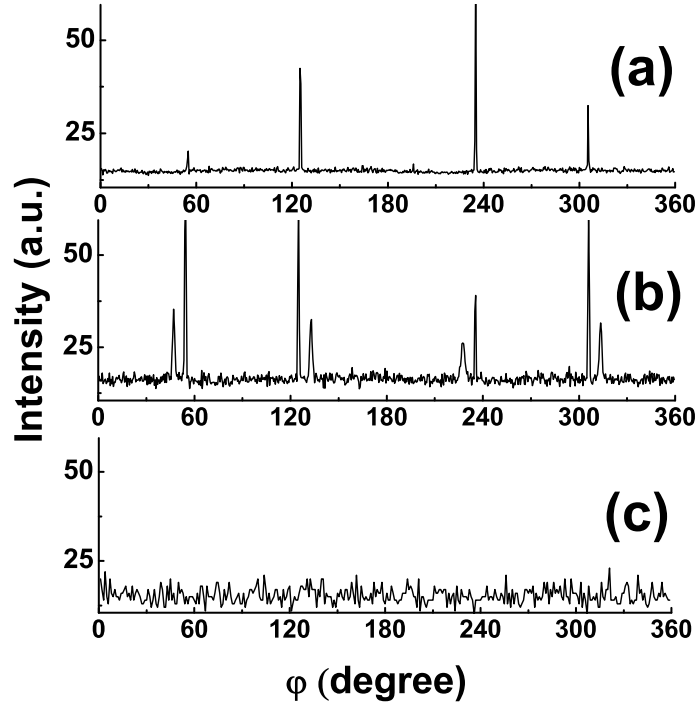} \centering \caption[$\phi$
scans of the PBCO - LSMO - YBCO heterostructure.]{$\phi$ scans of the PBCO - LSMO - YBCO heterostructure. Panels (a) and (b) show the (117)
$\phi$ scan for the (110) and (103) phases respectively, clearly indicating the presence of both (103)/(110) phases of YBCO. Panel (c) shows the
(109) $\phi$ scan for the (001) phase indicating the absence of the same.} \label{phi110} \end{figure}

\begin{figure}[t]
\centerline{\includegraphics[height=4.5in, angle=-90]{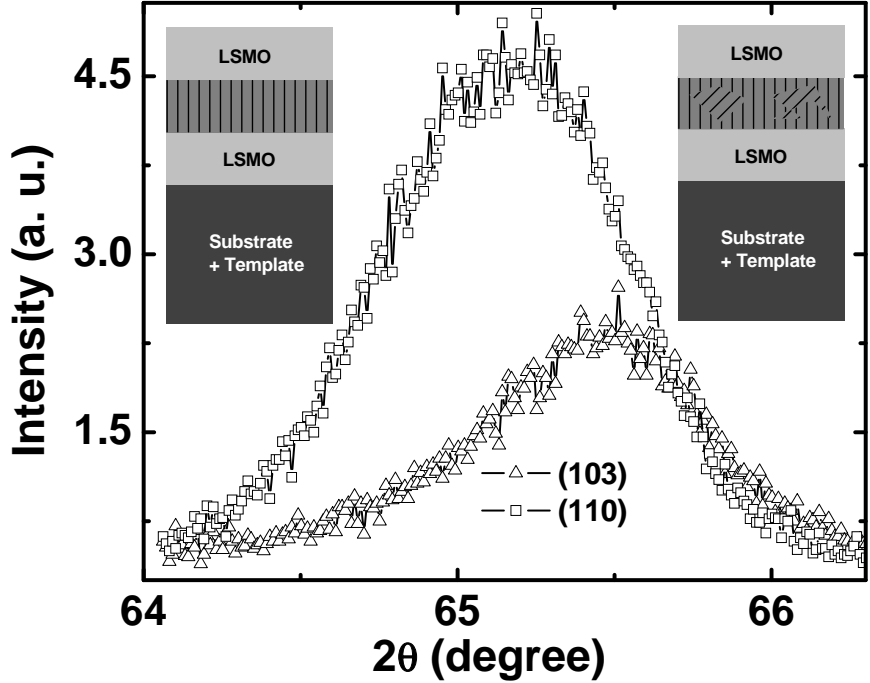}}\caption[Volume fraction measurement results on (110) trilayers]{Four circle
X-ray diffraction measurements of orientation distribution in the heterostructure. The figure shows the $\theta - 2\theta$ scans of (117) peaks
in (110) ($\square$) and (103) ($\triangle$) oriented grains. The inset shows the orientation of \cuo planes in a pure (110) film (a) and a
(110) film with (103) contamination (b)}\label{vol110} \end{figure}

\begin{figure} \includegraphics[width=4in, angle = 0]{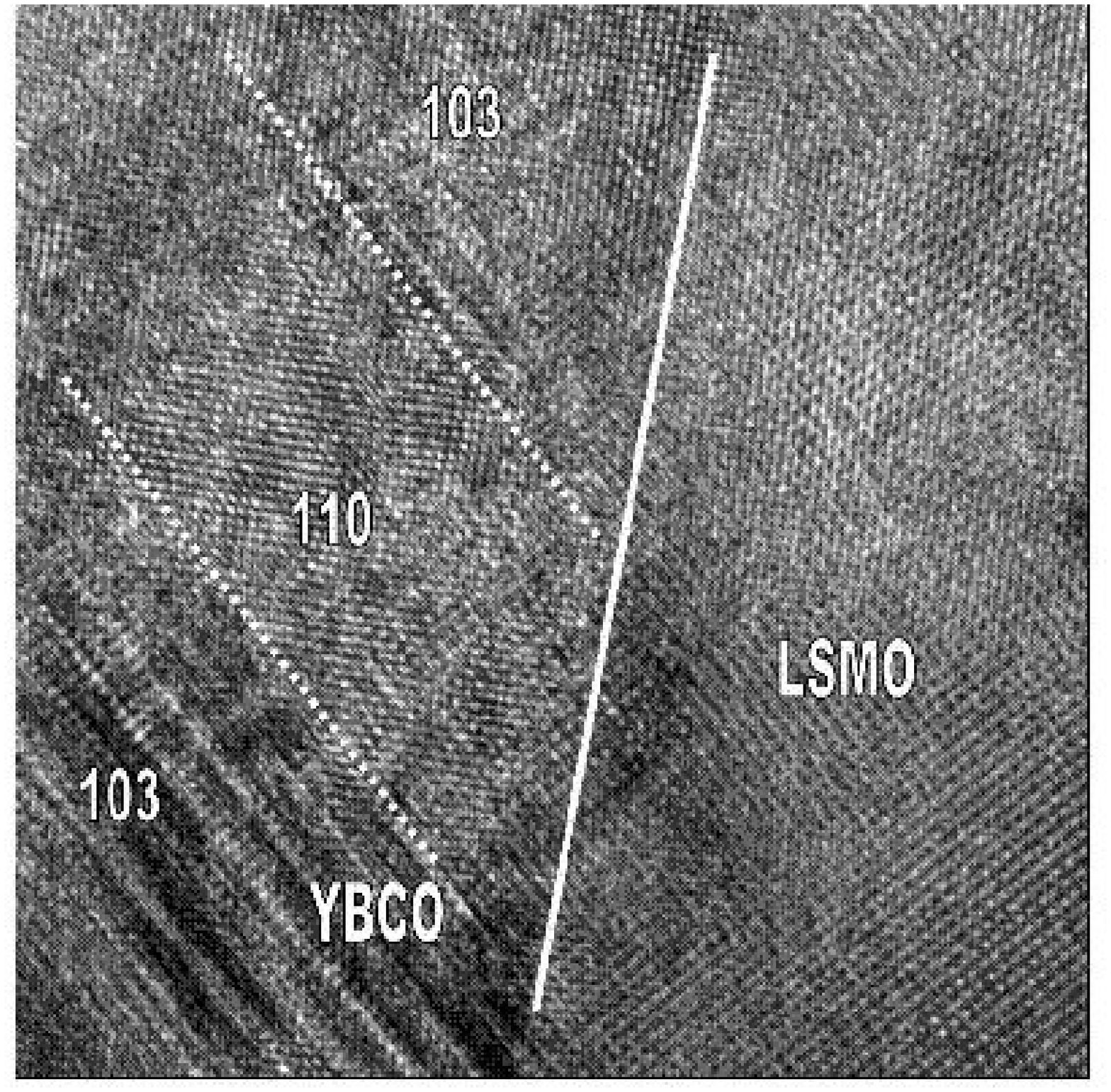} \centering
\caption[Cross-sectional TEM image of a templated bilayer]{Cross-sectional TEM image of a templated LSMO-YBCO bilayer. The image clearly shows
regions of (110) and (103) grains. \textit{Image taken by Y Zhu and group at Brookhaven National Laboratory, USA}} \label{cstem110} \end{figure}

Fig. \ref{phi110}  panels a, b and c respectively show the
$\phi$ scans of the (117) peak from the (110) and (103) phases of YBCO and the (109) $\phi$ scan for the (001) phase of YBCO. In panel (a), we
see four peaks while panel (b) shows eight. This is due to the fact that the (117) peaks from the (110) phase of YBCO and LSMO lie at the same
position while it is not the case when we are probing the (117) peaks from the (103) oriented phase. The absence of any peak in panel (c) rules
out the (001) phase. The volume percentage of (110) oriented grains calculated from the recipe of Westerheim et al \cite{West} comes out to be
$\gtrsim$65\% with the remaining volume is of (103) grains. While the growth is not 100\% (110) oriented, the remaining (103) grains still
permits injection of spin polarized carriers directly into the \cuo planes as these planes are oriented at 45$^o$ with respect to the substrate
(inset (b) in fig. \ref{vol110}). To demonstrate this point, in fig. \ref{vol110}, we have shown the result of four circle X-ray diffraction of
$\theta - 2\theta$ scans of the (117) peaks in (110) and (103) oriented grains. In the inset of fig. \ref{vol110} we have schematically drawn
the two cases i.e. when pure (110) YBCO is sandwiched between ferromagnetic layers (left hand side) and when a mixture of (110) and (103) phases
is present (right hand side).
 It is to be noted that the (110) percentage of YBCO directly grown on the
template under similar growth conditions is $\gtrsim$90\%. High-resolution TEM measurement done on the bilayers of templated LSMO-YBCO revealed
a mixture of (110) and (103) grains. In fig. \ref{cstem110} we have shown the HRTEM image of a bilayer.

\begin{figure} \centerline{\includegraphics[height=4in, angle = -90]{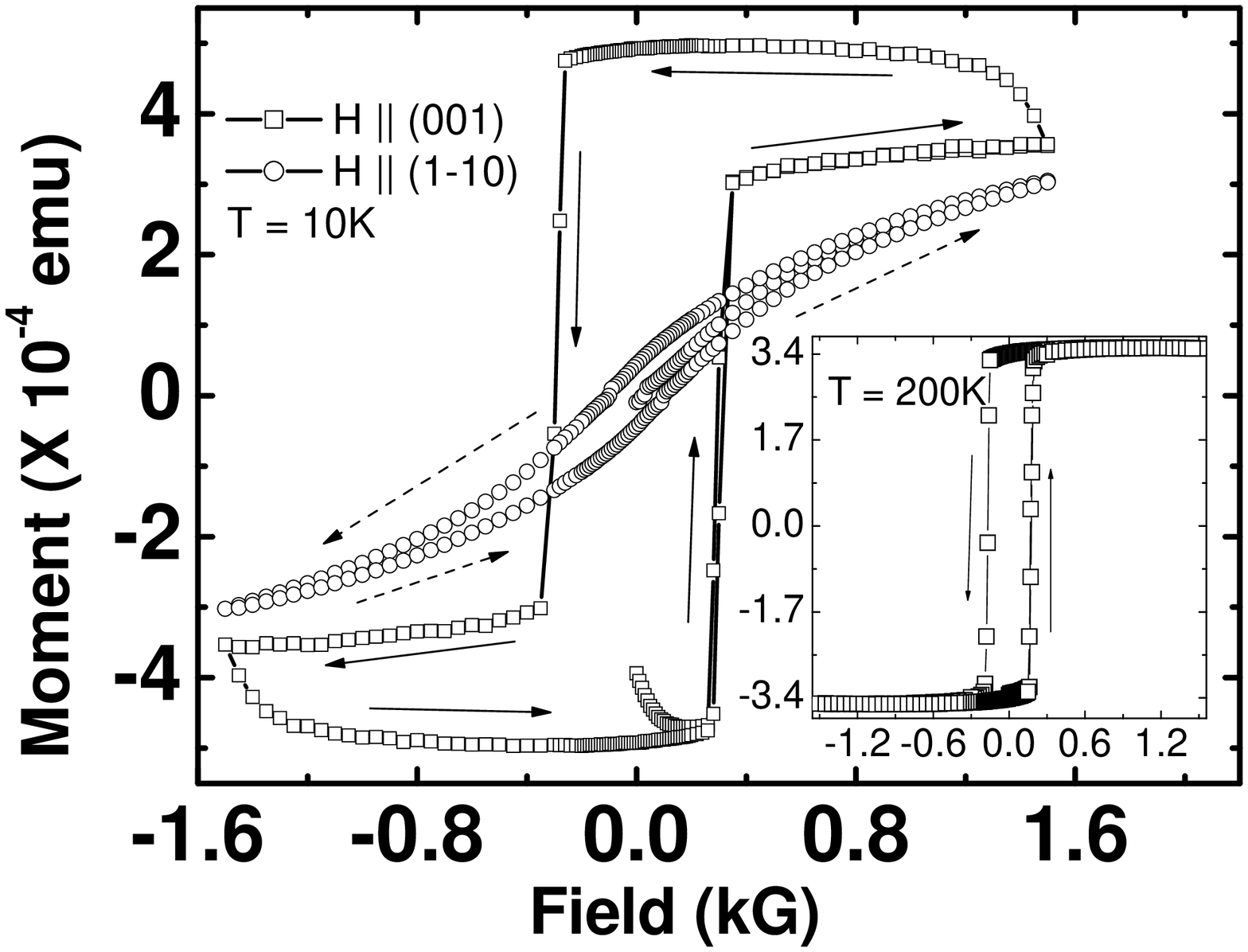}} \caption[M vs. H loops for PBCO - LSMO -
YBCO heterostructures taken at 10 K]{M vs. H loops for PBCO - LSMO - YBCO heterostructures taken at 10 K with the field $\parallel$ and $\perp$
to \onebar { } substrate edge. The inset shows the M - H loop with H along the magnetic easy axis at 200 K.} \label{mht110} \end{figure}

\begin{figure}
\centerline{\includegraphics[width=4in]{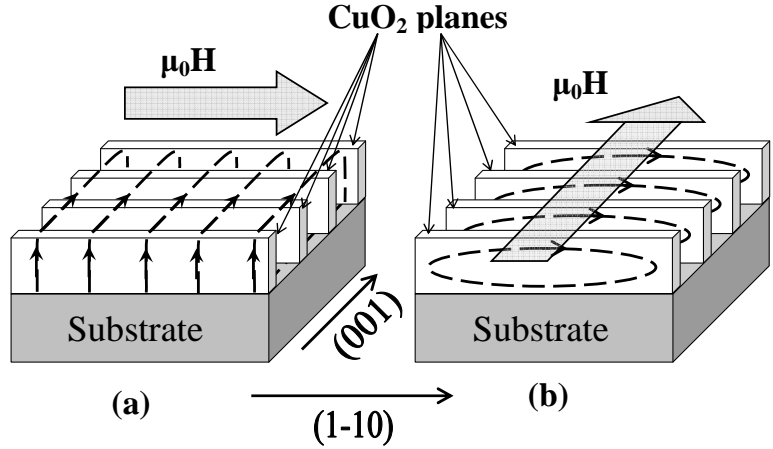}} \caption[Josephson tunneling controlled screening current in (110) heterostructure] {(a)
Josephson tunneling controlled screening current in (110) oriented heterostructure when the applied field H $\parallel$ \cuo planes. (b) In -
plane screening current in (110) oriented heterostructure when the applied field H $\perp$ \cuo planes.}\label{scr110}
\end{figure}

In fig. \ref{mht110} we have
shown the M vs. H loops for a PBCO - LSMO - YBCO heterostructure taken at 10 K with  field H parallel to the (001) and {\onebar} directions of
the substrate. The square hysteresis loop in the main figure (and in inset) when H $\parallel$ (001) as against a slowly saturating loop when H
$\parallel$ \onebar{ } clearly shows that (001) is the magnetic easy axis of the LSMO. We can also see distinctly the diamagnetic contribution
from the SC, which splits the field increasing and field decreasing arms of the loop beyond the saturation field. This splitting is because of
the SC-state is confirmed by the absence of the same in the M-H loop taken at 200 K (inset). It is possible to calculate the critical current
density (J$_c$) of the superconducting film from the diamagnetic splitting in the M-H loop of the film. The splitting in the M-H loop for a
circular film of radius R, thickness d and field perpendicular to surface, can be written as a function of the critical current density J$_c$ as
\cite{later} \begin{equation}m = \left( \frac{\pi}{3}\right)R^3 J_c d \label{beanjc}\end{equation} where m is the splitting of the M-H loop
divided by two. In our case the films are square with sides of length $w$ and thickness $t$. The field is parallel to the plane of the film. So
the surface perpendicular to field is $w \times t$. The radius of a circle with the same area will be given by $R = \sqrt{wt/\pi}$. In our case
$d$ in eq. \ref{beanjc} will be replaced by $w$. Hence, the final formula for J$_c$ comes out to be
\begin{equation}J_c = m \left(\frac{3}{\pi}\right) \left(\frac{\pi}{wt}\right)^{3/2} \frac{1}{w} \label{jcfinal}\end{equation} A calculation of
the critical current density (J$_c$) from the shift of M-H loop using eq. \ref{jcfinal} yields J${_c} \simeq 3.8 \times 10^{4}$ A/cm$^2$ at 10 K
and 350 Gauss field. A remarkable feature of these hysteresis loops is the near absence of the diamagnetic contribution when the magnetic field
is aligned along the magnetic hard axis \{\onebar\} of LSMO. These features can be understood if we visualize the way screening currents are
induced in the SC-film by the external field. As shown in fig. \ref{scr110},
when the external field is parallel to the \cuo planes (H $\parallel$
\onebar), the diamagnetic moment is produced by weak Josephson tunneling currents across the \cuo planes. However for H $\parallel$ (001), the
screening currents are confined to each \cuo plane. A large condensate density in the planes makes these currents strong and the diamagnetic
moment is distinctly visible in the M - H loop. The hysteresis loop shift due to superconductivity in La$_{0.7}$Ca$_{0.3}$MnO$_3$ - YBCO -
La$_{0.7}$Ca$_{0.3}$MnO$_3$ superlattices has also been seen by  Pe\~na et al. \cite{Pena}. Their calculation of J$_c$ on the basis of the Bean
model yields a suppression of the current by a factor of 20 at 5 K. However, the work of ref. \cite{Pena} is on c-axis oriented films, where the
suppression of J$_c$ due to the ferromagnetic proximity effect may be small as the c-axis coherence length $\xi_c$ of YBCO is only $\simeq$ 3
\AA.

\begin{figure}\centerline{\includegraphics[height=4in, angle = -90]{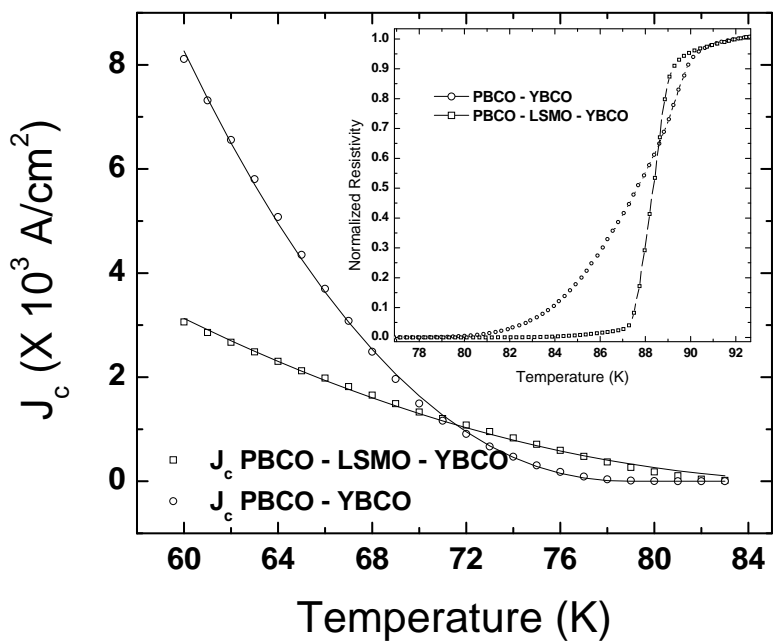}} \caption[Transport J$_c$(T) data for (110)
heterostructures]{Transport J$_c$(T) data for PBCO - YBCO and PBCO - LSMO - YBCO heterostructures taken with a voltage criterion of 10
$\mu$V/cm. The solid lines are the fitting using the formula $J_c = J{_o}(1-\frac{T}{T_{sc}})^\beta$. The inset shows the resistivity of the
same samples in the vicinity of T$_{sc}$. The resistivity has been normalized with respect to its value at 92 K.} \label{jct110}
\end{figure}

In fig. \ref{jct110}  we have shown the variation of transport $J_c$ with temperature for a PBCO - LSMO - YBCO (circle) and PBCO - YBCO (square)
heterostructure where PBCO is 500 \AA, LSMO is 300 \AA~ and YBCO is 2000 \AA~ thick, the LSMO layer being absent in the second structure while
other thicknesses are same. The SC  transition in these samples measured in a four probe geometry is shown in the inset of fig. \ref{jct110}. We
note that while this transition in the PBCO - LSMO - YBCO heterostructure is quite sharp ($\Delta$T$_{sc}\thickapprox 2.5$ K) as compared to the
(110) oriented PBCO - YBCO bilayer, its J$_c$ is greatly suppressed. The J$_c$(T) data have been fitted to the phenomenological relation $J_c =
J{_o}(1-\frac{T}{T_{sc}})^\beta$. The fitting parameters for the PBCO - LSMO - YBCO and PBCO - YBCO structures are $J_o = 2.0 \times 10^4$
A/cm$^2$ and $1.8 \times 10^5$ A/cm$^2$ respectively, while $\beta = 1.5$ and 2.16 respectively. In the Ginzburg - Landau description of
J$_c$(T), the prefactor J$_o$ is directly related to the condensate density. A highly suppressed J$_o$ in samples with ferromagnetic boundary
provides a strong indication of pair breaking by spin polarized carriers injected from the LSMO.

\subsection{(110) LSMO - YBCO - LSMO}
\begin{figure}[t] \centerline{\includegraphics[width=4in,angle=0]{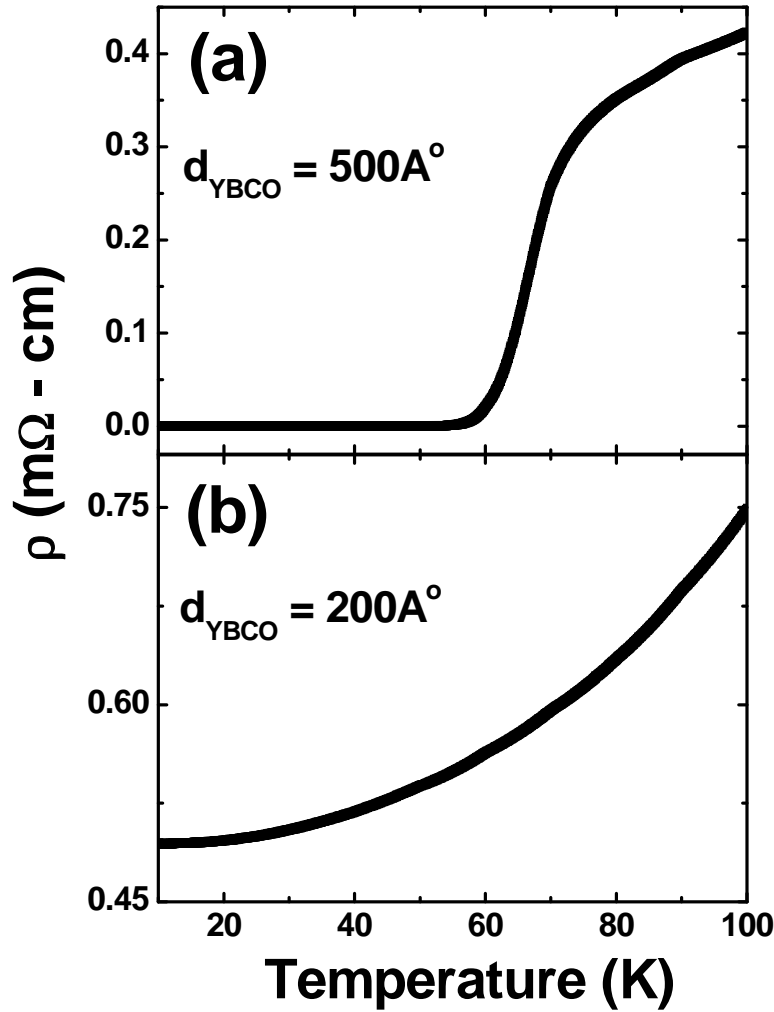}} \caption[Resistivity data
for (110) LSMO - YBCO - LSMO trilayer]{Resistivity data for (110) LSMO - YBCO - LSMO trilayer with d$_{YBCO}$ = 500{\AA} (top panel) and
200{\AA} (bottom panel). Here the films with 200 {\AA} thick YBCO layer are not superconducting while (001) trilayers films are superconducting
even for YBCO layer thickness of 50 {\AA}.} \label{rtt110}\end{figure}

With the optimized growth conditions, trilayers of (110) \LSMO-\YBCO-\LSMO were synthesized and their various magnetic and electronic properties
were measured. Fig. \ref{rtt110} shows the resistivity curves $\rho(T)$ for two trilayers. The upper panel
is the result for a trilayer with a 500 {\AA} YBCO spacer. The $\rho(T)$ curve is characterized by transition to a superconducting state which
starts at $\sim$80 K and completes when the temperature reaches $\sim$60 K. The bottom panel shows the resistivity for a similar structure with
a 200 {\AA} YBCO spacer. In this case the trilayer does not go into the superconducting state though it has a metallic behavior. It is
interesting to note that while the YBCO of thickness 200 {\AA} in the (110) trilayer shows no T$_c$, a superconducting transition can be seen
for YBCO thickness of even 50 {\AA} for the (001) trilayer \cite{Senapati}. This is presumably due to greater T$_c$ suppression in the case of
(110) films because of direct injection of spin polarized carriers in the superconducting \cuo planes of YBCO.
\begin{figure}\centerline{\includegraphics[width=4.5in,angle=0]{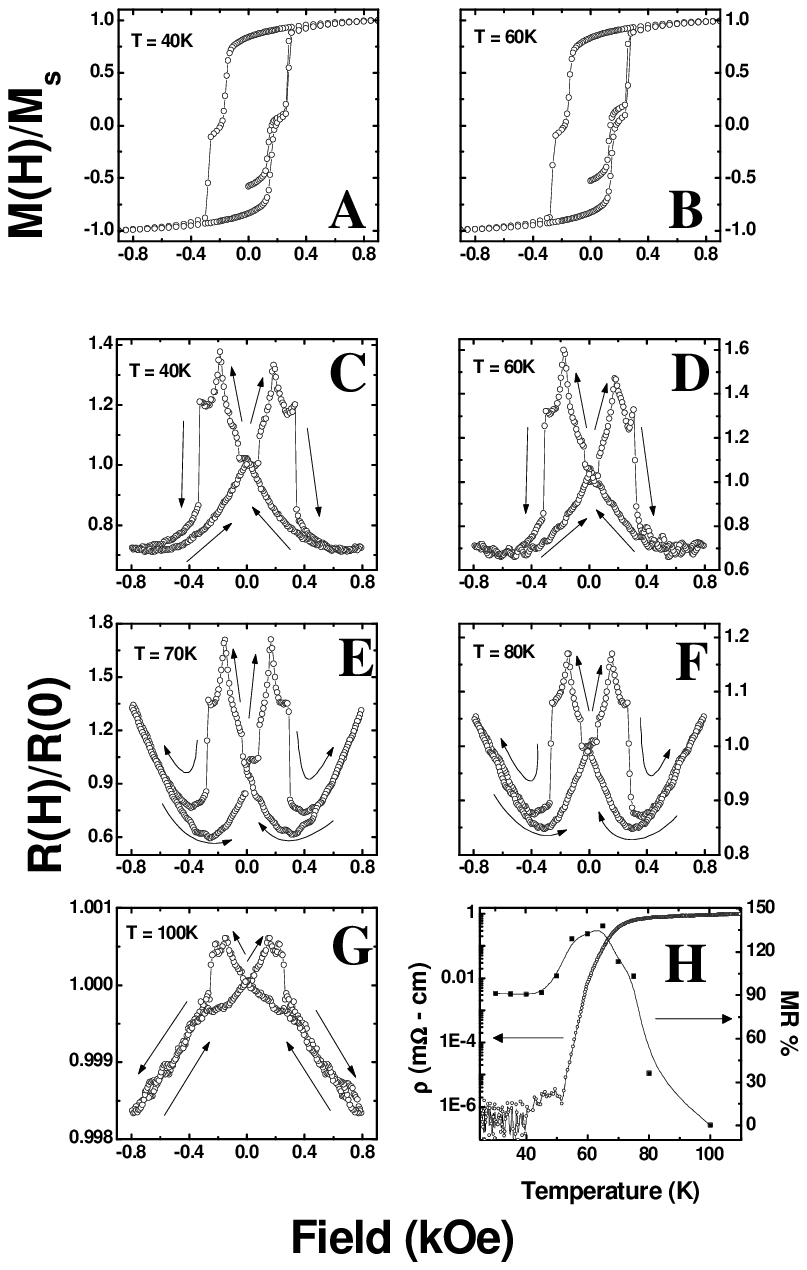}} \caption[Field dependence of resistivity for the
superconducting trilayer]{Panels A and B show the M-H loop of the trilayer at 40 K and 60 K. Panels C-G show the field dependence of resistivity
for the superconducting trilayer shown at a few representative temperatures across the transition temperature. The MR of the film in the
superconducting state is higher than that in the normal state. Panel H shows the comparison of MR\% with temperature and sample resistivity. A
clear peak seen near the transition temperature can be attributed to the unusual rise in the normal state properties of the superconductor near
the transition temperature. MR\% is the difference of MR between the highest and lowest points in the curve with respect to the resistance at a
zero applied field.} \label{rht110} \end{figure} In fig. \ref{rht110} A \& B we have shown the M-H loops of the trilayer with d$_{YBCO}$ = 500
{\AA}, at 40 and 60 K respectively. A plateau in the M-H loop near zero magnetization confirms the presence of an antiferromagnetic state. This
antiferromagnetic state is present in the normal state of the superconductor as well. Panels C through G in the same figure show the
magnetoresistance (MR) of the superconducting trilayer at a few temperatures across the transition. The MR in this case is defined as R(H)/R(0)
where R(H) is the resistance of the sample at applied field H. The field and current (I) in this case are coplanar but orthogonal to each other.
We first discuss panels C and D which present the data for the trilayer in the superconducting state at 40 K and 60 K respectively. Starting
from a fully magnetized state of LSMO layers at 800 Oe the MR first increases slowly as the field is decreased. At $\sim$400 Oe the rate of
increase becomes faster but remains continuous till the zero field. On reversing the field, a small step like jump is seen around $\sim$-50 Oe
and then the MR keeps rising to a peak value after, which a local minimum is attained followed by a sudden jump in the MR at $\sim$370 Oe to a
much lower value. Further increment of the field results in a gradual decrease in MR till a reversed field of 800 Oe is reached. This cycle
repeats itself once the field is decreased from -800 Oe and increased to 800 Oe. We have measured MR for the sample at a few more temperatures
below T$_c$. In all those measurements we found that the resistance ratio $R_{max}/R_{min}$ over the whole range of measurement is $\sim$2. The
current flowing through the sample in these measurements is $\sim$J$_c$ of the sample at that temperature. Panels E and F show the MR vs. H data
for 70 and 80 K respectively, where the YBCO layer in the trilayer is in the superconducting transition region (top panel of fig. \ref{rtt110}).
Here we can see that the high field negative magnetoresistance region, as seen in panels C and D in the field regime $\sim$ 400 - 800 Oe, is
replaced by a positive magnetoresistance which is completely opposite to the negative MR seen on LSMO films \cite{Revzin}. The resistance ratio
in panel E is $\sim$3 which is the highest over the whole range of measurement. This resistance ratio sharply drops once the film starts
entering the normal state as is evident from the panels showing the MR at 80 (panel F) and 100 K (panel G). Panel G shows MR data for 100K,
where negative magnetoresistance is seen in a high field, which is a characteristic feature of LSMO \cite{Revzin}. Even though in panel G the
resistance ratio is reduced due to the superconducting spacer entering into the normal state, however the first order jump in resistance near
H$\approx$350 Oe is still present clearly proving the fact that the resistance ratio is dependent on the spacer layer properties while the first
order transition is dependent on the FM layer properties. In panel H the MR\% plotted is defined as $\Delta R / R(0)$ where $\Delta R =
(R_{\uparrow\downarrow})_{max} - (R_{\uparrow\uparrow})_{min}$ and R(0) is the resistance at zero field at that temperature.
$(R_{\uparrow\downarrow})_{max}$ is the resistance at the peak position in the MR-H curve. $(R_{\uparrow\uparrow})_{min}$ is the minimum
resistance of the segment of MR-H curve where the magnetizations of both the FM layers are parallel to each other. A distinct behavior of MR\%
can be seen when the sample becomes superconducting. The sample in the normal state has a very low MR but once the sample starts moving into the
SC regime, the MR shoots up rapidly and then comes down to saturate at a constant value at low temperatures. The increase in MR in the vicinity
of T$_c$ can be attributed to the abnormal increase in the normal state properties of the superconductor \cite{Mishonov}.

Another important feature that is quite prominent in fig. \ref{rht110} is the presence of peaks in the MR data. A comparison of MR-H and M-H
plots (panels A and C, and panels B and D) shows that the peaks coincide with the region where the M-H curve has a plateau. The near zero
magnetization in the plateau suggests antiferromagnetic coupling between the magnetization vectors of the top and bottom LSMO layers. We can
estimate the exchange energy associated with the AF coupling in the following way. The free energy expression for two magnetic layers of the
same thickness coupled via the spacer can be written as \cite{Demokritov} \begin{equation} F = F_c + F_a - \vec{H}. \left(\vec{M_1} +
\vec{M_2}\right)t \label{enhs} \end{equation} where M$_1$ and M$_2$ are the magnetizations of the the top and bottom LSMO layers, $F_c$ is the
coupling energy per unit area and $t$ is the thickness of a single LSMO layer. The anisotropy part of the energy ($F_a$) is primarily dependent
on contributions from magnetocrystalline anisotropy as well as the in-plane uniaxial anisotropy of the film. Assuming a bilinear coupling, $F_c$
can be written as; \begin{equation} F_c = -J_1\left(\vec{M_1}.\vec{M_2}\right)\end{equation} where $\vec{M_1}$ and $\vec{M_2}$ are the unit
magnetization vectors, and J$_1 <$ 0 corresponds to antiferromagnetic coupling between FM layers. For a given external field, the minima of eq.
\ref{enhs} will yield the relative orientation of $\vec{M_1}$ and $\vec{M_2}$. If J$_1$ is positive, even in zero field $\vec{M_1} \|
\vec{M_2}$, so increasing the field does not change anything. If J$_1$ is negative then the minima is achieved when H = 0 and $\vec{M_1}$ and
$\vec{M_2}$ are antiparallel or antiferromagnetic in alignment. The anisotropic term in eq. \ref{enhs}, F$_a$ can be written as,
\begin{equation}F_a = KtM_{1,2} \end{equation} where M$_{1,2}$ is a function of $\vec{M_1}$ and $\vec{M_2}$ and $K$ is the anisotropy constant.
If $\mid J_1\mid \gg K$, then a second order reorientation transition and a smooth linear M-H dependence followed by saturation is predicted by
the theory. On the other hand if $\mid J_1\mid \ll K$ then the magnetization slowly increases in the low field, and then abruptly at some
critical field H$_s$, the system undergoes a first order transition with an abrupt jump to saturation magnetization. The critical field H$_s$
which is also known as saturation field or switching field can be written in terms of the magnetization density M$_s$, thickness t of one
ferromagnetic layer and coupling energy J$_1$ as \cite{Demokritov}, \begin{equation} H_s = - \left(\frac{J_1}{M_st}\right)
\label{eqji}\end{equation} It is to be noted that this equation is for the case where the two FM layers are of equal thickness $t$. The behavior
of the magnetization seen in fig. \ref{rht110} corresponds to this situation. \begin{figure}[!ht]
\centerline{\includegraphics[width=4in,angle=0]{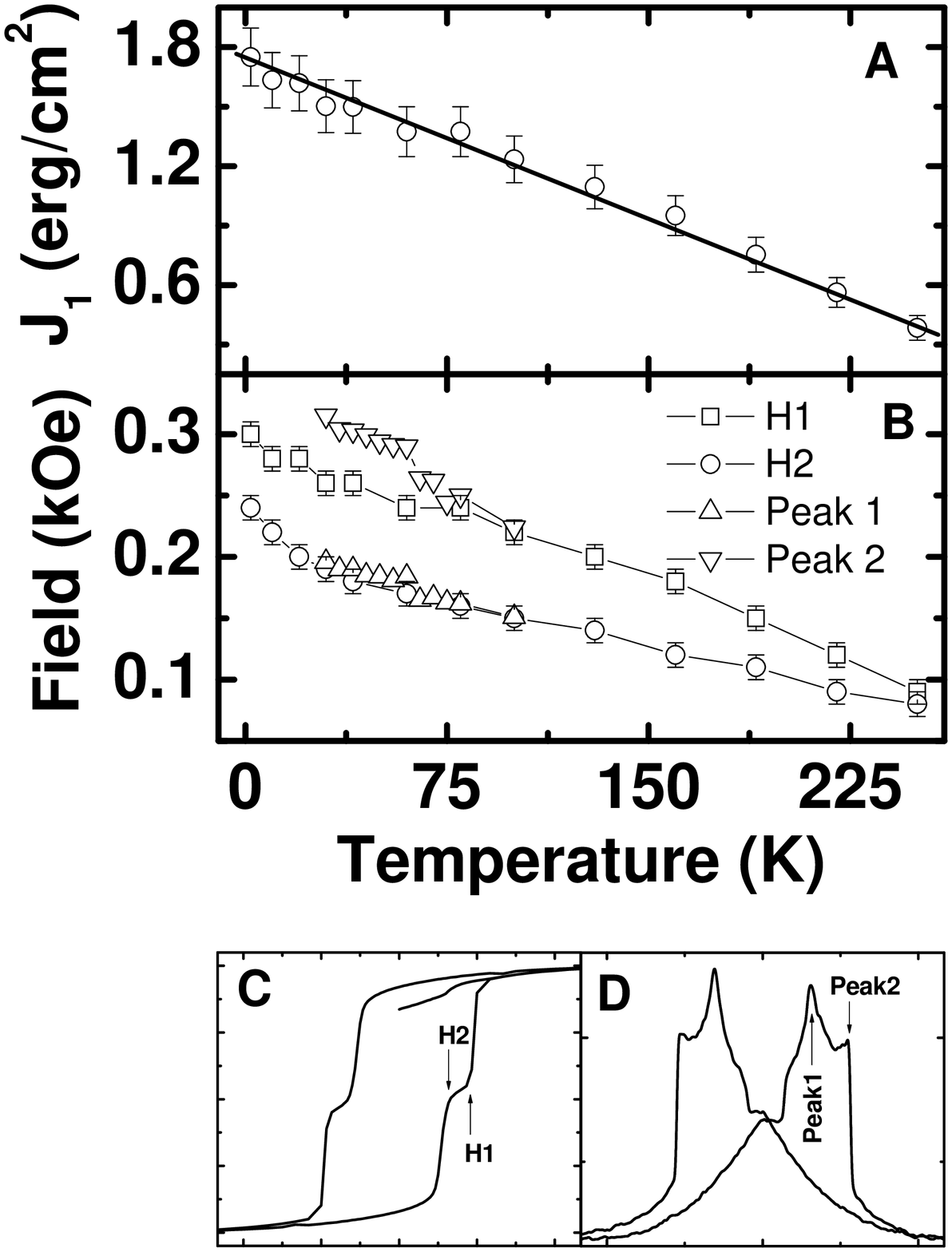}} \caption[Variation of coupling energy between two ferromagnetic layers.]{Panel A
shows the variation of coupling energy with temperature between two ferromagnetic layers. The coupling in this case is higher than that seen for
(001) layers (Details in text). Panel B shows the comparison between the peak position in MR data and the start and end points of the
antiferromagnetic phase in the MH data. Panels C and D show the position of the points on the MR and MH data plotted in the middle panel. The
agreement in the data points clearly shows the dependence of discontinuities in the MR data on the antiferromagnetic phase of the sample.}
\label{j1} \end{figure} In fig. \ref{j1}(a), we have shown the variation of J$_1$ as a function of the temperature calculated using eq.
\ref{eqji} for d$_{YBCO}$ = 500{\AA}. Panel B of fig. \ref{j1} shows a comparison between the peak position in the MR-H data and the start and
end points of the antiferromagnetic phase in MH data. Panels C and D show a typical MH and MR-H data for the sample. The arrows point to
positions of the points on the MR-H and MH data which have been plotted in panel B.

We now discuss the behavior of J$_1$ as seen in fig. \ref{j1}(panel A). The temperature dependence of the interlayer exchange coupling in
metallic multilayers has been worked out theoretically \cite{Edwards, Brunoprb}. The starting point for calculating the coupling is to calculate
the energy per unit area in the ferromagnetic and antiferromagnetic configuration. The difference of the two will give us the exchange coupling
of the system. The energy terms are functions of the reflection coefficients of the electrons in the spacer hitting the spacer-ferromagnet
interface calculated in the light of the free-electron model. Using the above method the dependence of linear exchange coupling J$_1$ with
temperature is given by \cite{Brunoprb} \begin{equation} J_1(T) = J_1(0)\left[\frac{\left(T/T_0\right)}{\sinh\left(T/T_0\right)}\right]
\label{j1t} \end{equation} where the characteristic temperature $T_0$ depends on the Fermi wave vector $k_F$ and spacer thickness $d_n$ through
the relation $T_0 = \hbar^2k_F/2\pi k_Bd_nm$, where $m$ is the free-electron mass and $\hbar$ and $k_B$ are the Planck and Boltzmann constants
respectively. In this case, since the transport is along the (110) axis, the relevant wave vector will be $k_{F_{(110)}}$. In fig. \ref{j1} it
is seen that $J_1$ increases linearly as the temperature decreases. This is different from the behavior expected from eq. \ref{j1t}. In general,
functions of the type x/sinh(x) saturate in the limit x $\rightarrow$ 0, but in this case we do not see any saturation of J$_1$ even at a
temperature of 2 K. The magnitude of $J_1$ is almost an order higher than what is seen for (001) oriented heterostructures \cite{Senapati}.
These high values are in line with the predictions of de Melo \cite{Melo} where he had pointed out that the coupling along the (110) direction
will be higher than that along the (100) or (001) directions. \begin{figure} \centerline{\includegraphics[width=4in,angle=0]{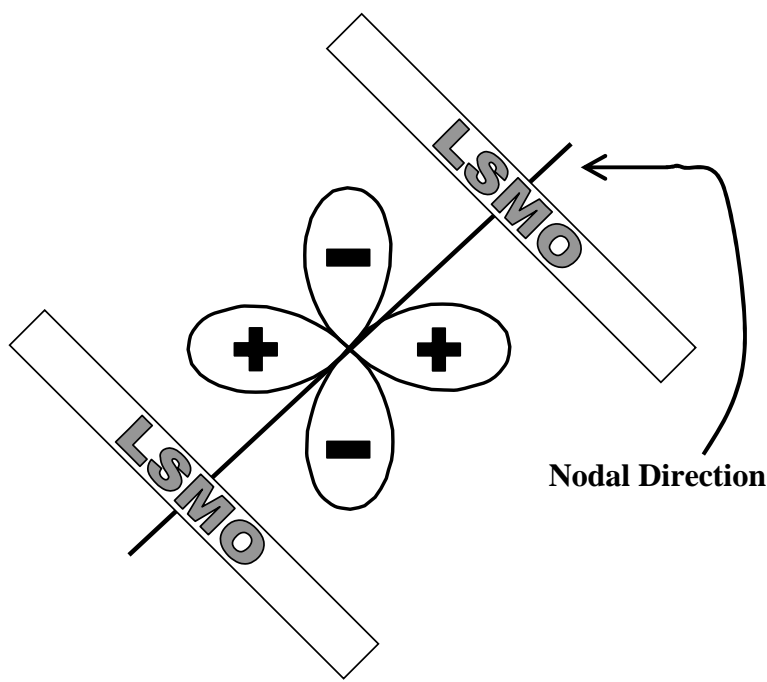}}
\caption[Schematic showing nodal direction in $d_{x^2-y^2}$-orbital.]{Schematic showing nodal direction in $d_{x^2-y^2}$-orbital. The LSMO
layers shown in the schematic signify the position of LSMO layers in a (110) trilayer. This schematic is for a trilayer where the spacer is a
$d$-wave superconductor.} \label{nodaldiro} \end{figure} In fig. \ref{nodaldiro}, we have shown the schematic of a (110) trilayer where the
spacer is a $d$-wave superconductor. It is clear from the figure that in this case the coupling, is mediated by the nodal quasiparticles whose
number density remains high even at T$\approx$0. This explains the large J$_1$ and the absence of any anamoly in J$_1$ near T$_c$. The middle
panel of fig. \ref{j1} shows the comparison between various critical points on the MH loop and MR data. Here we have plotted the starting and
end points of the plateau in the MH loop against temperature. From the MR(H) loop we have taken the points of discontinuity which are indicated
in the bottom right panel. The bottom left panel shows the position of points H1 and H2 on the MH loop. We can clearly see that both the
representative points agree well within experimental error with each other clearly demonstrating the fact that the discontinuities in MR the
data come from the antiferromagnetic regime of the sample.

\begin{figure} \centerline{\includegraphics[height=4in,angle=-90]{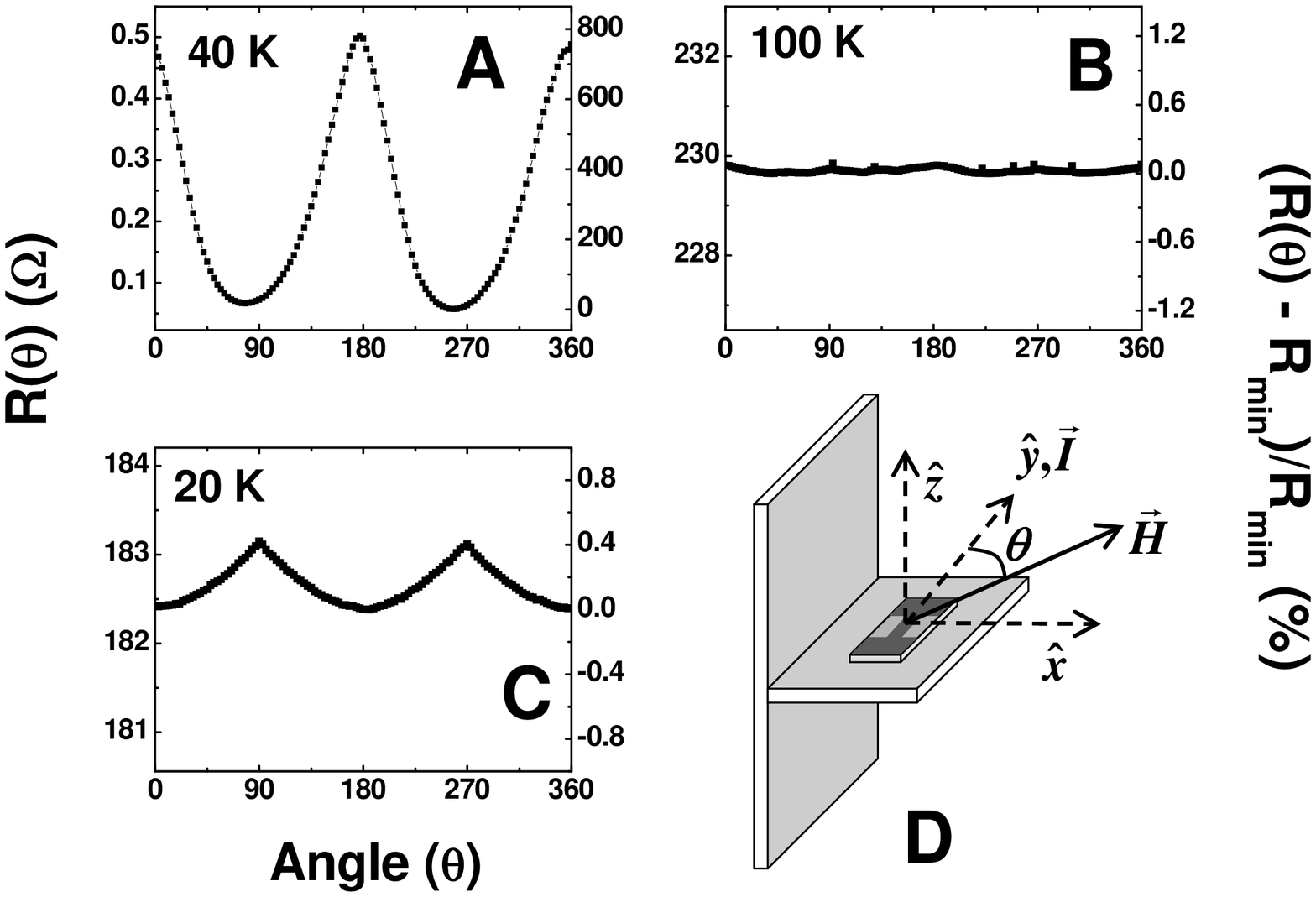}}
\caption[Temperature variation of angular dependence of MR]{Angular dependence of MR in the FM-SC-FM trilayer is plotted at T $<$ T$_c$ (panel
A), T $>$ T$_c$ (panel B) and a FM-NM-FM trilayer (panel C). The data clearly shows that a superconducting spacer in the superconducting state
enhances and modifies the AMR considerable. The AMR of the FM-NM-FM trilayer is mostly dependent on the AMR of the FM layer. Panel D shows the
schematic of the measurement geometry.} \label{tcom}\end{figure}

Fig. \ref{tcom} presents the AMR data on (110) trilayers. The x-axis defines the angle $\theta$ with respect to the current direction. The
left hand axis shows R$(\theta)$ and the right hand y-axis shows MR \% defined as (R$(\theta)$ - R$_{min}$)/R$_{min}$ where R$_{min}$ is the
minimum resistance of the sample over the whole range of measurement. Before we discuss the data in detail, let us explain the measurement
geometry which has been schematically shown in panel D of the diagram. The patterned sample is placed over a solid block of copper as shown in
the figure. The current in the sample is along \emph{\^y}. The applied field rotates in the $xy$-plane. The angle $\theta$ is measured with
respect to \emph{\^y}. The sample is patterned along \emph{\^y} in such a way that the \cuo planes of YBCO are in the $yz$-plane. In short,
\emph{\^x, \^y} and \emph{\^z} are parallel to (001), {\onebar} and (110) directions of the sample. Panels A and B show the AMR of the trilayer
at T = 40K ($<$ T$_c$) and T = 100 K ($>$ T$_c$) respectively on a film with d$_{YBCO}$ = 500 {\AA}. Panel C shows the AMR measurement at 20 K
on a film with d$_{YBCO}$ = 200 {\AA}. From panels A \& B, we can see that the trilayer shows a huge MR when the SC layer is in the
superconducting state. The same film hardly shows any AMR once the film moves into the normal state. This is also evident from the AMR data in
panel C. The trilayer in this case is non superconducting for all temperatures (bottom panel of fig. \ref{rtt110}). The angular dependence in
panel C is similar to the one seen for plane LSMO films (Results in Chapter 3). The dependence of AMR in the superconducting state is markedly
different from the one in the normal state. For our sample geometry, when the field is perpendicular to the current it is also perpendicular to
the \cuo planes resulting in maximum dissipation in the YBCO layer. So, the logical thing would be that the AMR is higher when the field is
perpendicular to the current, but what we see here is completely opposite. This can be explained as follows. We know that the dissipation in
YBCO when the applied field is perpendicular to the copper oxide planes is due to the formation of vortices and when the field is parallel, the
dissipation is mostly due to pair-breaking effects. In our geometry, the effective area of the sample exposed perpendicular and parallel to
field is equal to the thickness of the sample multiplied by the length and the breadth respectively. It is quite possible that the effect of
vortex formation in such small a area has lower dissipation than pair breaking effects. Hence, if we assume that pair breaking is causing larger
dissipation in the YBCO layer in these trilayers we can safely conclude that for fields parallel to the current (or copper oxide planes) will
show higher AMR.

\begin{figure}[!ht] \centerline{\includegraphics[height=6.5in,angle=0]{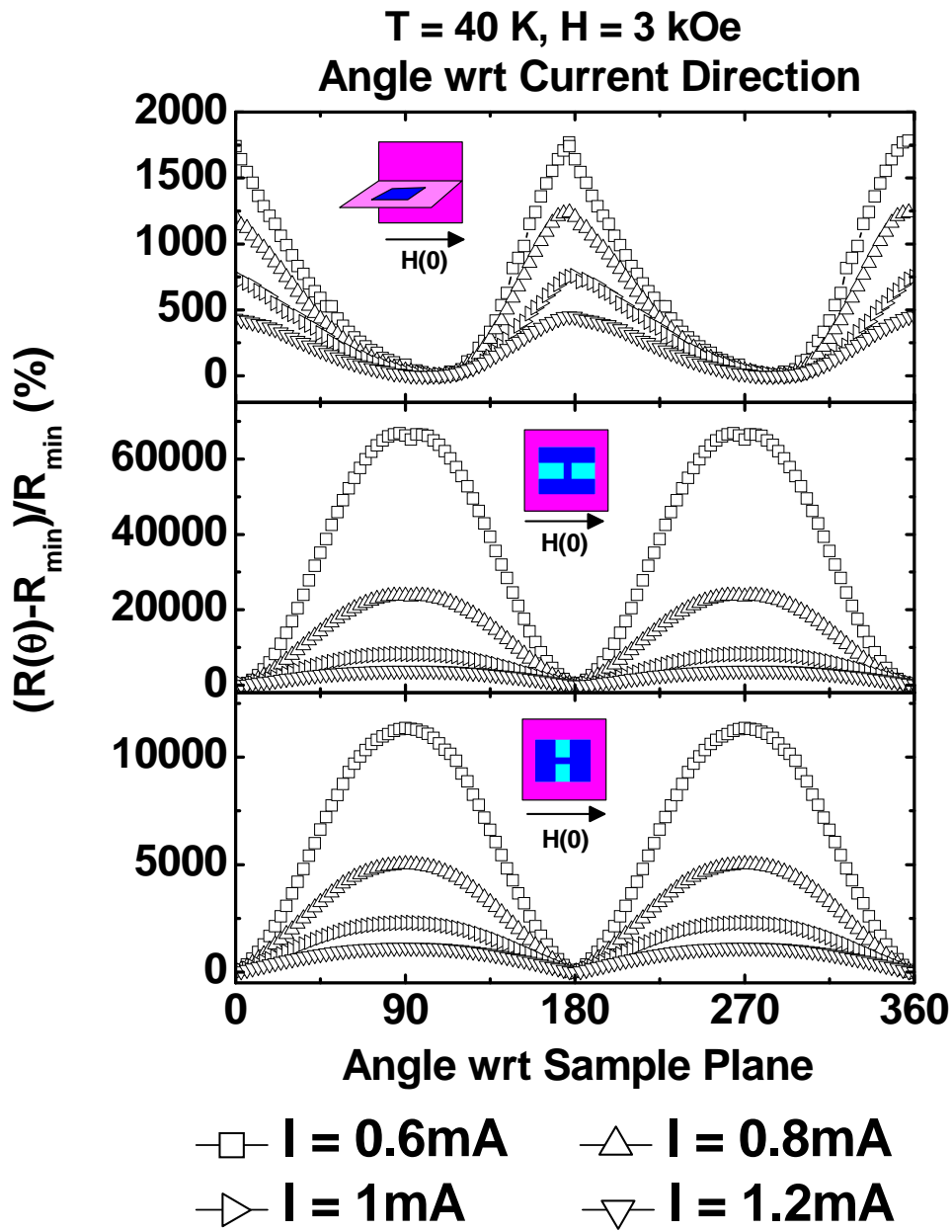}} \caption[Current dependence of AMR
measured at three different configurations.]{Current dependence of AMR measured at three different configurations. The AMR of the samples
increases as the current through the sample is decreased clearly proving the fact that a low resistance spacer enhances MR in these trilayers.}
\label{amri} \end{figure}

In fig. \ref{amri}  we have plotted the current dependence of the AMR for three different orientations of the sample as shown in the
figure. H(0) in the figure corresponds to the situation when the angle of the field with respect to the current or sample plane is zero. In the
top panel, the measurement geometry is the same as is shown in fig. \ref{tcom}(d). In the middle panel, the \cuo planes are perpendicular to
field when $\theta = 0$ and in the bottom panel the field always stays parallel to the \cuo planes. The top panel shows the dependence in the
coplanar configuration. We can see that as I is decreased AMR increases. For the same film, the middle and bottom panel show unusually large
AMR. It is to be noted that the AMR in the middle and bottom panel come from contributions from two different effects. First will be the AMR due
to field being parallel or perpendicular to the \cuo layer of the SC spacer and the second will be due to the out-of-plane field which gives
rise to a high resistance state. In the bottom panel, we can see that the applied field always stays parallel to the \cuo planes. The AMR seen
here essentially comes from the contributions of the field perpendicular to SC layer. In the middle panel, we can see that the AMR is much
higher than that of any other configuration. In case we assume that the contribution is only from the first case, as pointed out earlier, then
there should be no angular dependence in the bottom panel and if we assume that all the contribution is coming from the second case, then the
middle and bottom panels should show comparable AMR. The fact that the middle panel shows an AMR almost 6 times that of bottom panel points to
the fact that the AMR is coming from both the contributions which in this case, have same behavior in the configuration shown in middle panel
and hence the effect is additive. We can also see that in all the cases, reduction in the current results in an increase in the AMR vindicating
the hypothesis that low resistance in the spacer contributes to a higher magnetoresistance.

\subsection[(110) LSMO - YBCO and (110) YBCO thin films]{Anisotropic Magnetoresistance of (110) LSMO - YBCO and (110) YBCO thin films.}
\begin{figure} \centerline{\includegraphics[width=4in,angle=0]{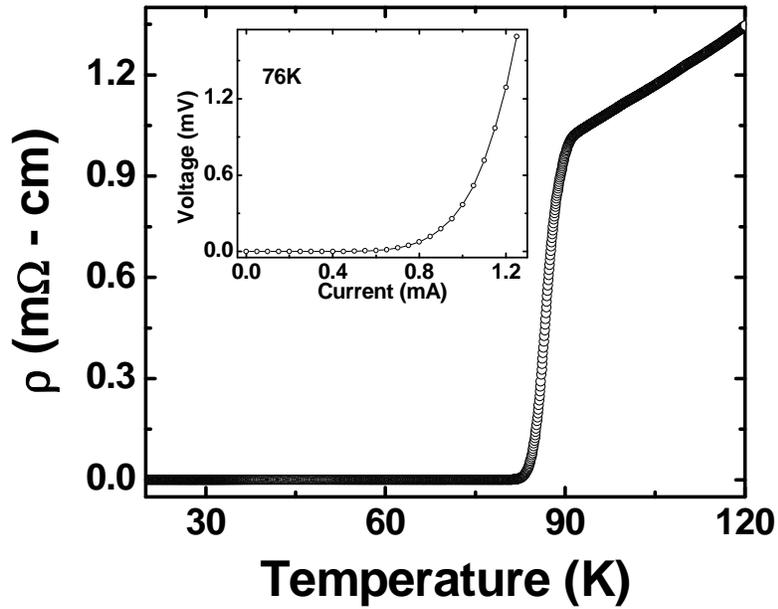}}
\caption[Resistivity data for a (110) LSMO - YBCO bilayer]{Resistivity data for a (110) LSMO - YBCO bilayer. The inset shows the current voltage
characteristic for the film at 76 K.} \label{rtivbi} \end{figure}

\begin{figure} \centerline{\includegraphics[width=4in,angle=0]{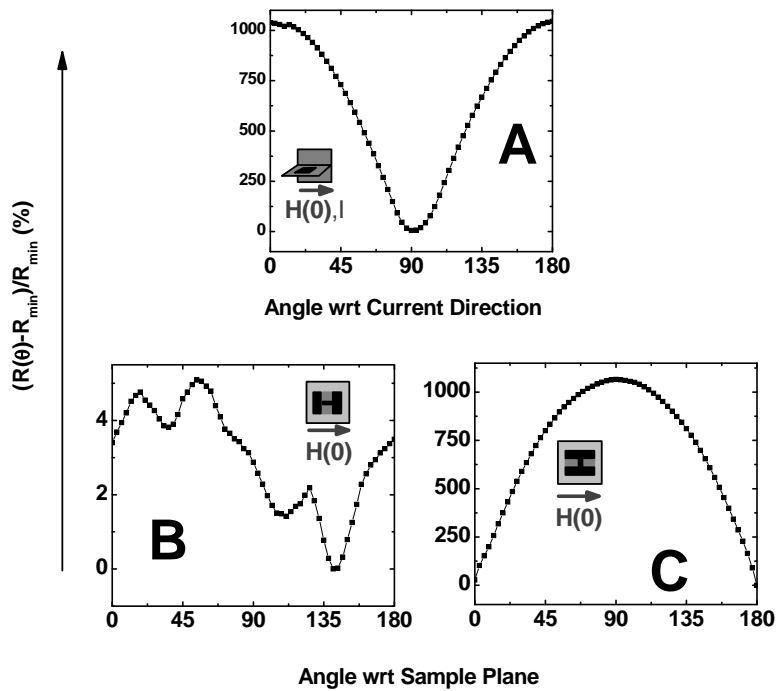}}
\caption[AMR for an LSMO-YBCO bilayer in three different configurations.]{AMR for an LSMO-YBCO bilayer in three different configurations. We can
see that the AMR in this case is almost two orders of magnitude smaller than that seen for the trilayer.} \label{amrb} \end{figure}

To verify our results of unusually high AMR in the trilayers we have done some control experiments involving an LSMO-YBCO bilayer and a YBCO
single layer film. The growth conditions for these films are exactly the same as the trilayer except for the fact that these films were made
with d$_{YBCO}$ = 1000{\AA}. Fig. \ref{rtivbi}
 shows the resistivity data for the bilayer. The inset shows a current voltage
characteristic for the bilayer at 76 K. In fig. \ref{amrb}, 
\begin{figure} \centerline{\includegraphics[width=3.8in,angle=0]{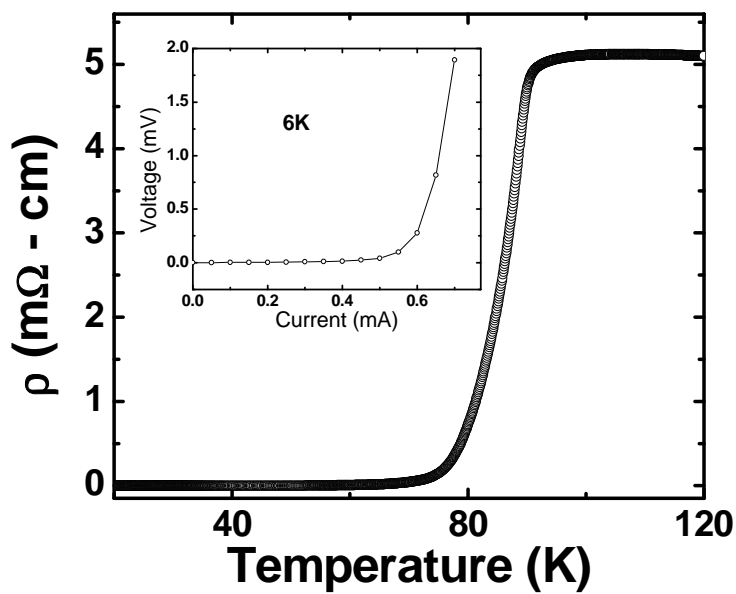}}
\caption[Resistivity data for a single (110) oriented YBCO layer]{Resistivity data for a single (110) oriented YBCO layer. The inset shows the
current voltage characteristic for the film at 6 K.} \label{rtivsl}
\end{figure} \begin{figure}
\centerline{\includegraphics[width=3.9in,angle=0]{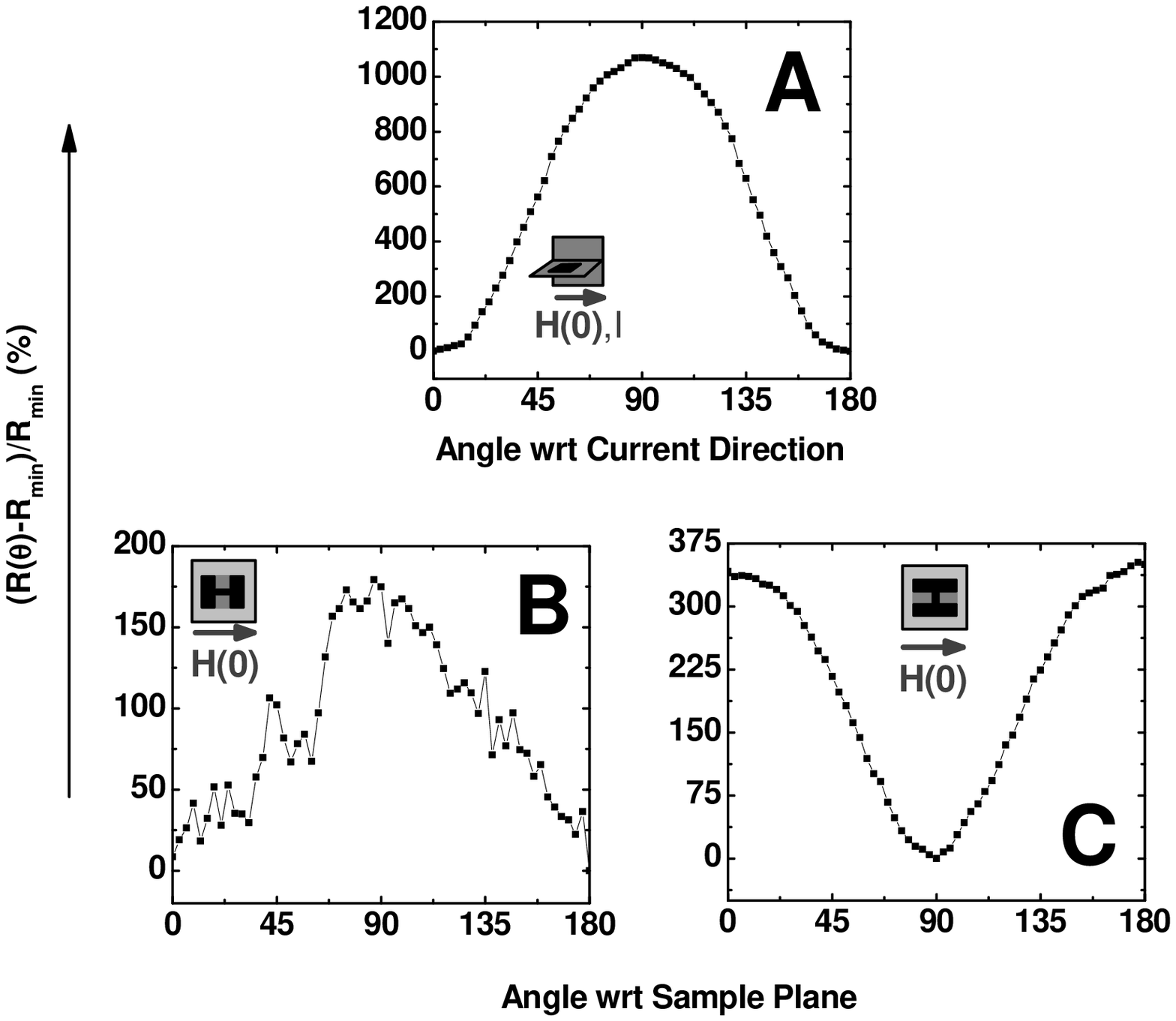}} \caption[AMR for (110) YBCO film in three different configuration.]{AMR
for (110) YBCO film in three different configuration. The AMR primarily arises from the orientation of magnetic field with respect to \cuo
planes.} \label{amrs} \end{figure} we have shown the AMR in this film at 3 kOe for three different configurations. We can see that AMR in this
film is much smaller as compared to that of the superconducting trilayer. Most of the contribution comes from the fact that the field moves from
parallel to perpendicular to the \cuo planes. The middle panel shows the AMR when the field stays in the \cuo plane. The dependence seen here
mostly comes from the fact that the field moves in and out of plane of the sample. In fig. \ref{rtivsl}, we have plotted the resistivity data
for a single layer of (110) YBCO. The inset shows the current voltage characteristic for the film at 6 K. Fig. \ref{amrs} shows the AMR for this
film at different configuration. Here we can clearly see that angular dependence comes primarily from the positioning of the field perpendicular
or parallel to the \cuo planes. But one look at panel C of figs. \ref{amrb} and \ref{amrs} tells us that the AMR for the bilayer is higher than
that for the single layer. This is explained by the presence of an FM magnetic layer near a superconducting layer. Earlier, people had seen a
higher change in the resistivity in a bilayer than on a single layer when the film was pushed into the superconducting state \cite{Petrashov}.

\section{Conclusion}
In conclusion, manganite - cuprate bilayers and trilayers where the \cuo planes are normal to the plane of the templated (110) \sto have been
synthesized and their various transport and magnetic properties have been studied. We find that the coupling between the two FM layers is higher
in this case than on that of the (001) bilayer as predicted by de Melo \cite{Melo}. We have also observed unusually high ($\sim72000\%$) angular
magnetoresistance in these trilayers. Some controlled experiments have been done to point out the fact that the unusually high AMR comes from
the coupling between the two ferromagnetic layers. The MR\% calculated from the magnetoresistance measurements shows a peak near the
superconducting transition temperature which has been attributed to the unusual increase in the normal state properties of the superconductor
near its transition temperature.

\chapter[LSMO-YPBCO-LSMO heterostructures]{\LSMO-\YPBCO-\LSMO heterostructures}
\section{Introduction}
In the introduction of this thesis, we discussed in detail the interesting physics issues of ferromagnet-superconductor (FM-SC) hybrids. The
simplest heterostructure that will show some of the exotic phenomena of FM-SC hybrids is a trilayer where a superconducting film is sandwiched
between two ferromagnetic layers. Interestingly, such a structure in the normal state of the SC also constitutes the well-known spin valve in
which two ferromagnetic layers sandwich a non-magnetic (NM) metallic spacer \cite{Binasch, Baibich, Dieny, book}. The giant negative
magnetoresistance seen in FM-NM-FM trilayers and multilayers is related to the asymmetric scattering of spin-up and spin-down electrons as they
criss-cross the spacer while diffusing along the plane of the heterostructure \cite{book}. This flow of spin polarized electrons is expected to
change profoundly when the spacer material becomes superconducting. Indeed, a large negative magnetoresistance has been observed by Pe\~na et
al. \cite{Pena2, Nemes} in La$_{0.7}$Ca$_{0.3}$MnO$_3$-\YBCO-La$_{0.7}$Ca$_{0.3}$MnO$_3$ trilayers in the narrow superconducting transition
region which they attribute to spin accumulation in YBCO when the FM layers are coupled antiferromagnetically. The accumulated spins presumably
cause depairing and hence a large resistance in accordance with the spin imbalance theory of Takahashi, Imamura and Maekawa \cite{spinimbalance,
spinimbalance2}.

The study of substitution of Y by Pr in YBCO has attracted a lot of attention because it leads to a non-trivial depression of T$_c$
\cite{Radousky, Xiong}. Although superconducting Y$_{1-x}$Pr$_x$Ba$_2$Cu$_3$O$_7$ (YPBCO) has the same orthorhombic structure as YBCO, yet in
contrast it retains the full stoichiometric oxygen content. By varying the Pr content, the superconducting properties can be influenced without
causing changes in the crystal symmetry. Hall effect studies on this system have shown a decrease in charge carriers as the Pr concentration is
increased suggesting a greater than +3 valency for Pr \cite{Matsuda}.  In another study, the Y$^{3+}$ ions in YPBCO were replaced by Ca$^{2+}$
ions, resulting in (Y$_{1-x-y}$Ca$_y$)Pr$_x$Ba$_2$Cu$_3$O$_7$, to create holes in the \cuo planes \cite{Neumeier}. The plot of T$_c$ vs. $y$ at
a fixed $x$ revealed that there are two contributions coming into play because of these substitutions:- i) the counteracting effects of the
generation and the filling of holes in the \cuo sheets by Ca$^{2+}$ and Pr$^{4+}$ ions respectively, and ii) the depairing of superconducting
electrons via exchange scattering of the mobile holes in the \cuo sheets by the Pr magnetic moments. But a later study by Fink et al.
\cite{Fink} found evidence against hole filling by Pr in YBa$_2$Cu$_3$O$_7$. Their electron energy loss spectroscopy study revealed that Pr does
not fill holes in the \cuo planes. It suppresses the mobility of holes leading to the suppression of superconductivity. Another interesting work
on Pr substitution is by Gaur et al \cite{Gaur}. Here they have tried to replace Ba with Pr. They found that the rate of T$_c$ suppression in
this case is much higher than that for Y substitution by Pr. They also reported a phase transition from orthorhombic to the tetragonal phase as
the Pr concentration is increased. Many more interesting results can be found on this system but the exact mechanism of T$_c$ suppression by Pr
is not yet clear. It is even more puzzling since the substitution of Y by other members of the same group from the periodic table, except for
Ce, Tb and Pr, does not affect T$_c$ \cite{Maple}.
\begin{figure} \centerline{\includegraphics[height=4in,angle=0]{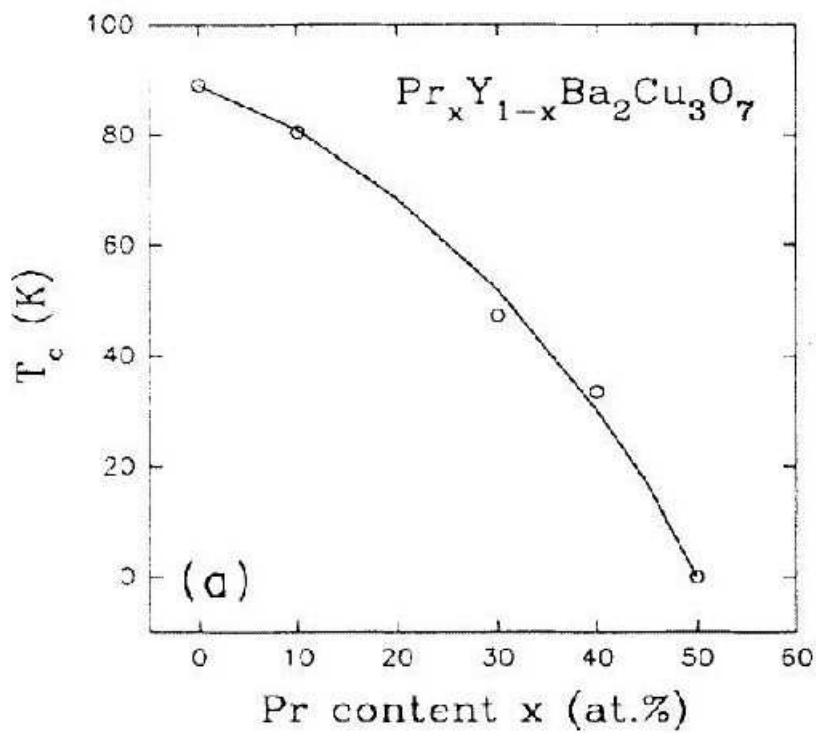}}
\caption[T$_c$ as a function of Pr concentration for YPBCO]{T$_c$ as a function of Pr concentration for Y$_{1-x}$Pr$_x$Ba$_2$Cu$_3$O$_7$
(Adapted from ref. \cite{Xiong})} \label{ypbcph} \end{figure}In this chapter we examine the relevance of pair breaking by the dipolar field and
injected spins in the low carrier density cuprate \YPBCO which has an insulating c-axis resistivity and hence a poor spin transmittivity. We
further address the issue of giant MR in three distinctly illuminating ways which involve:- i) a current density dependence of MR over a broad
range of temperature below T$_c$, ii) field dependence of MR when the magnetizations of LSMO layers $\vec{M_1}$ and $\vec{M_2}$ are parallel and
fully saturated, and iii) dependence of MR on the angle between the current and the field below and above the critical temperature. These
measurements permit the disentanglement of the contributions of flux flow and pair breaking effects in YPBCO, and the intrinsic anisotropic MR
of LSMO layers to GMR in FM-SC-FM trilayers, and further to establish a fundamental theorem which warrants diverging MR in the limit of
infinitely conducting spacer. In fig. \ref{ypbcph} we have shown the variation of T$_c$ with Pr concentration in the
Y$_{1-x}$Pr$_x$Ba$_2$Cu$_3$O$_7$ system. It is evident from the figure that for $x=0.4$ we have a weak superconductor with T$_c \sim$ 30 K. As
for single crystals \cite{Sandu}, T$_c$ decreases with Pr concentration and for $x \geq 0.55$, the system has an insulating and
antiferromagnetic ground state \cite{Radousky, Tournier, Lal, Felner, Cooke}. In short, we concentrate on the x = 0.4 film because of its low
carrier density and order parameter phase stiffness \cite{Peng}, both of which would enhance its susceptibility to pair-breaking by spin
polarized carriers injected from the LSMO.

\section{Results and Discussion}
Thin epitaxial trilayers of LSMO-YPBCO-LSMO were deposited on (001) SrTiO$_3$. A multitarget pulsed laser deposition technique based on KrF
excimer laser ($\lambda$ = 248 nm) was used to deposit the single layer films and heterostructures as described in our earlier work
\cite{Senapati1}. The thickness of each LSMO layer (d$_{LSMO}$) was left constant at 30nm whereas the d$_{YPBCO}$ was varied from 30 to 100nm.
The interfacial atomic structure of the trilayers was examined with high resolution transmission electron microscopy (TEM) at Brookhaven
National Laboratory, USA.

\begin{figure}
\centerline{\includegraphics[height=6in,angle=-90]{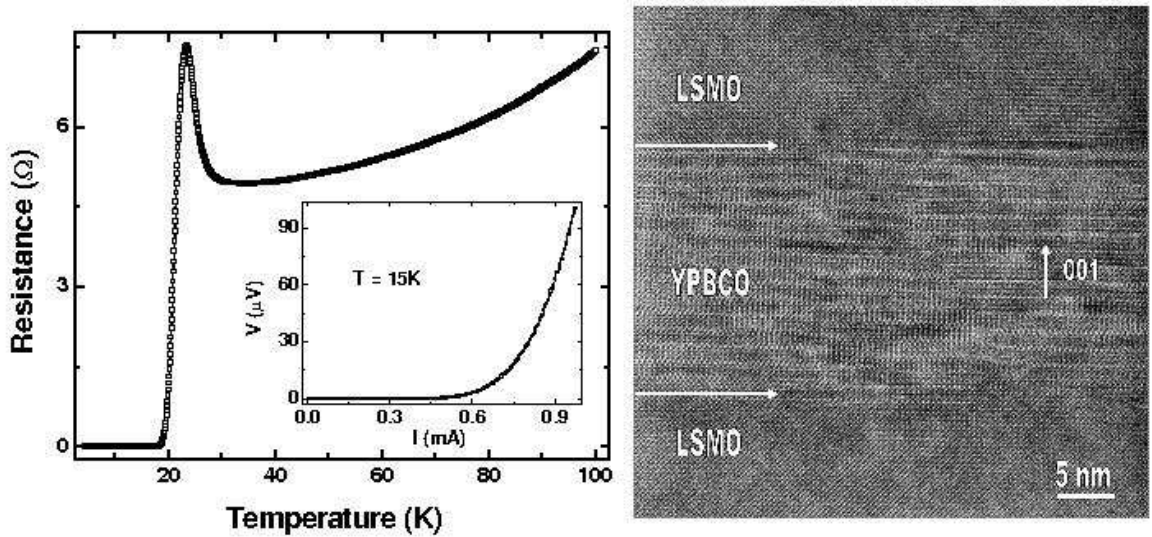}} \caption[The R vs. T curve \& HRTEM for an LSMO - YPBCO - LSMO trilayer]{The R
vs. T curve for an LSMO - YPBCO - LSMO trilayer with a 100 nm YPBCO sandwiched between two 30 nm LSMO layers is shown in the left panel. The
rise in resistance at 30 K suggests some structural disorder in the YPBCO film which presumably localizes the charge carriers before
superconductivity sets in at $\approx$ 24 K. The inset of the left panel shows a typical I vs. V characteristic of the trilayer at 15 K. The
right panel shows a high resolution cross-sectional TEM of the trilayer. A careful examination of the lattice image of the superconducting layer
shows the presence of stacking faults. \textit{Image taken by Y Zhu and group at BNL, Brookhaven, USA}} \label{rtsem}
\end{figure}

Fig. \ref{rtsem} shows the resistance R(T) of a trilayer where the Y$_{0.6}$Pr$_{0.4}$Ba$_2$Cu$_3$O$_7$ film thickness is $\approx 100$ nm and
the LSMO layers are each 30 nm thick. The R(T) curve is characterized by a steep increase in the resistivity near $\approx 30 K$ before the
superconductivity sets in at still lower temperatures. While the Y$_{1-x}$Pr$_x$Ba$_2$Cu$_3$O$_7$ film with $x \geq 0.5$ show superconductor -
insulator transition due to carrier localization with a $\rho(T)$ similar to that seen in fig. \ref{rtsem}, for the composition used here
(x=0.4), the resistivity is expected to remain metallic \cite{Covington1}. The semiconductor-like resistivity seen in fig. \ref{rtsem} in the
temperature window of $\rm{T}_c < \rm{T} \leq 30 \rm{K}$ is likely to be due to the structural disorder present in these sandwiched films, some
evidence of which comes from high resolution TEM imaging. The right panel of fig. \ref{rtsem} shows a cross-sectional transmission electron
micrograph of the heterostructure. We can clearly see a sharp interface between LSMO and YPBCO in the atomic resolution image. While the
manganite layers are free of growth defects, we do see distinct stacking faults in the YPBCO layers which can be related to the disorder due to
the difference in the ionic radii of Y$^{3+}$ and Pr$^{3+}$. The inset of fig.\ref{rtsem} shows a typical current (I) - voltage (V)
characteristic of the structure at 15 K. The critical current density (J$_c$) inferred from this measurement is $\sim 5 \times 10^3$ A/cm$^2$.
\clearpage

\begin{figure}[t]
\centerline{\includegraphics[height=14cm,angle=-90]{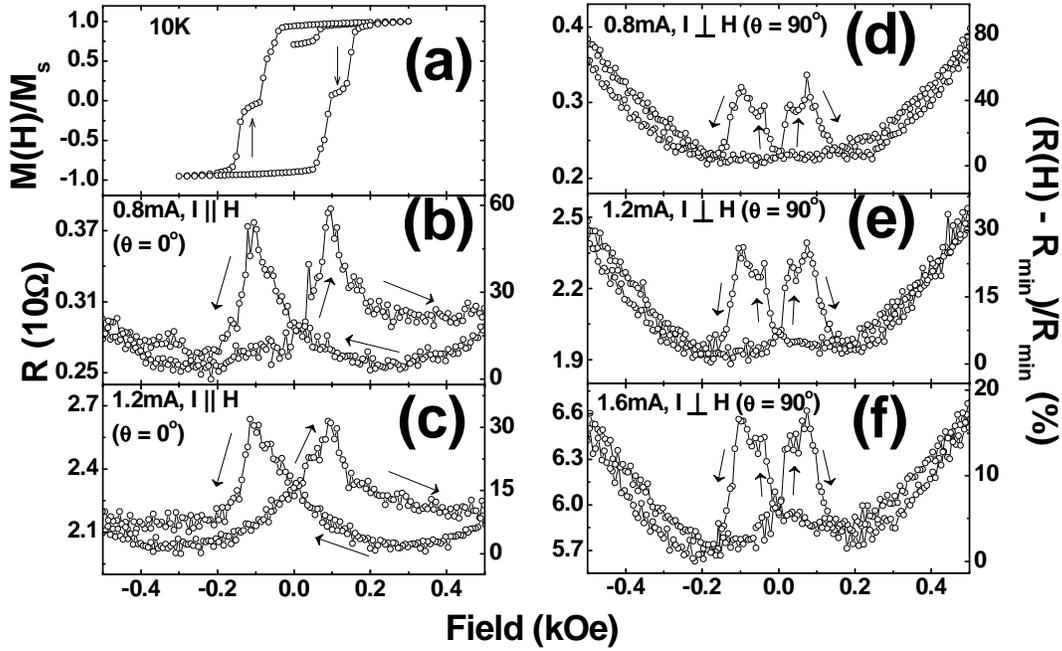}} \caption[M vs. H loop and MR data for LSMO - YPBCO - LSMO trilayer]{Panel
(a) shows the M vs. H loop of the trilayer taken at 10 K. The two symmetric small plateaus (indicated by arrows), with zero magnetization show
antiferromagnetic coupling between the two FM layers. Panels (b) to (f) show MR measured at 15 K. In panels (b) \& (c), the current was parallel
to the field with values 0.8 and 1.2 mA respectively, whereas for the remaining three panels, the in-plane field was orthogonal to the current
($\theta = 90^o$) which takes three vales; 0.8, 1.2 and 1.6 mA for (d), (e) and (f) respectively. The left hand side of the y-axis shows the
resistance in units of 10$\Omega$ and the right hand y-axis shows $(R(H) - R_{min})/R_{min}$ in percent.} \label{rh15k}
\end{figure} Fig. \ref{rh15k} (panel a) shows the magnetic field ($\vec{H}$) dependence of the magnetization ($\vec{M}$) at 10 K with $\vec{H}$ in
the plane of the heterostructure and parallel to the easy axis, (110) of LSMO. Starting from a fully saturated magnetic state at H$\simeq$ 180
Oe, the magnetization reaches a plateau over a field range of -90 to -130 Oe on field reversal. This is indicative of antiferrromagnetic (AF)
alignment of the magnetization vectors of the top and bottom LSMO films. The full cancellation of the moment (M$\simeq$0) seen in the plateau
also suggests that the two layers have equal saturation magnetizations (M$_s$). While a plateau in $\vec{M}$ symmetric about the y-axis can also
result due to a difference in the switching fields of the top and bottom layers, either because of a difference in their thicknesses or due to
the pinning of $\vec{M}$ by the substrate, a perfect cancellation of the moments at the plateau indicates an antiferromagnetic interlayer
exchange mediated by the cuprate spacer. We have demonstrated earlier that the poor c-axis conductivity of YBCO actually quenches the
oscillatory part of the interlayer exchange interaction and only an exponentially decaying AF-exchange remains in the LSMO-YBCO-LSMO system
\cite{Senapati}. In the remaining five panels of fig. \ref{rh15k}, we show the in-plane resistance of the trilayer as a function of $\vec{H}$
coplanar with the measuring current (I). Two values of the angle between I and H have been chosen; in one case  $\theta = 0^o$ (fig. \ref{rh15k}
b \& c) and for the other three panels (d, e \& f), $\theta = 90^o$ but the magnitude of I is different. While these measurements have been
performed at several currents, only a few representative field scans of the MR are shown in fig. \ref{rh15k}. The conventional way of measuring
the MR is to calculate the ratio $\Delta R/R_0$, where $\Delta R = R_H - R_0, R_0$ being the resistance at zero applied field and $R_H$ the
resistance when the applied field is H. Here we have used a slightly different definition. We have replaced R$_0$ by R$_{min}$, which is the
minimum resistance seen in R vs. H curves. The magnetoresistance for both $\theta = 90^o$ and $\theta = 0^o$ configurations has two distinct
regimes of behavior. Starting from a fully magnetized state at 500 Oe in the $\theta = 90^o$ configuration, the MR first drops to a minimum as
the field is brought to zero following a dependence of the type $\sim \alpha$H + $\beta$H$^2$, where $\alpha = -6.8 \times 10^{-6}$ and $\beta =
7.7 \times 10^{-8}$ for I = 0.8mA. The MR shows a step-like jump at the reversed field of $\sim40$ Oe where the magnetization switches to AF
configuration and remains high till $\vec{M_1}$ and $\vec{M_2}$ become parallel again. On reversing the field towards the positive cycle, a
mirror image of the curve is seen in the positive field quadrant. A remarkable feature of the MR seen in fig. \ref{rh15k} is its dependence on
the current I. The peak MR at 500 Oe and I $\bot$ H drops from $\sim$80\% to 17\% on increasing the current by a factor of two. The height of
the MR curves remain nearly the same when the magnetic field is rotated from $\theta = 90$ to $\theta = 0$ with some differences in the detailed
shape of the curves. The pertinent factors which affect the MR of such structures are:- i) the behavior of MR in the normal state of YPBCO, ii)
the explicit role of superconductivity which is suppressed by the dipolar and exchange fields of the FM layers and by the spin polarized
electrons injected from the FM layers, and iii) a parasitic non-zero tilt of the sample away from the parallel configuration which will result
in a high concentration of vortices in the superconducting spacer even at very low fields.
\begin{figure}[t] \centerline{\includegraphics[height=14cm,angle=90]{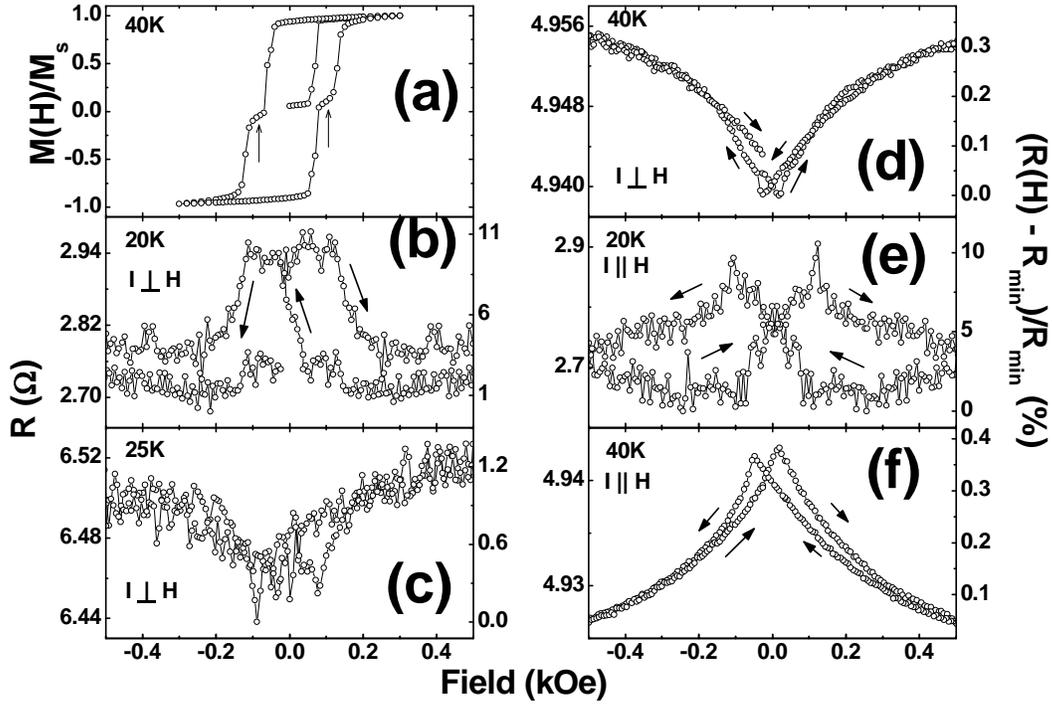}}
\caption[M vs. H loop at 40 K and MR data at several T.]{(a) M vs. H loop for the trilayer taken at 40 K. Two symmetric small plateaus
(indicated by arrows) with zero magnetization show antiferromagnetic coupling between the two FM layers. Panels (b), (c) \& (d) show the field
dependence of the magnetoresistance in $\theta = \pi/2$ ($\vec{I} \bot \vec{H}$) configuration at 20, 25 and 40 K respectively. Panels (e) \&
(f) show the field dependence of the magnetoresistance in $\theta = 0$ ($\vec{I} \| \vec{H}$) configuration at 20 and 40 K respectively. The
left hand side of the y-axis shows the resistance ($\Omega$) and the right hand y-axis shows $(R(H) - R_{min})/R_{min}$ in percent.}
\label{rhvart} \end{figure} These factors are addressed with the help of fig. \ref{rhvart} where we have plotted the M(H) loop at 40 K (normal
state). The overall shape of this curve is not different from what is seen in the SC state (fig. \ref{rh15k}) except for a temperature dependent
change in the switching fields and M$_s$. The AF alignment of $\vec{M_1}$ and $\vec{M_2}$ in the vicinity of the zero-field persists in the
normal state as well. Fig. \ref{rhvart} also shows the field dependence of magnetoresistance in  $\theta = 90^o$ and $\theta = 0^o$
configurations at a few representative temperatures as the sample is taken from the superconducting to the normal state. A striking drop in the
MR on approaching the normal state is evident in addition to a noticeable change in its field dependence. At 40 K and $\theta = 90^o$, it drops
monotonically on reducing the field from full saturation till the reverse switching field is reached where it shows a small but discernible
step-like increase followed by an unremarkable field dependence in the negative field side. For the I$\|$H configuration (fig. \ref{rhvart}(f))
the R(H) curve is an inverted image of fig. \ref{rhvart}(d) reflecting the anisotropic magnetoresistance (AMR) of LSMO films. The 20 K data of
figs. \ref{rhvart} (b) \& (e) show that the MR values are similar with the exception that the antiferromagnetic regime is clearly seen for I
$\bot$ H.

It becomes clear from figs. \ref{rh15k} and \ref{rhvart} that the field dependence of the MR in these FM-SC-FM trilayers can be divided into two
field regimes, one covering the range  -150 Oe $<$ H $<$ 150 Oe where the reorientation of $\vec{M_1}$ and $\vec{M_2}$ is the deciding factor
and the other at higher fields where $\vec{M_1} \| \vec{M_2}$ and it goes as $\sim \alpha$H + $\beta$H$^2$. While the MR in these regimes is
intimately linked with the superconductivity of YPBCO, its mechanism appears to be remarkably different. We first concentrate on the low-field
regime where we define the MR as (R$_{\uparrow\downarrow}$ - R$_{min}$)/ R$_{\uparrow\downarrow}$ where R$_{\uparrow\downarrow}$ and R$_{min}$
are the resistances of the trilayer when $\vec{M_1}$ and $\vec{M_2}$ are antiparallel, i.e. the plateau region and R$_{min}$ is the minimum
resistance as defined earlier respectively. \begin{figure}[t] \centerline{\includegraphics[width=4.5in,angle=0]{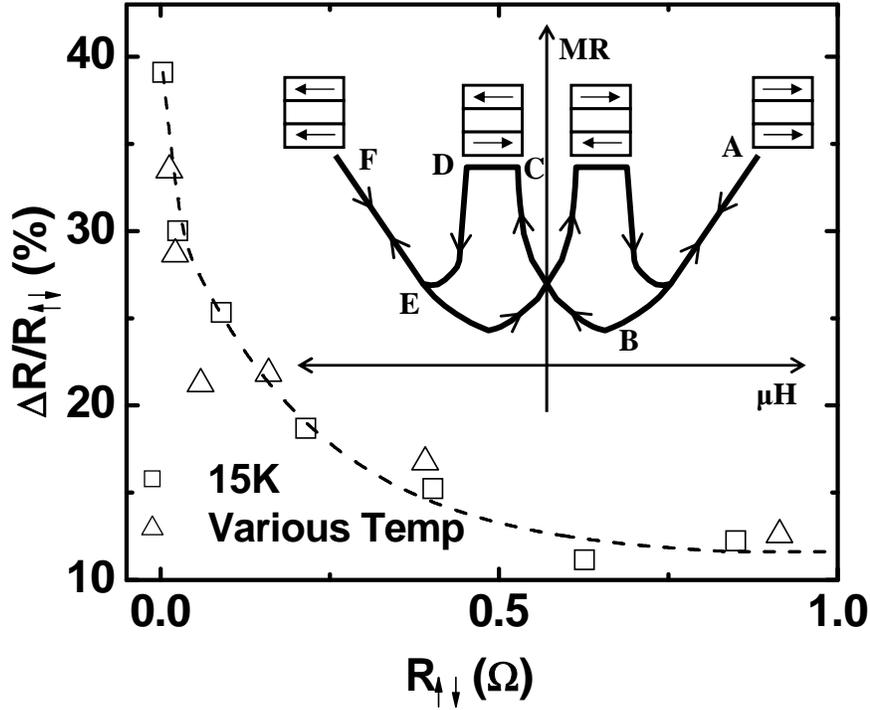}} \caption[Dependence
of MR on R$_{\uparrow\downarrow}$ for LSMO-YPBCO-LSMO trilayers]{Dependence of MR on R$_{\uparrow\downarrow}$. This figure contains the MR data
collected at 15 K with variable current and at several temperatures across the transition at constant current. A remarkable universality of the
dependence of MR on the ground state resistance of the structure emerges. The inset shows a typical sketch of the MR vs. H curve in the
superconducting state and identifies some critical points where $\vec{M_1} \& \vec{M_2}$ change their orientation (details in the text).}
\label{mrap}\end{figure} The variation of MR with R$_{\uparrow\downarrow}$ at a fixed temperature (15 K) with variable current and at several
temperatures across the transition at constant current is shown in fig. \ref{mrap}. A remarkable universality of the dependence of MR on the
ground state resistance of the structure emerges. The magnetoresistance starts with a negligibly small value at T $>$ T$_c$ but then diverges on
entering the superconducting state. While an enhancement in the MR has been seen in spin valves of cleaner spacers \cite{book, Binasch,
Baibich}, the regime of diverging MR is only accessible with a superconducting spacer. Unlike the case of free--electron--like metal spacers,
where the strength of the MR is attenuated by spin flip scattering processes in the interior of the spacer and at the spacer - ferromagnet
interfaces \cite{book, Levy}, the physics of transport of spin polarized carriers in FM-SC-FM structures is much more challenging. Here we
identify various factors which can contribute to MR and then single out the ones which perhaps are truly responsible for the behavior seen in
fig. \ref{mrap}. In the inset of fig. \ref{mrap} we sketch a typical MR vs. H curve at T $<$ T$_c$ and identify some critical points on the
curve where the orientation of $\vec{M_1}$ and $\vec{M_2}$ and the effective magnetic field seen by the SC layer change significantly. We first
consider the behavior of the MR in a field regime very close to the origin in fig. \ref{mrap}(inset). For the AF configuration of $\vec{M_1}$
and $\vec{M_2}$ (point C), the dipolar field in the spacer cancels out but for the parallel alignment  (point B) it adds up. Thus, strictly from
the angle of pairbreaking by the dipolar field, the SC layer should have a lower resistance in the AF configuration. Moreover, a much stronger
effect of the exchange field of ferromagnetic layers on superconductivity when $\vec{M_1}$ and $\vec{M_2}$ are parallel should make the AF state
less resistive \cite{Gennes}. Both these effects are inconsistent with the observation of a higher resistance in the AF state. However, before
we rule out the effects of the dipolar field altogether in influencing MR, a careful examination of the MR curve along the path A $\rightarrow$
B $\rightarrow$ C $\rightarrow$ D $\rightarrow$ E $\rightarrow$ F of the inset traced on reducing the field from the parallel alignment of
$\vec{M_1}$ and $\vec{M_2}$ needs to be made. At point B the dipolar field of ferromagnetically aligned $\vec{M_1}$ and $\vec{M_2}$ in the
superconductor completely cancels out the positive external field leading to a minimum in resistance. As the field is reduced to zero and then
made negative (between points B and C) the net field seen by the superconductor increases. At the negative coercive field H$_c$, just before
$\vec{M_1}$ and $\vec{M_2}$ become antiparallel (point C), the internal field in the SC is $B_{\uparrow\uparrow} = -\mu_oH_{ext} - \mu_o(m_1^d
+m_2^d)$, where $m_1^d$ and $m_2^d$ are the dipolar contributions to the magnetization in the superconductor. However, just beyond H$_c$, i.e.
{$\mid\rm{H}\mid > \rm{H}_c$} in the AF state, the internal field (B$_{\uparrow\downarrow}$) is only $\mu_oH_{ext}$ (assuming $m_1^d \sim
m_2^d$). While this sudden reduction in B$_{int}$ at H$_c$ could be responsible for the plateau (segment CD) seen in the R(H) vs. H curve in the
AF state, the higher resistance in the AF state still remains a puzzle. Although one could attribute it to the piling up of spin polarized
quasiparticles in the SC spacer, such an interpretation would require a deeper understanding of the c-axis transport in these structures where
the \cuo planes are parallel to the magnetic layers. The observation of this effect in a low carrier density cuprate of the present study is
much more intriguing because its c-axis resistivity is insulator-like in the normal state \cite{Tournier}. The precipitous drop in resistance
from point D to E also points towards the critical role of the net internal field in the SC layer and its influence on the MR  because at point
D, the $\vec{M_1}$ and $\vec{M_2}$ vectors switch to a parallel configuration leading to an additive dipolar field in the superconductor
pointing 180$^o$ away from the direction of the external field.

\begin{figure} [t] \centerline{\includegraphics[width=12cm,angle=0]{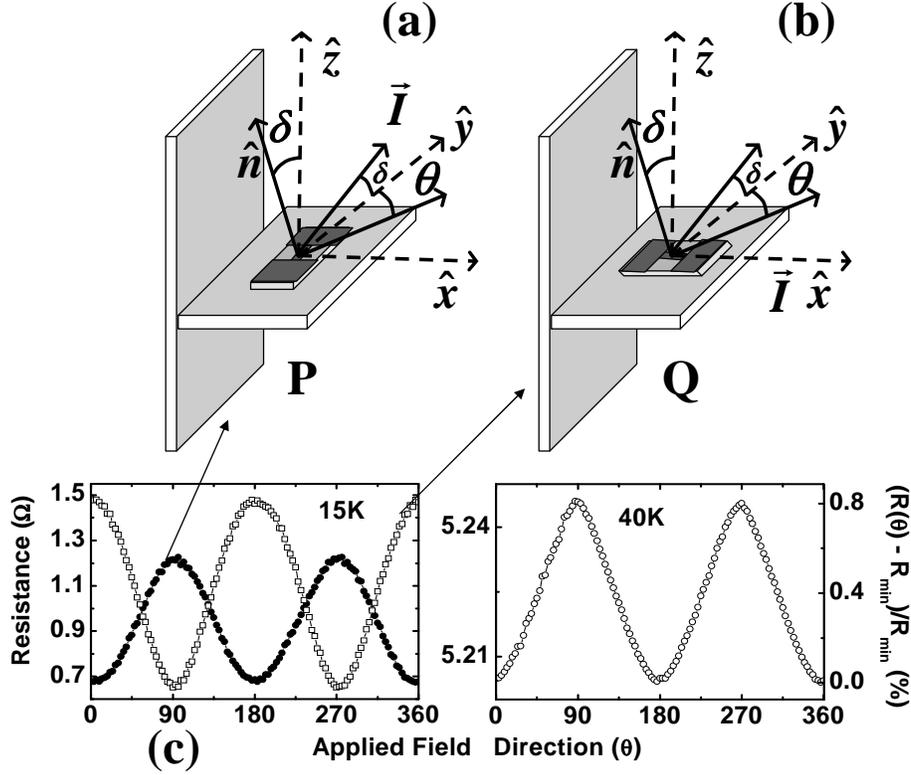}} \caption[Sample mounting in the cryostat and its comparison with
AMR data.]{(a) \& (b) respectively show the configurations P and Q of the sample mounting in the cryostat. The sample stage has a non-zero tilt
($\delta$) with respect to the x-y plane. (c) shows the  AMR of the trilayer measured at 15 K in configurations P and Q and at 40 K in
configuration P.} \label{tilt} \end{figure}

We now discuss the large positive magnetoresistance in the ferromagnetic configuration of $\vec{M_1}$ and $\vec{M_2}$. The field dependence of
MR in this regime derives contributions from the pair-breaking effects of spin polarized electrons injected from LSMO and of the net field seen
by the YPBCO. Moreover, a parasitic normal component of the field due to misalignment will introduce vortices and a large dissipation due to
flux flow. Here a small negative contribution to R is also expected due to the intrinsic MR of LSMO which would vary as $M^2$. We have estimated
the contribution of the sample tilt by measuring its resistance in two configurations P and Q as shown in figs. \ref{tilt} (a\&b).
 We assume that the platform on which the film is mounted for measurement, instead of being on the
x-y plane, has a small tilt $\delta$ away from the y-axis. In configuration P, the sample is mounted in such a manner that the stripe of film
along which the current flows is nominally along \^y. Fig. \ref{tilt}(b) shows the 90$^o$ geometry such that the stripe is now along \^x. This
is labeled as configuration Q. We rotate $\vec{H}$ in the xy-plane and measure R as a function of the angle $\theta$ between \^y and the field
direction. We expect three distinct contributions to R($\theta$) coming from:- i) vortex dissipation due to the normal component of the field
[($\Delta$R)$_{\upsilon\bot}$], ii) the Lorentz force on Josephson vortices in the plane of the film [($\Delta$R)$_{\upsilon\|}$], and iii) the
anisotropic magnetoresistance of the LSMO layers ($\Delta$R)$_{AMR}$ which peaks when I is perpendicular to the in-plane field \cite{Suzuki}.
While all these contributions to R are periodic in $\theta$ with a periodicity of $\pi$, in configuration P, [($\Delta$R)$_{\upsilon\bot}$] will
peak at $\theta = 0$ and $\pi$ whereas the peaks in [($\Delta$R)$_{\upsilon\|}$] and ($\Delta$R)$_{AMR}$ will appear at $\theta = \pi/2$ and
$3\pi/2$. Since the resistivity of the sample in configuration P peaks at $\pi/2$ and $3\pi/2$ (see fig. \ref{tilt}(c)), it is evident that
($\Delta$R)$_{\upsilon\bot} <$ (($\Delta$R)$_{\upsilon\|} + $ ($\Delta$R)$_{AMR}$). For configuration Q on the other hand,
($\Delta$R)$_{\upsilon\bot}$, ($\Delta$R)$_{\upsilon\|}$ and ($\Delta$R)$_{AMR}$ are all in phase with the peak value appearing at $\theta = 0$
and $\pi$ as seen in Fig. \ref{tilt}(c). Clearly, the difference in the peak heights at $\theta$ = 0 of Q and $\theta$ = $\pi/2$ of P gives us
the flux flow resistance due to the motion of vortices which nucleate because of a non-zero tilt. Its contribution to the resistance is $\sim$
10\% at 15 K and a 3 kOe nominally parallel field. Of course, its strength will also vary with the current. It is clear that a much larger
contribution to +ve MR comes from the in-plane field and its attendant effects. \clearpage

\section{Conclusion}
In summary, we have seen an exceedingly large magnetoresistance in a \LSMO - Y$_{0.6}$Pr$_{0.4}$Ba$_2$Cu$_3$O$_7$ - \LSMO  trilayer in the
superconducting transition region of the cuprate. The significant feature of these results is the divergence of the MR as the resistance of the
spacer goes to zero. We identify the key contributing factors to the MR in the superconductor based spin valves. These are:- (i) dipolar and
exchange fields in the SC layer, (ii) depairing by the accumulation of spin polarized electrons in the superconductor in the antiferromagnetic
state of the spin valve, and (iii) the contribution of vortex motion to resistance. We establish that the dipolar field of the LSMO layer in the
superconductor plays a crucial role in setting the scale of the MR in the field regime where the magnetization vectors of the FM layers switch
from an antiparallel to a parallel configuration.

\chapter{Conclusion and Future Scope}
This concluding chapter contains all the key results of this thesis along with the identification of some relevant issues which could be
addressed in the future. In this thesis, we have successfully studied the magnetic and transport properties of two polytypes of \LSMO [(001) and
(110) oriented epitaxial films], (110) oriented \LSMO-\YBCO-\LSMO trilayers, and (001) oriented \LSMO-\linebreak \YPBCO-\LSMO trilayers. From
our literature survey, we find that the research in FM-SC-FM hybrids involving cuprates and manganites is very interesting but in most studies
the \cuo planes are parallel to manganite layer which does not permit the direct injection of spin polarized carriers from the manganite into
the superconducting \cuo planes of the cuprate. We have been successful in preparing hybrids where the \cuo planes are perpendicular to the
manganite layer thus enabling us to study the injection of spin polarized carriers in the superconducting planes of the cuprate. Our main focus
in this thesis was on the hybrids of \LSMO and \YBCO. The absence of a detailed study of the relative magnetic anisotropy in (001) and (110)
LSMO prompted us to have a look into this problem before pursuing the case of (110) hybrids. We have successfully worked out the H-T phase space
for both polytypes of LSMO, defining the pinned and the depinned states. With this knowledge of the LSMO anisotropies, we addressed the
galvanomagnetic properties of (110) LSMO-YBCO-LSMO hybrids where we find an unusually high MR when the applied field is rotated in the plane of
the sample. In the last part of the thesis, we have investigated the galvanomagnetic properties of (001) LSMO-YPBCO-LSMO hybrids. A comparison
of the results shows that the MR in the case of (110) hybrids is an order of magnitude higher than that of (001) hybrids. For the (001) hybrids,
we were able to show that the MR in these systems is directly related to the spacer layer resistance. Below we give a detailed description of
our key results in this thesis.
\section[Anisotropic Magnetoresistance in LSMO]{Anisotropic Magnetoresistance in \LSMO (001) and (110)}
We have carried out a comparative study of the isothermal magnetoresistance of (001) and (110) oriented epitaxial \LSMO films as a function of
the angle between the current  and the coplanar magnetic field at several temperatures between 10 and 300K. The magnetic easy axes of the (001)
and (110) films are along the (110) and (001) directions respectively. In view of their similar texture, which could otherwise contribute to
shape anisotropy, we conclude that the easy axis of magnetization is fundamentally related to the orientation of the Mn-O-Mn bonds on the plane
of the substrate. The isothermal resistance $\rho_\bot$ and $\rho_\|$ for $\vec{I} \bot \vec{H}$ and $\vec{I} \| \vec{H}$ configurations
respectively of these two types of films obeys the inequality $\rho_\bot > \rho_\|$ for all fields and temperatures. However, $\rho(\theta)$
shows a deviation from the simple $\cos^2\theta$ dependence at low fields due to the pinning of the magnetization vector $\vec{M}$ along the
easy axis. This effect manifests itself as a discontinuity in $\rho(\theta)$ at some $\theta > \pi/2$ and a concomitant hysteresis on reversing
the angular scan. We establish a magnetization reorientation phase transition in this system and extract the H-T phase space where $\vec{M}$
remains pinned. A robust pinning of the magnetization seen in (110) films suggests strong in-plane anisotropy as compared to the (001) films. We
have carried out a full fledged analysis of the rotational magnetoresistance of the two types of epitaxial LSMO films in the framework of the
D\"oring theory \cite{Doring} of anisotropic magnetoresistance in metallic ferromagnet single crystals. We have found that for (110) films, the
dependence is mostly on even powers of $\cos\psi$, where $\psi$ is the angle between the  magnetization and the (001) axis. For the same kind of
analysis, the (001) samples showed a $\cos\psi\sin\psi$ type of dependence with measurable weightage on the even powers of $\cos\psi$ as well.
In this case, $\psi$ is the angle between the applied field and the easy axis of the sample and $\psi = \theta - \pi/4$ where $\theta$ is the
angle between the applied field and the current direction. We note that a strong deviation from the predicted angular dependence exists in the
irreversible regime of magnetization. A simple estimation of the orbital MR in these films suggest that the RMR is dominated by spin-orbit
interaction dependent anisotropic magnetoresistance.

\section[LSMO-YBCO-LSMO heterostructures]{\LSMO-\YBCO-\LSMO heterostructures}
\subsection[Growth recipe]{Growth Recipe}
We have been successful in synthesizing (110) oriented bilayers and trilayers of \linebreak \LSMO and YBa$_2$Cu$_3$O$_7$. To the best of our
knowledge, this constitutes the first successful attempt by any group to synthesize such structures where the copper oxide planes are
perpendicular to the plane of the heterostructure. While the growth of (110) oriented YBCO films has been achieved by many groups in the past,
the synthesis of (110) heterostructures consisting of aan LSMO layer on top of the YBCO is quite non-trivial. The optimized growth conditions
for the bilayers paved the way for the synthesis of (110) LSMO - YBCO - LSMO trilayers on which a variety of transport and magnetic property
measurements were performed. The bilayers of LSMO-YBCO thus deposited were examined by four circle x-ray diffractometer to calculate the volume
fraction of (110) grains. The result shows $\sim$65\% of the grains are (110) with the rest being (103) oriented. The superconducting transition
temperature of these films were found to be $\sim$90 K. Measurement of the critical current density revealed a strong suppression in the case of
films where the YBCO was in the proximity of an LSMO layer.

\subsection[Results on (110) trilayers]{Transport and Magnetic Properties of (110) trilayers}
We have measured the magnetic and transport properties of (110) trilayers. We find that the coupling between the two FM layers is higher in the
case of the (110) trilayer than the (001) trilayer as predicted by de Melo \cite{Melo}. This result is a key result which has validated
explicitly a long standing prediction in cuprate-manganite hybrids. We have also observed an unusually high ($\sim72000\%$) angular
magnetoresistance in these trilayers. We have done some controlled experiments to point out the fact that this unusually high AMR comes from the
coupling between the two ferromagnetic layers. The MR\%, defined as $\Delta R / R(0)$ where $\Delta R = (R_{\uparrow\downarrow})_{max} -
(R_{\uparrow\uparrow})_{min}$, R(0) is the resistance at zero field at that temperature, $(R_{\uparrow\downarrow})_{max}$ is the resistance at
the peak position in the MR-H curve and $(R_{\uparrow\uparrow})_{min}$ is the minimum resistance of the segment of the MR-H curve where the
magnetizations of both the FM layers are parallel to each other, calculated from the magnetoresistance measurements, shows a peak near the
superconducting transition phase which can be attributed to the unusual increase in the normal state properties of the superconductor near the
transition temperature of the superconductor.

\section[LSMO-YPBCO-LSMO heterostructures]{\LSMO-\YPBCO \linebreak -\LSMO heterostructures}
In this chapter of the thesis, we have studied magnetotransport and magnetic coupling in $c$-axis oriented trilayers. Here we concentrate on
films of \YPBCO because of its low carrier density and order parameter phase stiffness \cite{Peng}, both of which would enhance its
susceptibility to pair-breaking by spin polarized carriers injected from the LSMO. We have seen an exceedingly large magnetoresistance in \LSMO
- Y$_{0.6}$Pr$_{0.4}$Ba$_2$Cu$_3$O$_7$ - \LSMO  trilayer in the superconducting transition region of the cuprate. The salient feature of these
results is the divergence of the MR as the resistance of the spacer goes to zero. We identify the key contributing factors to MR in
superconductor based spin valves. These are (i) dipolar and exchange fields in the SC layer, (ii) depairing by the accumulation of spin
polarized electrons in the superconductor in the antiferromagnetic state of the spin valve, and (iii) the contribution of vortex motion to the
resistance. We establish that the dipolar field of the LSMO layer in the superconductor plays a crucial role in setting the scale of the MR in
the field regime where the magnetization vectors of the FM layers switch from antiparallel to parallel configuration. \clearpage

\section{Scope for further research}
The results presented in this thesis, specially those in chapters IV and V open several new possibilities in the study of manganite-cuprate
heterostructures. Some interesting issue which could be studied in the near future are;

\begin{enumerate}
\item The recipe developed in Chapter 4 for the growth of (110) manganite -- cuprate hybrids reveals that under proper growth conditions, it is
possible to grow these structures with a majority of grains being oriented in the direction of interest. We have only done a limited study on
these hybrids with fixed thicknesses for the FM and SC layers. A careful study of these hybrids by varying the FM and SC layer thickness may
reveal the nature of exchange coupling present in this system. Earlier studies have revealed oscillatory critical current in $c$-axis oriented
trilayers. It needs to be seen if this is true for (110) hybrids as well.

\item Extension of the ideas presented in Chapter 4 can also be applied to YPBCO systems. It needs to be seen if a stable recipe can be evolved
for growing (110) oriented LSMO-YPBCO-LSMO structures. Once possible, similar studies as done on (110) LSMO-YBCO-LSMO systems can be carried out
by varying the Pr concentration in the YPBCO layer.

\item The results presented in Chapter 5 show that the MR is directly dependent on the spacer resistance. Scope exists for similar studies with
varying spacer layer thickness and Pr concentration.

\item Apart from these, the effect of putting average bandwidth manganite such as; La$_{1-x}$Ca$_x$MnO$_3$ in these systems may lead to some interesting
results.
\end{enumerate}

\begin{singlespace}

\end{singlespace}


\begin{thebibliography}{}
\bibitem{Easy}
The easy axis of both [110] and [001] oriented films were determined by measuring M
vs H, by applying fields at different angles with respect to a known
crystallographic direction in the plane of the film.
\end{thebibliography}
\end{document}